\pdfoutput=1

\documentclass[a4paper,11pt]{article}
\usepackage{jheppub}

\usepackage{graphicx}
\usepackage{placeins}

\usepackage{xcolor}

\usepackage{amsmath}
\usepackage{amssymb}
\usepackage{bm}
\usepackage{slashed}
\usepackage{xspace}

\usepackage{fontawesome5}

\usepackage{multirow}
\usepackage{array}
\newcolumntype{L}[1]{>{\raggedright\let\newline\\\arraybackslash\hspace{0pt}}m{#1}}
\newcolumntype{C}[1]{>{\centering\let\newline\\\arraybackslash\hspace{0pt}}m{#1}}
\newcolumntype{R}[1]{>{\raggedleft\let\newline\\\arraybackslash\hspace{0pt}}m{#1}}

\newcommand{\linkCode}{https://github.com/sbaum90/AIMforGW}
\newcommand{\TMN}{``$5 \times \left(\text{MAGIS-1\,km}\right)$''\xspace}

\title{Gravitational Wave Measurement in the Mid-Band with Atom Interferometers}

\author[a,b,c]{Sebastian~Baum,}
\author[a,b]{Zachary~Bogorad,}
\author[a,b]{and Peter~W.~Graham}

\affiliation[a]{Stanford Institute for Theoretical Physics, Stanford University, Stanford, CA 94305, USA}

\affiliation[b]{Kavli Institute for Particle Astrophysics and Cosmology, Department of Physics, \\ Stanford University, Stanford, CA 94305, USA}

\affiliation[c]{Institute for Theoretical Particle Physics and Cosmology, RWTH Aachen University, \\ D-52056 Aachen, Germany}

\emailAdd{sbaum@physik.rwth-aachen.de}
\emailAdd{zbogorad@stanford.edu}
\emailAdd{pwgraham@stanford.edu}

\abstract{
Gravitational Waves (GWs) have been detected in the ${\sim}\,100\,$Hz and nHz bands, but most of the gravitational spectrum remains unobserved. A variety of detector concepts have been proposed to expand the range of observable frequencies. In this work, we study the capability of GW detectors in the ``mid-band'', the ${\sim}\,30\,{\rm mHz} - 10\,$Hz range between LISA and LIGO, to measure the signals from and constrain the properties of ${\sim}\,1-100\,M_\odot$ compact binaries. We focus on atom-interferometer-based detectors. We describe a Fisher matrix code, \texttt{AIMforGW}, which we created to evaluate their capabilities, and present numerical results for two benchmarks: terrestrial km-scale detectors, and satellite-borne detectors in medium Earth orbit. Mid-band GW detectors are particularly well-suited to pinpointing the location of GW sources on the sky. We demonstrate that a satellite-borne detector could achieve sub-degree sky localization for any detectable source with chirp mass $\mathcal{M}_c \lesssim 50\,M_\odot$. We also compare different detector configurations, including different locations of terrestrial detectors and various choices of the orbit of a satellite-borne detector. As we show, a network of only two terrestrial single-baseline detectors or one single-baseline satellite-borne detector would each provide close-to-uniform sky-coverage, with signal-to-noise ratios varying by less than a factor of two across the entire sky. We hope that this work contributes to the efforts of the GW community to assess the merits of different detector proposals.
}

\preprint{TTK-23-24}

\begin{document} 
\maketitle
\flushbottom

\section{Introduction} \label{sec:Introduction}
In the years since the first direct discovery of a gravitational wave (GW) signal~\cite{LIGOScientific:2016vlm}, observations in the 20\,Hz -- few kHz regime have become almost routine, with more than 90 signals from mergers of black holes and neutron stars observed to date~\cite{LIGOScientific:2018mvr,LIGOScientific:2020ibl,LIGOScientific:2021usb,KAGRA:2021vkt}. In the nHz regime, Pulsar Timing arrays are observing growing evidence of a stochastic GW signal~\cite{NANOGrav:2020bcs,Goncharov:2021oub,Chen:2021rqp,Antoniadis:2022pcn,NANOGrav:2023gor,EPTA:2023fyk,Reardon:2023gzh,Xu:2023wog}. In the next decades, various collaborations aim to extend the frequency coverage of GW detectors from nHz to kHz and above, with the greatest sensitivity possible, similarly to what has been achieved in the electromagnetic spectrum, where observatories operate from the $10\,$MHz range up to gamma-ray detectors that have observed PeV photons. Achieving a broad GW frequency coverage will require a variety of different detector techniques: proposals include astrometry~\cite{Pyne:1995iy,Schutz:2010lmv,Book:2010pf,Klioner:2017asb,Moore:2017ity,Park:2019ies,Wang:2020pmf,Fedderke:2022kxq,Wang:2022sxn}, ranging between asteroids~\cite{Fedderke:2021kuy}, studying orbital perturbations of astrophysical binaries~\cite{Blas:2021mqw,Blas:2021mpc}, future atomic~\cite{Dimopoulos:2007cj,Hogan:2011tsw,Canuel:2017rrp,Graham:2017pmn,Zhan:2019quq,Tino:2019tkb,AEDGE:2019nxb,Badurina:2019hst,MAGIS-100:2021etm} or laser~\cite{TianQin:2015yph,TianQin:2020hid,LISA:2017pwj,Armano:2016bkm,Baker:2019nia,Maggiore:2019uih,Reitze:2019iox,Sesana:2019vho,Kawamura:2020pcg} interferometers on Earth or in space, atomic clocks in space~\cite{Kolkowitz:2016wyg,Alonso:2022oot}, lunar GW detectors~\cite{LSGreport,Paik:2009zz,Coughlin:2014boa,LGWA:2020mma,LSGA,Jani:2020gnz}, and various types of terrestrial detectors with coverage at higher (MHz -- GHz) frequencies; see, e.g., Ref.~\cite{Aggarwal:2020olq} for a recent review.

In this work, we focus on the frequency band between the mHz range targeted by the future space-based LISA laser interferometer mission and the kHz band where terrestrial laser interferometers such as the LIGO-VIRGO-KAGRA (LVK) network operate; we will refer to this $f_{\rm GW} \sim 30\,{\rm mHz} - 10\,$Hz frequency range as the ``mid-band''. We study the ability of mid-band GW detectors to observe the signal from and reconstruct the parameters of compact binaries such as binary black hole, binary neutron star, or black hole-neutron star systems. This detailed study of GW detectors in the mid-band serves as a starting point for studies of the physics case of such detectors, e.g., to perform population studies of compact binary coalescences or for measuring the expansion history of our Universe via various types of ``standard siren'' approaches~\cite{Schutz:1986gp,Schutz:2001re,Holz:2005df,DelPozzo:2011vcw} (see also Refs.~\cite{Graham:2017lmg, Cai:2021ooo, Yang:2021xox, Liu:2022rvk, Yang:2022iwn} for related work). 

At the same time, this study allows us to explore trade-offs in design choices of future mid-band GW detectors, such as the number and location of multiple terrestrial detectors that could be operated as a network (similar to LVK), or between different orbits of space-based detectors. Understanding the relative advantages of these options is critical before mid-band GW detectors can be constructed, but has received little attention in the literature to-date.

While the ability to measure the signals from and constrain the properties of GW sources has been explored extensively in the mHz as well as in the kHz bands (see, for example, Refs.~\cite{Finn:1992xs,Cutler:1994ys,Poisson:1995ef,Balasubramanian:1995bm,Cutler:1997ta,Flanagan:1997kp,Pai:2000zt,Berti:2004bd,Arun:2004hn,Cornish:2005qw,Allen:2005fk,Vallisneri:2007ev,MockLISADataChallengeTaskForce:2007iof,Ajith:2009fz,Yunes:2009ke,MockLISADataChallengeTaskForce:2009wir,Babak:2012zx,LIGOScientific:2013yzb,Berry:2014jja,LIGOScientific:2016ebw,Nitz:2017svb,Thrane:2018qnx,Ashton:2018jfp,Liu:2020eko,Vitale:2020aaz,LISA:2022yao,Baghi:2022ucj,Christensen:2022bxb}), the mid-band has received much less attention~\cite{Nakamura:2016hna,Graham:2017lmg,Isoyama:2018rjb,Nair:2018bxj,Sedda:2019uro,Zhao:2021bjw,Cai:2021ooo,Liu:2021dcr,Yang:2021xox,Liu:2022mcd,Kang:2022nmz,GilChoi:2022nhs,Yang:2022iwn, Liu:2022rvk, Yang:2022fgp}. We consider signals from systems similar to those observed by LVK: compact binaries with (component) masses ranging from that of neutron stars, $\sim 1\,M_\odot$, to $\sim 100\,M_\odot$. A GW detector in the mid-band would observe the {\it inspiral} signal from such a binary, and the lifetime of the GW signals from such compact binaries in the mid-band ranges from months to $\sim 200\,$years. The long lifetime of signals lends GW detectors in the mid-band a unique capability for localizing the GW sources on the sky \cite{Graham:2017lmg}. 

Fundamentally, there are two methods to infer a GW source's sky-position: First, by measuring the GW signal with high signal-to-noise-ratio along several different spatial directions, the sky position can be inferred directly from the details of the GW signal. This method is not effective for most of the GW sources any detector will observe, however, as it requires extremely large signal-to-noise ratio signals for good localization.

Second, the sky-position can be inferred by effectively aperture-synthesizing a GW ``telescope''. This can be achieved by measuring the difference in arrival time of the signal between multiple detectors, or by measuring the signal in a single detector while that detector moves (in an accelerated way, or else it is merely a contribution to redshift) relative to the source. Both methods of aperture-synthesis yield an angular resolution given by the familiar diffraction limit, i.e., the ratio of the (GW) wavelength to the effective aperture of the ``telescope,'' combined with the Signal-to-Noise Ratio (SNR). For a network of GW detectors localizing GW sources via the difference of signal-arrival time between detectors, the effective aperture is approximately the distance between the detectors. Thus, for a network of terrestrial detectors in the kHz regime such as LVK, the effective aperture is comparable to Earth's diameter. The corresponding angular resolution is, for SNRs of order a few, tens of degrees. At frequencies below ${\sim}\,1$\,Hz, however, GW signals from ${\sim}\,1 - 100\,M_\odot$ compact binaries can be measured over months-to-years timescales. For a single detector moving on an (accelerated) trajectory while the GW signal is observed, the effective aperture is given by the size of that trajectory. Terrestrial or space-based detectors\footnote{In geocentric orbit or in a ${\sim}\,1\,$AU-radius heliocentric orbit, as proposed for virtually all GW detector concepts using man-made test masses.} complete a significant fraction of their (accelerated) paths around the Sun over these timescales, leading to an effective aperture size of order AU. In the mHz band, where LISA will operate, the corresponding GW wavelengths are so large that one again finds angular resolutions of tens of degrees for SNRs of a few. In the mid-band, however, signals are still sufficiently long-lived to allow a single detector to achieve order AU effective apertures, but the wavelength is much shorter than in the mHz band. Thus, GW detectors in the mid-band offer a unique possibility to localize all visible ${\sim}\,1 - 100\,M_\odot$ GW sources with sub-degree precision.

The qualitative arguments for unique science opportunities in the GW mid-band have been made before~\cite{Buchmueller:2023nll}, with first numerical results also shown in, for example, Ref.~\cite{Graham:2017lmg} by one of us. Here, we present a detailed study of the ability of GW detectors in the mid-band to infer the properties of ${\sim}\,1 - 100\,M_\odot$ GW sources. Of course, detectors in the mid-band could also observe GW signals from sources that do not reach the LVK band above ${\sim}\,10$\,Hz, including white dwarf binaries and black hole binaries with masses $\gtrsim 100\,M_\odot$. For these sources, the merger event itself could be observed in the mid-band. We leave considerations of such signals for future work.

While our numerical results are derived using sensitivity curves from proposed ``MAGIS'' detectors, our results are not specific to these detector concepts but readily carry over to any other atom-interferometry based GW detector (e.g.~AEDGE~\cite{AEDGE:2019nxb, Bertoldi:2021rqk}, AION~\cite{Badurina:2019hst}, MIGA~\cite{Canuel:2017rrp}, SAGE~\cite{Tino:2019tkb}, VLBAI~\cite{Schlippert:2019hzx}, or ZAIGA~\cite{Zhan:2019quq}) with quantitative adjustments corresponding to changes in the sensitivity curves. We expect many of our results to carry over to other detector techniques in the mid-band, e.g. laser-interferometer proposals such as DECIGO~\cite{Kawamura:2020pcg}, although beyond the changes stemming from different sensitivity curves there are additional differences to atom interferometers from, e.g., DECIGO envisaging Michelson-type interferometers rather than a single-baseline detectors, and heliocentric orbits lacking the fast (i.e., hour-timescale) detector reorientation timescale common to the proposed terrestrial or medium Earth orbit atom-interferometer detectors. 

The remainder of this paper is organized as follows: in Section~\ref{sec:MidbandGWDetection}, we briefly review the GW sources, signals, and detectors in the mid-band. Section~\ref{sec:ParameterReconstruction} presents the Fisher Matrix framework we use in this work to explore the capabilities of future atom-interferometer GW detectors to measure the GW signal from and reconstruct the parameters of compact binary coalescences. In Sec.~\ref{sec:ResultsDiscussion}, we present numerical results for the ability of atom-interferometer GW detectors in the mid-band to measure the signals from and constrain the properties of compact binaries in the ${\sim}\,1 - 100\,M_\odot$ mass range. We focus on three particular quantities: the signal-to-noise ratio, the ability to measure the luminosity distance of the GW source, and the ability to measure the sky-localization of the GW source. We present results for these quantities in two planes: chirp mass vs luminosity distance (Sec.~\ref{sub:ChirpMassLumiDistScaling}), and sky-maps spanned by the right ascension and declination of the GW source (Sec.~\ref{sub:SkyLocationScaling}). Our results demonstrate the unique potential of GW detectors in the mid-band to localize essentially all visible GW sources in the sky with sub-degree precision. In Sec.~\ref{sub:ObservationTimeScaling} we present additional results demonstrating the early-warning ability of GW detectors in the mid-band to predict the sky-location and the merger time hours to days before the merger, enabling electromagnetic telescopes to train their sights on the merger event. We reserve Sec.~\ref{sec:Conclusion} for our conclusions. 

Appendix~\ref{app:Signal} presents a more detailed review of the GW signal from coalescing compact binaries. We then discuss atom-interferometer GW detectors in more detail in App.~\ref{app:Detectors}. Additional numerical results, and a discussion thereof, are collected in Appendix~\ref{app:SupplementalPlots}. 

We make the numerical code used to produce the numerical results in this work, \texttt{AIMforGW}, publicly available here\footnote{\href{\linkCode}{\url{\linkCode}}}: \href{\linkCode}{\faGithubSquare}. This is, to our knowledge, the first publicly-available code designed to perform Fisher matrix forecasting in the mid-band, and has been optimized with this frequency band in mind. Moreover, unlike at least some of Fisher forecast codes previously used in the mid-band~\cite{Graham:2017lmg, Cai:2021ooo, Yang:2021xox, Yang:2022iwn}, \texttt{AIMforGW} calculates the detector strain signal in the time-domain, allowing us to avoid simplifying assumptions about how to include effects from (changing) Doppler shifts and time-delays of the GW signal due to detector motion. 

\section{Gravitational Waves and Detectors in the Mid-band}\label{sec:MidbandGWDetection}

In this section, we provide a brief introduction to the sources, signals, and detectors relevant to the detection of gravitational waves in the mid-band. For more detailed discussions of these topics, see Appendix \ref{app:Signal} (for a discussion of sources and signals) and Appendix \ref{app:Detectors} (for a discussion of atom interferometer-based detectors), as well as the references therein.

We begin by summarizing some key properties of the gravitational wave signals seen by detectors in the mid-band. In this work, we focus on signals produced by the inspiral phases of their source binaries, such that they are well described by the leading terms of the post-Newtonian expansion of General Relativity; this restricts us to source chirp masses $\mathcal{M}_c \lesssim 300\,M_\odot$ for our space-based atom-interferometer GW detector benchmark and $\mathcal{M}_c \lesssim 50\,M_\odot$ for our terrestrial detector benchmark.  Our numerical computations employ 3.5/3.0 PN order (frequency evolution/amplitude correction) waveforms; see, for example, Refs.~\cite{Maggiore:2007ulw,Buonanno:2009zt,Blanchet:2013haa,Mishra:2016whh,Isoyama:2020lls} for the corresponding expressions. This section, however, will present only leading-order expressions, which are almost always sufficient for a qualitative understanding of our results. Note that we neglect the possibility of non-zero spins of binary components and non-zero orbital eccentricity in this work, although we hope to return to these in future work. Subject to these assumptions, the signals we consider are fully described by the nine parameters in Table \ref{tab:Parameters}.

\begin{table}
    \centering
    \renewcommand{\arraystretch}{1.1}
    \begin{tabular}{R{4cm} L{3cm} | C{4.5cm}} 
        \multicolumn{2}{c |}{Parameter} & Benchmark value \\ \hline\hline
        Chirp Mass: & $\mathcal{M}_c = \frac{(m_1 m_2)^{3/5}}{(m_1 + m_2)^{1/5}}$ & --- \\ \hline
        Mass ratio: & $q = m_1 / m_2$ & 1.15 \\ \hline
        Luminosity distance: & $d_L$ & --- \\ \hline
        Binary phase at $t_{\rm ref}$: & $\Phi_0$ & $0$ \\ \hline
        Time of merger: & $t_c$ & solar equinox \\ \hline
        Inclination angle: & $\iota$ & $45^\circ$ \\ \hline
        Polarization angle: & $\psi$ & $60^\circ$ \\ \hline
        Right ascension: & $\alpha$ & $60^\circ$ \\ \hline
        Declination: & $\delta$ & $6.6^\circ$ ($30^\circ$ above ecliptic)
    \end{tabular}
    \caption{The nine parameters used to describe GW sources in this work. The benchmark values listed are the values used for the numerical results presented in this work unless a different value is specified explicitly. Note that $\Phi_0 = \Phi(t_{\rm ref})$ is the phase of the GW signal at a reference time, see text below Eq.~\eqref{eq:Amplitude}.}
    \label{tab:Parameters}
\end{table}

The time evolution of a GW's frequency is given by (see, for example, Refs.~\cite{Cutler:1994ys, Poisson:1995ef, Holz:2005df})
\begin{align}
    \frac{d f_{\rm GW}}{dt} &= \frac{96}{5}\pi^{8/3}\mathcal{M}_c^{5/3}f_{\rm GW}^{11/3} \;, \label{eq:FrequencyEvolution} 
\end{align}
where
\begin{equation}
    \mathcal{M}_c \equiv \frac{(m_1 m_2)^{3/5}}{(m_1 + m_2)^{1/5}}
\end{equation}
is the chirp mass for source binaries with component masses $m_{1,2}$. Note that we use detector-frame masses throughout this work, unless noted otherwise.

The GW's amplitudes for the $+$ and $\times$ polarizations (precisely defined in Appendix \ref{app:Signal}), meanwhile, are given by
\begin{equation} \begin{split}
    \label{eq:Amplitude}
    h_+(t) &= \frac{2 \mathcal{M}_c^{5/3} \pi^{2/3} f_{\rm GW}(t)^{2/3}}{d_L} \left[ 1 + \left(\hat{\bm L} \cdot \hat{\bm n}\right)^2 \right] \cos\Phi(t) \;, \\
    h_\times(t) &= \frac{4\mathcal{M}_c^{5/3} \pi^{2/3} f_{\rm GW}(t)^{2/3}}{d_L} \left( \hat{\bm L} \cdot \hat{\bm n} \right) \sin\Phi(t) \;,
\end{split} \end{equation}
where $\Phi(t) = \int_{t_{\rm ref}}^t dt'~2\pi f_{\rm GW}(t') + \Phi_0$ is the GW phase with $\Phi_0 = \Phi(t_{\rm ref})$ the phase at some reference time $t_{\rm ref}$, ${\bm n}$ is the sky position of the source, and ${\bm L}$ is the angular momentum of the binary. We use heliocentric equatorial coordinates in this work, such that this sky position is described by a luminosity distance $d_L$, a right ascension $\alpha$, and a declination $\delta$. We will usually prefer to work in terms of an angle $\iota$, defined by $\cos\iota = \hat{\bm L} \cdot \hat{\bm n}$, with $\psi$ the (usually uninteresting) second angle specifying $\hat{\bm L}$.

\begin{figure}
    \centering
    \includegraphics[width=0.7\linewidth]{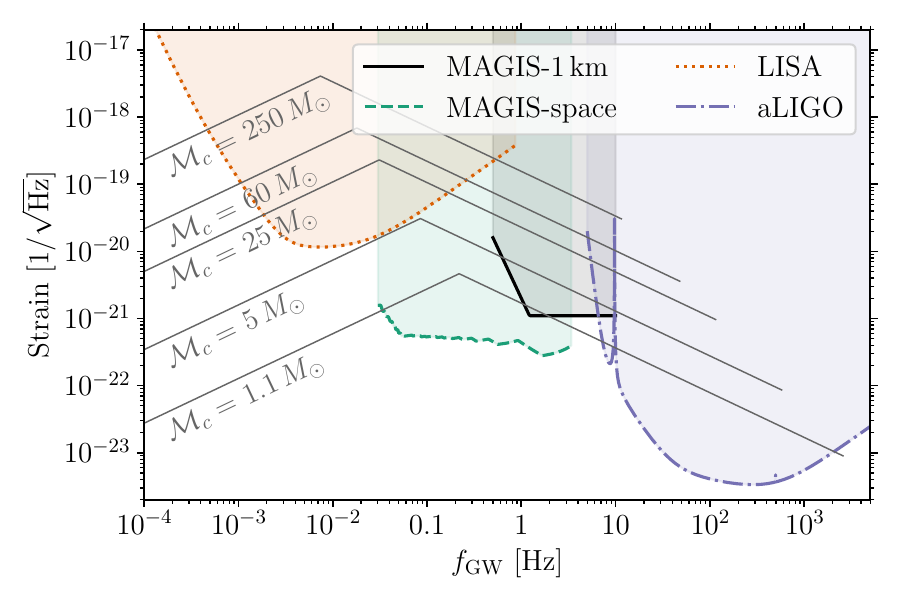}
    \caption{Projected strain sensitivity to GW signals for the two atom-interferometer detector benchmarks considered in this work: a terrestrial detector with atom sources separated by 1\,km (``MAGIS-1\,km'', adapted from Ref.~\cite{MAGIS-100:2021etm}) and a satellite-based detector in medium Earth orbit (``MAGIS-space'', adapted from Ref.~\cite{Graham:2017pmn}). For comparison, we also show the projected sensitivity curves for two laser-interferometers: the future satellite-based LISA detector~\cite{Babak:2021mhe}, and the terrestrial aLIGO detector~\cite{aLIGO:2018}. The gray lines show the signal strength [defined as $2 |\widetilde{h}(f)| \sqrt{f}$, see Eq.~\eqref{eq:ScalarProduct}] for the five benchmark choices of the chirp mass used throughout this work; we set the luminosity distance of all five sources to $100\,$Mpc (note that we will later use different distances for our benchmark sources). The turnover in the slope of the signals strengths comes from us restricting the maximal observation time of the GW signal to $1\,$yr; i.e., the position of the peak in signal strength marks the GW frequency one year before merger. We terminate the line of each signal at $f_{\rm GW}^{\rm max} = 2 \nu_{\rm ISCO} = \frac{1}{6\sqrt{6} \pi} \frac{c^3}{G M}$, where $M = \mathcal{M}_c \left[ \left(1+q^2\right)/q \right]^2$ and we use our benchmark value of $q=1.15$; this is a rough estimate of the merger frequency as discussed in Sec.~\ref{app:Signal}.}
    \label{fig:NoiseCurve}
\end{figure}

The resulting GW strains of several example sources are plotted (in the frequency domain rather than the time domain, for visual clarity) in Fig. \ref{fig:NoiseCurve}. One additional subtlety is reflected in this figure however: finite observation time. Fig. \ref{fig:NoiseCurve}, like almost all of the discussion in this work, considers signals observed for no more than one year before their merger. This reduces the observed strain at low frequencies, which occurs primarily more than one year before the merger, leading to the turnovers visible in Fig. \ref{fig:NoiseCurve}.

We now turn to summarizing some key features of the detectors considered in this work. For concreteness, we focus on atom-interferometer based GW detectors, although our results will qualitatively apply to any other detector technology in the mid-band. In a nut-shell, the idea of atom interferometers as GW detectors is to use two (or) more atom interferometers at different points in space to measure changes to the light-travel time of common laser pulses driving both interferometers~\cite{Dimopoulos:2007cj}. Conceptually, this idea is similar to using two atomic clocks to measure the light-travel time of laser pulses between the clocks as a means to measure their distance. Using atom interferometers driven by common laser pulses, rather than directly measuring light travel times between atomic (lattice) clocks, helps to suppress several sources of noise; see Appendix~\ref{app:Detectors}.

We will focus on two proposed detector benchmarks: first, we consider terrestrial vertical km-scale atom-interferometer GW detectors. Such detectors would be most sensitive in the frequency range of $\sim 1 - 10\,$Hz, with the sensitivity at frequencies below ${\sim}\,1\,$Hz limited by the effects of Newtonian Gravity Gradient noise. Note that since atom-interferometers use inherently free-floating atom clouds as test masses, they are not sensitive to vibrational noise from Earth's seismic activities or from human activity, which limits the sensitivity of terrestrial laser-interferometers such as LVK below ${\sim}\,10\,$Hz. We will pay particular attention to how well a network of multiple km-scale atom-interferometer based detectors on Earth could constrain the parameters of binary mergers compared to a single detector, and what the effect of possible locations of such detectors on Earth would have on the parameter reconstruction capabilities. For our numerical results, we will assume a sensitivity curve adapted from Ref.~\cite{MAGIS-100:2021etm}; see the line labeled ``MAGIS-1\,km'' in Fig.~\ref{fig:NoiseCurve}.

Second, we will consider a space-based atom interferometer detector comprised of two satellites in medium Earth orbit. Space-based detectors are not subject to significant Newtonian Gravity Gradient noise sourced by seismic waves, which limits the sensitivity of terrestrial detectors below $\sim 1\,$Hz. Instead, the lowest frequencies a space-based atom-interferometer GW detector will be sensitive to are controlled by the technically achievable free-fall time of the atom clouds. It appears achievable to have free-fall times on the order of tens of seconds or longer~\cite{Graham:2016plp, Hogan:2011tsw, Dimopoulos:2008sv, Dimopoulos:2007cj}, allowing a space-based detector to maintain its best strain sensitivity down to frequencies in the tens-of-mHz regime. Furthermore, a detector in space allows for much larger separation between the individual atom interferometers than terrestrial detectors which are limited by the length of (vertical) shafts on Earth. Since atom-interferometers (like most GW detectors) fundamentally measure the {\it position} or {\it acceleration} of test masses, the sensitivity to the GW {\it strain}, i.e.~the fractional change in the distance between the test masses, improves with larger test-mass separation\footnote{This holds as long as one remains in a regime where the error in the position/acceleration measurement does not increase linearly with the distance between the test masses. It would not hold, for example, in a laser interferometer if the separation between the test masses surpassed the Rayleigh length of a laser used for the distance measurement.}. For our numerical results, we will employ a sensitivity curve adapted from Ref.~\cite{Graham:2017pmn}; see the line labeled ``MAGIS-space'' in Fig.~\ref{fig:NoiseCurve}.

Finally, let us comment on the physical locations we assume for the detectors, controlling their antenna functions discussed in App.~\ref{subapp:AntennaFuns}. For the terrestrial ``MAGIS-1\,km'' benchmark, we consider five possible locations: the Homestake Mine in South Dakota, USA; Vale's Creighton Mine in Sudbury, Ontario, Canada; the Renstr\"{o}m mine near Boliden, Sweden; the TauTona Mine in Carletonville, South Africa; and Zaoshan Mountain, China, the proposed site of ZAIGA~\cite{Zhan:2019quq}. In all cases, we will assume that the detectors are oriented vertically, i.e., $\hat{\bm \ell}$ in Eq.~\eqref{eq:AntennaFun} points outwards seen from Earth's center. While we will study the sensitivity for different combinations of such detectors, many of our results will assume that detectors with the ``MAGIS-1\,km'' benchmark sensitivity are operated at each of these five locations. We will refer to this network as \TMN.

For the ``MAGIS-space'' detector benchmark, we will assume that two atom-in{\-}ter{\-}fer{\-}om{\-}e{\-}ters are operated on satellites in medium Earth orbit with a radius of $2 \times 10^4\,$km, corresponding to an orbital period of $7.8\,$h. We assume that the satellites are trailing each other in the same orbit, 130$^\circ$ apart, leading to a separation of the atom-interferometers of ${\sim}\,3.6 \times 10^4\,$km. We will focus on three choices for the orbits: in the plane of the ecliptic, $45^\circ$ inclined relative to the ecliptic plane, and $90^\circ$ inclined relative to the ecliptic plane.

\begin{table}
    \centering
    \renewcommand{\arraystretch}{1.1}
    \begin{tabular}{C{2.4cm} | C{2cm} | C{3.1cm} | C{2cm} | C{3.1cm}} 
        Chirp Mass & \multicolumn{2}{c|}{\TMN} & \multicolumn{2}{c}{``MAGIS-space''} \\
        $\mathcal{M}_c$ & lifetime & $d_{L}(\rho \sim 5)$ & lifetime & $d_{L}(\rho \sim 5)$ \\ 
        \hline
        \hline
        $1.1\,M_\odot$ & 40\,days & 30\,Mpc & 200\,yr & 100\,Mpc \\ \hline
        $5\,M_\odot$ &3\,days & 100\,Mpc & 16\,yr & 500\,Mpc ($z \sim 0.1$) \\ \hline
        $25\,M_\odot$ & 5\,h & 400\,Mpc ($z \sim 0.1$) & 1\,yr & 3\,Gpc ($z \sim 0.5$) \\ \hline
        $60\,M_\odot$ & 1\,h & 600\,Mpc ($z \sim 0.1$) & 3\,months & 5\,Gpc ($z \sim 1$) \\ \hline
        $250\,M_\odot$ & $7\,$min & --- & 8\,days & 20\,Gpc ($z \sim 2$) 
    \end{tabular}
    \caption{Lifetime in frequency band and largest luminosity distance to which GW signals from sources with different benchmark values of the chirp mass could be observed. For the ``MAGIS-1\,km'' benchmark, the ``lifetime'' is the time it takes the GW signal to evolve from $f_{\rm GW} = 0.5\,\text{Hz}$ to $f_{\rm GW} = 10\,$Hz, while for the ``MAGIS-space'' detector benchmark, ``lifetime'' refers to the time to evolve from $f_{\rm GW} = 30\,{\rm mHz}$ to $3\,$Hz. The largest luminosity distance, $d_L(\rho=5)$, is the distance at which a signal with all other parameters fixed to the benchmark values in Tab.~\ref{tab:Parameters} is observable with a signal-to-noise ratio of $\rho = 5$ assuming the sensitivity curves shown in Fig.~\ref{fig:NoiseCurve}, where for the \TMN benchmark we assume a network of five terrestrial detectors. Note that for the ``MAGIS-1\,km'' benchmark, we do not quote a $d_L(\rho \sim 5)$ value for the $\mathcal{M}_c = 250\,M_\odot$ source: such a binary would merge within the sensitivity band, hence, estimating the signal-to-noise ratio of the signal would require going beyond the inspiral GW waveforms we use in this work.}
    \label{tab:LifetimeHorizon}
\end{table}

\begin{figure}
   \includegraphics[height=6.5cm, trim=0cm 0cm 3.8cm 0cm, clip]{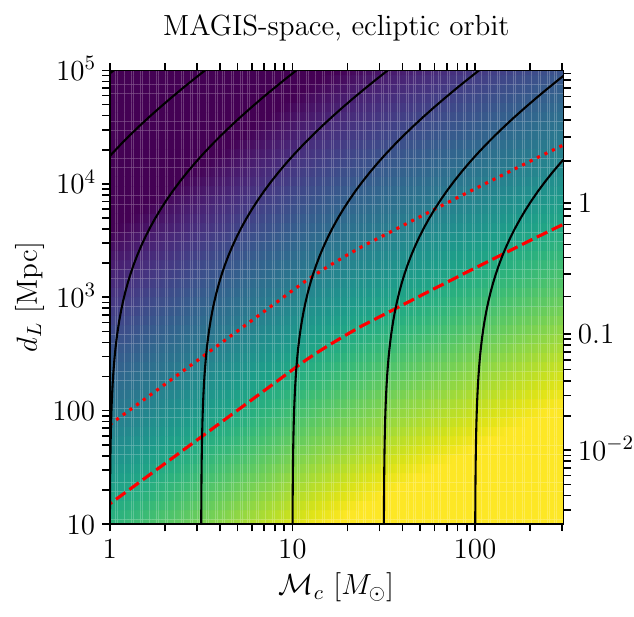}
   \hspace{0.3cm}
   \includegraphics[height=6.5cm, trim=0.9cm 0cm 0cm 0cm, clip]{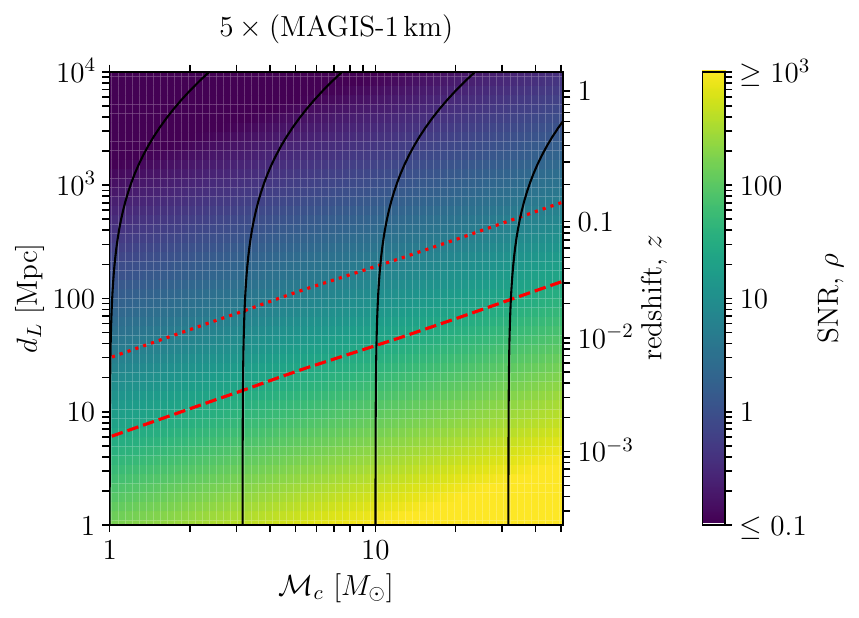}
   \caption{{\bf Signal-to-noise ratio (SNR)} in the plane of the (detector frame) chirp mass and luminosity distance of the GW source, all other parameters are set to the benchmark values given in Tab.~\ref{tab:Parameters}. The dotted (dashed) red lines are $\rho=5$ ($\rho=25$) contours of SNR. The right $y$-axes show the corresponding cosmological redshift, $z$, assuming a reference cosmology with a Hubble constant of $H_0 = 67.4\,$km/Mpc/s and a matter density of $\Omega_m = 0.315$. The solid black lines show lines of constant source-frame chirp-masses corresponding to the detector-frame chirp-masses shown on the $x$-axis. {\it Left:} Results for a single ``MAGIS-space'' detector in an ecliptic orbit. {\it Right:} Results for a network of five terrestrial ``MAGIS-1\,km'' detectors, located at Homestake, Sudbury, Renstr\"om, Tautona, and Zaoshan; see App.~\ref{app:Detectors}. Note that the range of $d_L$ shown on the vertical axes differs between the two panels.}
   \label{fig:ChirpMass_SNR_Scans}
\end{figure}

With these two benchmarks in mind, Tab. \ref{tab:LifetimeHorizon} illustrates the characteristic lifetimes (within the observable frequency band) of several example source masses for each benchmark. It also shows approximate luminosity distances for each source to reach an SNR of $\rho = 5$ (see Eq. \eqref{eq:SNR}). More generally, Fig. \ref{fig:ChirpMass_SNR_Scans} illustrates the SNR observed by each of the two benchmarks as a function of both chirp mass and luminosity distance, keeping the other source parameters set to the default values in Table \ref{tab:Parameters}.

\section{Parameter Reconstruction Method}\label{sec:ParameterReconstruction}
In Sec.~\ref{sec:FisherForm}, we summarize the Fisher-matrix approach to parameter estimation forecasts we use here; for a more complete discussion of this formalism, see, for example, Refs.~\cite{Cutler:1994ys, Finn:1992wt, Finn:1992xs}. In Sec.~\ref{sec:Code}, we discuss the technical implementation of this formalism in the numerical code we use to derive the results shown in this work, including some of the subtleties that arise when considering GW signals in the mid-band.

\subsection{Fisher Matrix Forecast}
\label{sec:FisherForm}

Let the time-stream strain signal seen by the detector be
\begin{align}
    s(t) = h(t, {\bm \lambda}) + n(t) \;,
\end{align}
where $h(t, {\bm \lambda})$ is the signal arising from a GW source parameterized by a set of parameters ${\bm \lambda}$ (see Tab.~\ref{tab:Parameters}) and $n(t)$ is the noise present in the detector. We will denote frequency-domain equivalents of time-domain functions with a tilde:
\begin{equation}
    \widetilde{h}(f) = \int_{-\infty}^\infty dt\,e^{2\pi i f t} h(t) \;.
\end{equation}
It is useful to characterize the detector's noise by its one-sided power spectral density (PSD), $S_h(f)$. Then, one can define the (noise-weighted) scalar product between two signals $r$ and $s$,
\begin{align} \label{eq:ScalarProduct}
    \left\langle r,s \right\rangle \equiv 2 \int_{-\infty}^{\infty} df\frac{\widetilde{r}(f)\widetilde{s}(f)^*}{S_h(f)} = 4{~\rm Re}\left[ \int_{0}^{\infty} df\frac{\widetilde{r}(f)\widetilde{s}(f)^*}{S_h(f)} \right] \,.
\end{align}
The latter expression holds for real-valued (time-domain) signals, e.g., strain signals. 

The squared signal-to-noise ratio (SNR) of a signal $h$ is
\begin{align} \label{eq:SNR}
    \rho^2 \equiv \left\langle h,h \right\rangle \;.
\end{align}
Let us stress than throughout this work, we call $\rho = \sqrt{\langle h,h \rangle}$ the SNR of a signal, not $\rho^2$.

The Fisher information matrix for a signal $h(t, {\bm \lambda})$ is obtained by taking the scalar products between derivatives of the signal with respect to the parameters ${\bm \lambda}$,
\begin{align}
    \Gamma_{ij} \equiv \left\langle \frac{\partial h}{\partial \lambda_i}, \frac{\partial h}{\partial \lambda_j} \right\rangle \,.
    \label{eq:FisherElementGeneric}
\end{align}
For a signal with true parameters ${\bm \lambda}$, the probability for the parameters being reconstructed to ${\bm \lambda} + \Delta {\bm \lambda}$ in a particular measurement is, in the limit of large SNR, given by
\begin{align}
    p(\Delta \boldsymbol{\lambda}) \propto e^{-\frac{1}{2} \sum
    \limits_{ij} \Gamma_{ij}\Delta\lambda_i \Delta\lambda_j} \;,
    \label{eq:FisherPosteriorGaussian}
\end{align}
as can be seen from the central-limit theorem. Thus, the variance-covariance matrix is obtained by inverting the Fisher information matrix,
\begin{align} \label{eq:Cmat}
    \mathcal{C}_{ij} = (\Gamma^{-1})_{ij} \;.
\end{align}
In particular, the uncertainty with which a particular parameter $\lambda_i$ can be measured is estimated by
\begin{align} \label{eq:errorDef}
    \sigma_i \equiv \sqrt{\mathcal{C}_{ii}} \;.
\end{align}

In the case of the angles describing the sources sky location and orientation of the binary, we will often present uncertainties on the solid angle, rather than the coordinate-system dependent individual angles. This is defined, for the sky location, by \cite{Cutler:1997ta}
\begin{align}
    \sigma_{\Omega_n} = 2\pi\cos \delta \sqrt{\mathcal{C}_{\alpha\alpha} \mathcal{C}_{\delta\delta} - \mathcal{C}_{\alpha\delta}^2} \;,
    \label{eq:OmegaNUncertaintyDefinition}
\end{align}
and, for the binary orientation, by\footnote{The $\cos\delta$ vs $\sin\iota$ difference in Eqs.~\eqref{eq:OmegaNUncertaintyDefinition} and~\eqref{eq:OmegaLUncertaintyDefinition} is due to the different definitions of the respective coordinate systems. The declination angle runs from $\delta \in \left[ -\pi/2, \pi/2 \right]$, while the inclination angle of the binary has a range $\iota \in \left[0, \pi \right]$.}
\begin{align}
    \sigma_{\Omega_L} = 2\pi\sin\iota\sqrt{\mathcal{C}_{\iota\iota} \mathcal{C}_{\psi\psi} - \mathcal{C}_{\iota\psi}^2} \;.
    \label{eq:OmegaLUncertaintyDefinition}
\end{align}

This formalism is straightforward to extend to multiple detectors. In general, the noise PSD is promoted to a matrix in the space of detectors, which we denote with Latin indices, $\left[ S_h (f) \right]_{ab}$. Similarly, any strain signal, $h(t)$, gets promoted to a vector in the space of detectors, ${\bm h}(t)$: for a GW signal, the difference between the entries $h_a(t)$ corresponding to the response of various detectors is due to the different antenna functions of the detectors as well as the difference in arrival time of the signal between detectors. The scalar product for signals ${\bm r}$ and ${\bm s}$ is then defined via contracting the noise PSD matrix with the signal vectors:
\begin{align}
\left\langle {\bm r}, {\bm s} \right\rangle \equiv \sum_{a,b} 4{~\rm Re}\left\{ \int_{0}^{\infty} df \; \widetilde{r}_a(f) \left[ S_h^{-1} (f) \right]_{ab} \widetilde{s}_b(f)^* \right\} \,.
\end{align}
The SNR and the Fisher information matrix are computed from the scalar product as for a single detector.

When considering networks of detectors in this work, we will assume their noise to be uncorrelated. Then, the noise PSD matrix is diagonal, and the scalar product simplifies to 
\begin{align}
    \left\langle {\bm r}, {\bm s} \right\rangle \equiv \sum_a 4{~\rm Re}\left\{ \int_{0}^{\infty} df \; \widetilde{r}_a(f) \left[ S_h^{-1} (f) \right]_{aa} \widetilde{s}_a(f)^* \right\} \,.
\end{align}

As we noted above, the Fisher-matrix formalism yields a good approximation for the expected covariance matrix only in the limit of large SNR~\cite{Vallisneri:2007ev}. While we will generally assume that this holds throughout this paper, leaving finite-SNR corrections to future work, we make one significant exception: compact binaries are parametrized by four angles (two for their location in the sky, and two for the direction of their angular momentum), which can take values only in finite intervals ($\alpha, \psi \in [0, 2\pi]$, $\delta \in [-\pi/2, \pi/2]$, $\iota \in [0, \pi]$). For some of the sources we consider, and depending on the SNR, the uncertainty on one or more of these angles, as calculated above, is much larger than $\mathcal{O}(\pi)$. This can lead to large, unphysical uncertainties on other parameters due to degeneracies between them. In a Bayesian Markov chain Monte Carlo analysis, such effects can be controlled by using appropriate priors for these parameters, restricting them to their physical ranges. Such an restriction cannot immediately be made in the Fisher matrix approach. Here, we instead include what effectively amounts to a Gaussian prior on these parameters by adding a matrix whose only non-vanishing entries are $1/\pi^2$ for the diagonal entries corresponding to angles with a range of $\pi$ and $1/(2 \pi)^2$ for angles with a range of $2\pi$ to the Fisher information matrix defined in Eq.~\eqref{eq:FisherElementGeneric}. Note that this is a rather conservative implementation -- the full width of this ``prior'' corresponds to their true range. While inexact, this approach eliminates the large errors discussed above and should introduce at most order-unity errors in our projections.

\subsection{Implementation of Fisher Matrix Forecast in \texttt{AIMforGW}}
\label{sec:Code}

Before we move on to discussing the numerical results of our parameter estimation forecast for GW signals from compact binaries with atom-interferometer detectors in the mid-band, let us describe some of the technical details of our implementation of the Fisher matrix analysis. \texttt{AIMforGW}, the code used to produce all numerical results in this work, is publicly available at this repository\footnote{\href{\linkCode}{\url{\linkCode}}}: \href{\linkCode}{\faGithubSquare}. The reader not interested in these details should directly skip to Sec.~\ref{sec:ResultsDiscussion}.

We parameterize the GW signal from a merging binary by nine parameters, see Tab.~\ref{tab:Parameters}: the chirp mass ($\mathcal{M}_c$), the mass ratio ($q$), the luminosity distance ($d_L$), a reference phase of the binary ($\Phi_0$), the time of merger ($t_c$), two angles describing the position of the binary on the sky ($\alpha$, $\delta$), and two angles describing the orientation of the binary's angular momentum ($\iota$, $\psi$). 

Here arises a subtlety: the detectors are moving on accelerated paths. Terrestrial detectors rotate on the surface of Earth while space-based detectors orbit Earth, and in both cases, the detectors follow Earth's orbit around the Sun. However, the GW waveforms from merging binaries we use are defined in inertial frames. We choose an inertial frame (co-moving with the solar system) located at the position of the center of Earth at the mid-point in time between the signal entering and leaving the sensitivity band of the detector to compute the GW signal. Strictly speaking, $\mathcal{M}_c$ and $t_c$ are defined in this frame. For the measurement of $t_c$ with a detector (network), this induces a degeneracy between $t_c$ and the sky-position of the sources: a detector measures a quantity related to $t_c$ but defined in its co-moving reference frame. Relating this quantity to $t_c$ as defined in our setup involves a frame transformation that requires knowing the sky-position of the source. Parametrically, for a source whose signal is measured over the time-span of a year in a detector orbiting Sun with the Earth, this induces an uncertainty in $t_c$ of order $1\,{\rm AU}/\sqrt{\sigma_{\Omega_n}}$, e.g., for $\sqrt{\sigma_{\Omega_n}} \sim 1^\circ$, this amounts to an uncertainty on $t_c$ of $\sim 1\,$s from the frame transformation.

We compute polarization-basis inspiral waveforms at 3.5/3.0 PN order (frequency evolution/amplitude correction) in the time-domain using the expressions in Ref.~\cite{Blanchet:2013haa} in the inertial frame discussed above. From these waveforms, we then compute the response of the detector using the appropriate (time-dependent) antenna functions and shifting the signal by the time-dependent time-delay from the inertial frame, in which we compute the waveforms, to the detector frame. This time-shift incorporates two effects: first, for a network of detectors, it encodes the difference in arrival time of the signal between different detector locations. Second, it encodes the time-dependent Doppler shift of the GW signal for a detector that follows an accelerated trajectory while measuring the signal.

From the detector-response in the time domain, we then compute the frequency-domain response by Fourier-transforming the signal. This necessitates windowing the signal; we use a Planck-tapered window function~\cite{McKechan:2010kp} for our numerical results. 

With the frequency-domain detector response in hand, we can compute the signal-to-noise ratio of the detector (network) using Eq.~\eqref{eq:SNR}. In order to compute the elements of the Fisher-information matrix, we need to compute the derivatives of the detector response with respect to the parameters controlling the signal, see Eq.~\eqref{eq:FisherElementGeneric}. We compute these derivatives as one-sided first-order finite difference derivatives, using the same algorithm as described above to compute the detector response. As discussed above, we add a conservative prior constraining angular variables to their physical range to the Fisher information matrix. Finally, we compute the covariance matrix by inverting the Fisher information matrix; see Eqs.~\eqref{eq:Cmat} and~\eqref{eq:errorDef}.

Let us stress that computing the detector-response in the time domain allows us to include the full effect of moving detectors without having to resort to any approximations. This is in contrast to other Fisher Matrix codes previously used in the mid-band~\cite{Graham:2017lmg, Cai:2021ooo, Yang:2021xox, Yang:2022iwn} as well as recent publicly available Fisher Matrix codes (primarily aimed at other frequency bands) such as \texttt{GWBENCH}~\cite{Borhanian:2020ypi}, \texttt{GWFISH}~\cite{Dupletsa:2022scg}, or \texttt{GWFAST}~\cite{Iacovelli:2022mbg}: these codes compute the polarization basis waveforms $\widetilde{h}_{+,\times}(f)$ directly in the frequency domain. The different codes then use various approximations for the time-frequency relation $t(f)$ of the signal that are typically not fully consistent with the employed waveform approximant. They account for the rotation of the detector via $t(f)$-dependent antenna functions, and for the motion of the detector via a ``Doppler-phase'' $\phi_D(f) \equiv 2\pi f \left(\hat{\bm{n}} \cdot \bm{r}_D\right)$ where $\bm{r}_D = \bm{r}_D[t(f)]$ is the time-dependent position of the detector relative to the frame in which the $\widetilde{h}_{+,\times}$ and $t(f)$ are computed. The frequency-domain detector response is then obtained via
\begin{equation}
    \widetilde{h}_D(f) = e^{i \phi_D} \times \left\{ F_+[t(f)] \times \widetilde{h}_+(f) + F_\times[t(f)] \times \widetilde{h}_\times(f) \right\} \,. 
\end{equation}
While this approximation does capture the leading effect of moving and rotating detectors, our approach in \texttt{AIMforGW} of calculating the detector response in the time-domain described above does not force us to make any approximations about the effects of moving detectors. Of course, the cost of our approach is that it restricts us to using time-domain waveform approximants and that it is computationally expensive.

\section{Numerical Results and Discussion} \label{sec:ResultsDiscussion}

In this section, we present numerical results for the ability of atom-interferometer GW detectors in the mid-band to measure the signals from and constrain the properties of compact binaries in the ${\sim}\,1 - 100\,M_\odot$ mass range. We consider two detectors benchmarks (see Sec.~\ref{sec:MidbandGWDetection}): a terrestrial (network of) ``MAGIS-1\,km'' detector(s), and a single satellite-borne ``MAGIS-space'' detector. We begin by considering these detectors' SNRs and their abilities to determine the luminosity distance and sky locations of sources in Sec.~\ref{sub:ChirpMassLumiDistScaling}. We then turn to the dependence of SNR and sky localization uncertainty on the location of GW sources in the sky in Sec.~\ref{sub:SkyLocationScaling}. Finally, Sec.~\ref{sub:ObservationTimeScaling} addresses the question of how far before a merger mid-band detectors can detect events as well as measure their sky location and merger time, which are crucial questions for multimessenger astronomy.

Additional results for parameter estimation forecasts are presented and discussed in App.~\ref{app:SupplementalPlots}, see Figs.~\ref{fig:spaceOrbits}--\ref{fig:Lmap_dL_Ground}. 

\subsection{Dependence on Chirp Mass and Luminosity Distance} \label{sub:ChirpMassLumiDistScaling}

Let us begin by considering the dependence of the SNR on the chirp mass of and luminosity distance to the binary. We show the SNR in the $\mathcal{M}_c$--$d_L$ plane in Fig.~\ref{fig:ChirpMass_SNR_Scans}, both for the ``MAGIS-space'' detector (left panel) and for a network of five ``MAGIS-1\,km'' detectors at Homestake, Sudbury, Renstr{\"o}m, Tautona and Zaoshan (see Sec.~\ref{sec:MidbandGWDetection}) which we refer to as \TMN. As one can immediately see from the expression for the GW strain amplitude in Eq.~\eqref{eq:Amplitude}, the SNR depends inversely on the luminosity distance, $\rho \propto 1/d_L$. 

In order to obtain the approximate scaling of the SNR ratio with $\mathcal{M}_c$, we can estimate the frequency-domain strain from Eqs.~\eqref{eq:FrequencyEvolution} and~\eqref{eq:Amplitude} using the stationary phase approximation,
\begin{equation}
    \widetilde{h}(f_{\rm GW}) \propto \sqrt{\frac{1}{d f_{\rm GW}/d t}} \, h(t) \propto \mathcal{M}_c^{5/6} f_{\rm GW}^{-7/6} \;;
\end{equation}
note that the $\sqrt{df_{\rm GW}/d t}$ factor can be understood as arising from the time the signal can be measured at any particular $f_{\rm GW}$.

For the \TMN network as well as for the ``MAGIS-space'' detector at $\mathcal{M}_c \gtrsim 25\,M_\odot$, we can indeed observe $\rho \propto \mathcal{M}_c^{5/6}$ in Fig.~\ref{fig:ChirpMass_SNR_Scans}. For $\mathcal{M}_c \lesssim 25\,M_\odot$, the scaling of $\rho$ with $\mathcal{M}_c$ becomes faster for ``MAGIS-space''. This is because the GW signals from binaries with $\mathcal{M}_c \lesssim 25\,M_\odot$ take more than one year to develop from the lower frequency edge of the ``MAGIS-space'' detector's sensitivity band to the upper edge; however, in our SNR and parameter estimation forecasts we include only the signal in the last year before the source leaves the detector's sensitivity band (see also Tab.~\ref{tab:LifetimeHorizon}). 

Let us stress that throughout this work, except where stated explicitly, we use only the last year that the GW signal remains in the detector's sensitivity band to forecast the SNR and the parameter estimation ability of mid-band GW detectors for long-lived signals. Observing the signal over a one-year period during an earlier phase of the binary's evolution would lead to smaller SNR and larger degeneracies, due to the smaller PN corrections early in the inspiral.  Of course, if one were to observe a source for more than the last year in the band, the SNR would increase and the parameter estimation would improve. Here, we use just the last year as a benchmark estimate.

One figure of merit we can read off from Fig.~\ref{fig:ChirpMass_SNR_Scans} is the effective horizon of the detectors. The dotted (dashed) red lines in Fig.~\ref{fig:ChirpMass_SNR_Scans} mark curves of constant $\rho=5$ ($\rho=25$). For example, a ``MAGIS-space'' detector could observe GW signals from binary neutron stars with $\mathcal{M}_c \sim 1$ out to $d_L \sim 100\,$Mpc, while the signal from a binary black hole system with $\mathcal{M}_c \sim 250\,M_\odot$ could be observed at distances as large as $20\,$Gpc, or at cosmological redshifts of $z \sim 2$. The \TMN network could measure GW signals from binary neutron stars out to distances of ${\sim}\,30\,$Mpc, while the same detector network could observe signals from binary black holes
with $\mathcal{M}_c \sim 50\,M_\odot$ out to distances of $d_L \sim 0.5\,$Gpc, corresponding to sources with cosmological redshifts of $z \sim 0.1$. Note that the horizon of a detector would improve linearly with an improvement in strain sensitivity: $d_L(\rho = {\rm const.}) \propto 1/\sqrt{S_h(f)}$.

\begin{figure}
   \includegraphics[height=6.5cm, trim=0cm 0cm 4.1cm 0cm, clip]{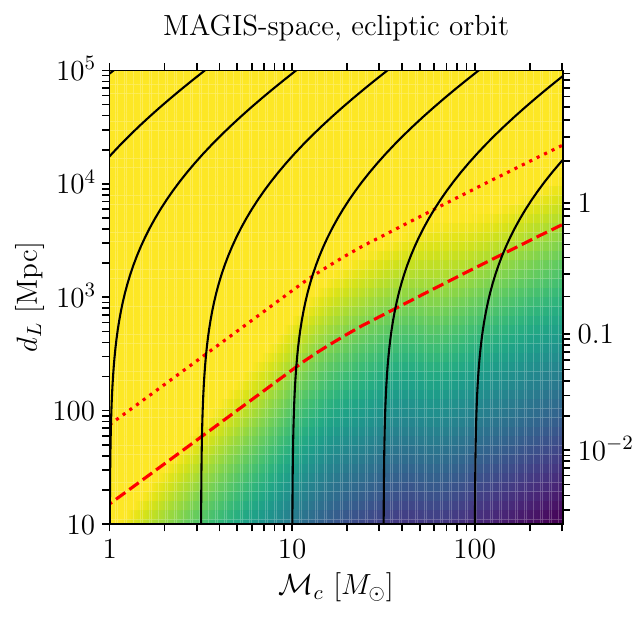}
   \hspace{0.3cm}
   \includegraphics[height=6.5cm, trim=0.9cm 0cm 0cm 0cm, clip]{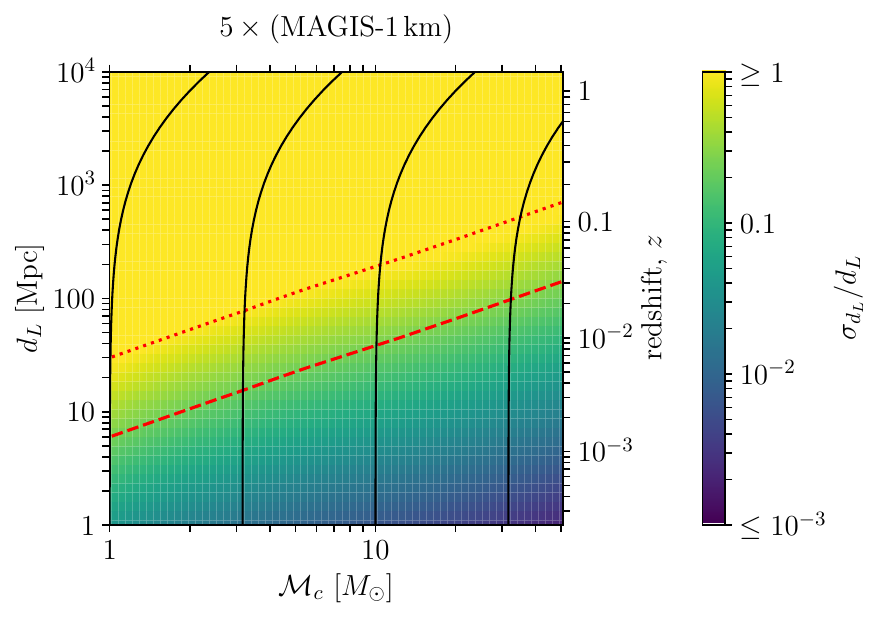}
   \caption{{\bf \boldmath Relative luminosity distance error [$\sigma(d_L)/d_L$]} in the plane of the (detector frame) chirp mass and luminosity distance of the GW source, all other parameters are set to the benchmark values given in Tab.~\ref{tab:Parameters}. The dotted (dashed) red lines are $\rho=5$ ($\rho=25$) contours of SNR. Note that the range of $d_L$ shown on the vertical axes differs between the two panels. The right $y$-axes show the corresponding cosmological redshift, $z$, assuming a reference cosmology with a Hubble constant of $H_0 = 67.4\,$km/Mpc/s and a matter density of $\Omega_m = 0.315$. The solid black lines show lines of constant source-frame chirp-masses corresponding to the detector-frame chirp-masses shown on the $x$-axis. {\it Left:} Results for a single ``MAGIS-space'' detector in an ecliptic orbit. {\it Right:} Results for a network of five terrestrial ``MAGIS-1\,km'' detectors, located at Homestake, Sudbury, Renstr\"om, Tautona, and Zaoshan; see Sec.~\ref{sec:MidbandGWDetection}.} 
   \label{fig:ChirpMass_dL_Scans}

   \vspace{2cm}
   
   \includegraphics[height=6.5cm, trim=0cm 0cm 4.1cm 0cm, clip]{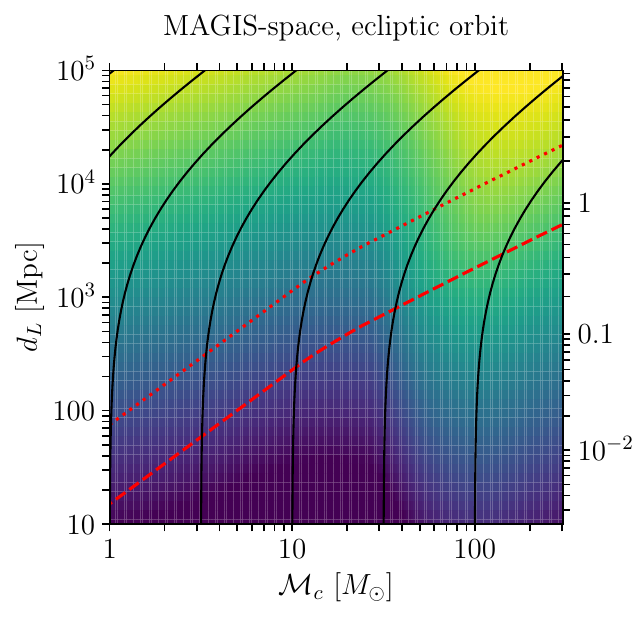}
   \hspace{0.3cm}
   \includegraphics[height=6.5cm, trim=0.9cm 0cm 0cm 0cm, clip]{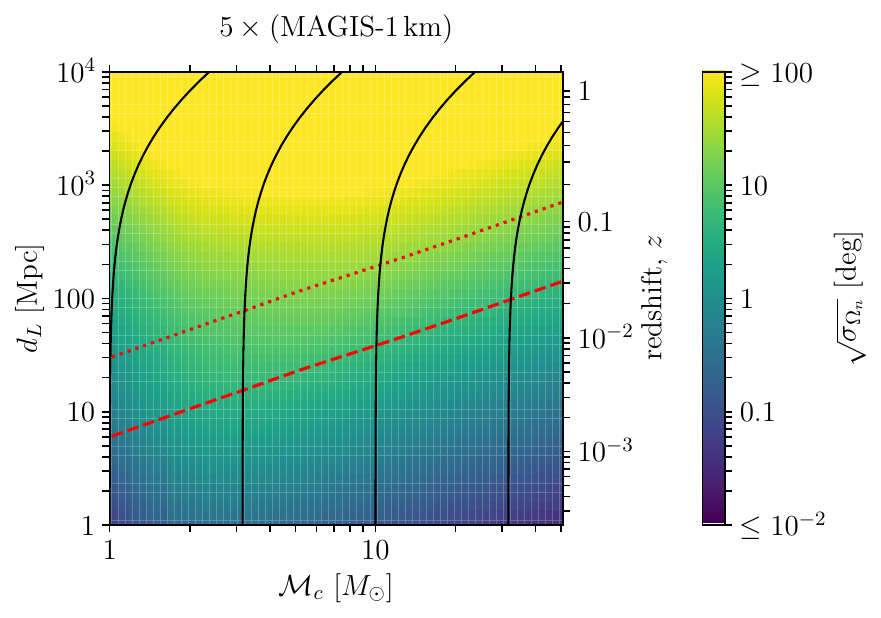}
   \caption{{\bf \boldmath Sky-localization error ($\sqrt{\sigma_{\Omega_n}}$)}  in the plane of the (detector frame) chirp mass and luminosity distance of the GW source, otherwise the same as Figs.~\ref{fig:ChirpMass_SNR_Scans} and~\ref{fig:ChirpMass_dL_Scans}.
   }
   \label{fig:ChirpMass_Omegan_Scans}
\end{figure}

In Fig.~\ref{fig:ChirpMass_dL_Scans}, we show the precision with which the luminosity distance could be constrained in the $\mathcal{M}_c$--$d_L$ plane. Let us begin by discussing the right panel, showing the results for the \TMN network. By comparing the color gradient with the red lines dotted (dashed) lines, which mark constant $\rho=5$ ($\rho=25$), we can note that the precision with which the luminosity distance of a source can be measured scales inversely with the SNR. Similarly, we can note that $\sigma_{d_L}/d_L \propto d_L$. This behavior is straightforward to understand by recalling that the amplitude of the GW signals scales as $h(t) \propto 1/d_L$, see Eq.~\eqref{eq:Amplitude}. In the absence of degeneracies with other parameters, the fact that the sole effect of a change in $d_L$ is to change the amplitude of the signal means that one could expect $\sigma_{d_L}/d_L = 1/\rho$. However, in the right panel of Fig.~\ref{fig:ChirpMass_dL_Scans} we can note that $\sigma_{d_L}/d_L$ is substantially larger than this bound, suggesting (partial) degeneracies of $d_L$ with other parameters. Indeed, as we discuss in App.~\ref{app:Signal}, inspiral waveforms show a characteristic $d_L$--$\iota$ degeneracy, which is most pronounced for $\iota = 0$ or $\pi$.

The left panel of Fig.~\ref{fig:ChirpMass_dL_Scans} is instead for a ``MAGIS-space'' detector. Most prominently, we observe that for $\mathcal{M}_c \lesssim 10\,M_\odot$, the $\sigma_{d_L}/d_L$ color gradient no longer follows the lines of constant SNR but the uncertainty increases faster. This behavior is mainly due to an increasing $d_L$--$\iota$ degeneracy which is more pronounced the deeper in the inspiral phase a detector observes a GW signal. We can also observe somewhat less pronounced deviations of the color gradient showing $\sigma_{d_L}/d_L$ from the lines showing constant SNR for $\mathcal{M}_c \gtrsim 50\,M_\odot$. As we will discuss in the next paragraph, the ability of a ``MAGIS-space'' detector to measure the sky-position of sources from a change in the Doppler shift is reduced for sources with $\mathcal{M}_c \gtrsim 50\,M_\odot$. Since the sky-position of a source affects the amplitude of the GW signal in the detector via the antenna functions, $d_L$ and the sky-position are partially degenerate.

Let us turn to Fig.~\ref{fig:ChirpMass_Omegan_Scans}, where we show the precision with which the sky-location of GW sources could be reconstructed. We start by considering the ``MAGIS-space'' detector benchmark, shown in the left panel. For sources with $\mathcal{M}_c \lesssim 50\,M_\odot$, we can see that for, any source close enough to be detected by a ``MAGIS-space'' detector (see the dotted red line marking $\rho = 5$), its position on the sky could be measured with sub-degree angular resolution. This is smaller than, for example, the field of view of the Zwicky Transient Facility camera or the Vera~C.~Rubin Telescope. For $\mathcal{M}_c \gtrsim 50\,M_\odot$, the uncertainty on the sky-position grows markedly. For example, for a source with $\mathcal{M}_c \sim 100\,M_\odot$, even at a $d_L$ where it would give rise to a $\rho \sim 25$ signal, we find $\sqrt{\sigma_{\Omega_n}} \sim 10^{\circ}$. 

In order to understand these results, recall that the ability of GW detectors to measure the position of GW sources on the sky primarily stems from various ways of aperture-synthesizing a telescope: if the detector follows an accelerated trajectory during the time it measures the signal, one can use the changing Doppler shift of the GW signal in the detector to infer its sky location. Then, the effective aperture $L$ is controlled by the size of the detector's trajectory. Alternatively, one can use the difference in arrival times of the signal in a network of detectors to infer the sky-position. In this case, $L$ is given by the separation of the detectors in the network. It is straightforward to show that, in both cases, the resulting angular resolution with which the source can be localized on the sky is parameterically given by the diffraction limit, $\sqrt{\sigma_{\Omega_n}} \propto 1/(\rho f_{\rm GW} L )$.

For sources with $\mathcal{M}_c \lesssim 50\,M_\odot$, a ``MAGIS-space'' detector could observe the signal for a few months or longer; see also Tab.~\ref{tab:LifetimeHorizon}. During this time, a ``MAGIS-space'' detector in geocentric orbit, as we assume here, would complete a significant part of the orbit of Earth around the Sun. If we set $\rho \sim 5$, the effective frequency where most of the signal's SNR would be accumulated to $f_{\rm GW} \sim 0.1\,$Hz, and the effective baseline to the size of Earth's heliocentric orbit, i.e. $L \sim 1\,$AU, we indeed find $1/(\rho f_{\rm GW} L) = \mathcal{O}(0.1^\circ)$. Since the sky-location is effectively measured from the changing Doppler shift and not from any details of the GW amplitude or frequency evolution, there is practically no degeneracy between the sky position and other parameters describing the binary in this regime of long-lived signals.

The ability of a ``MAGIS-space'' detector to localize GW sources on the sky depreciates for larger $\mathcal{M}_c$, because the GW signal becomes increasingly short-lived with respect to the time over which the detector (and Earth) orbit the Sun. Then, the ability of the detector to measure the sky-position of a source is driven by a combination of it aperture-synthesizing a telescope with effective aperture given by its orbit around Earth ($L \sim 10^4\,$km) and its ability to measure the signal along many different spatial directions due to the reorientation of the detector's baseline during the observation, which is always much longer than the reorientation time.

For a network of terrestrial ``MAGIS-1\,km'' detectors, shown in the right panel of Fig.~\ref{fig:ChirpMass_Omegan_Scans}, we find that typically, GW sources with $\rho \sim 25$ can be localized with a precision of $\sqrt{\sigma_{\Omega_n}} \sim 10^\circ$. Since the sensitivity of terrestrial ``MAGIS-1\,km'' detectors to signals below ${\sim}\,1\,$Hz is constrained by seismic (and other) Newtonian Gravity Gradient noise, most GW signals can only be measured for relatively short times such that the detectors cannot make use of the accelerated orbit of Earth around the Sun. However, for a network of five terrestrial detectors spread over the surface of Earth, as we assume in the right panel of Fig.~\ref{fig:ChirpMass_Omegan_Scans}, the sky localization can be inferred using the differences in arrival time of the signal between the detectors. While the separation of terrestrial detectors is smaller than the size of the ``MAGIS-space'' orbit, our ``MAGIS-1\,km'' detector benchmark is sensitive to somewhat larger GW frequencies than ``MAGIS-space''. Observing the signal at larger $f_{\rm GW}$ somewhat compensates the smaller $L$, since $\sqrt{\sigma_{\Omega_n}} \propto 1/(\rho f_{\rm GW} L )$.

\begin{figure}
  \centering
    \includegraphics[height=7.0cm, trim=0cm 0cm 0cm 0cm, clip]{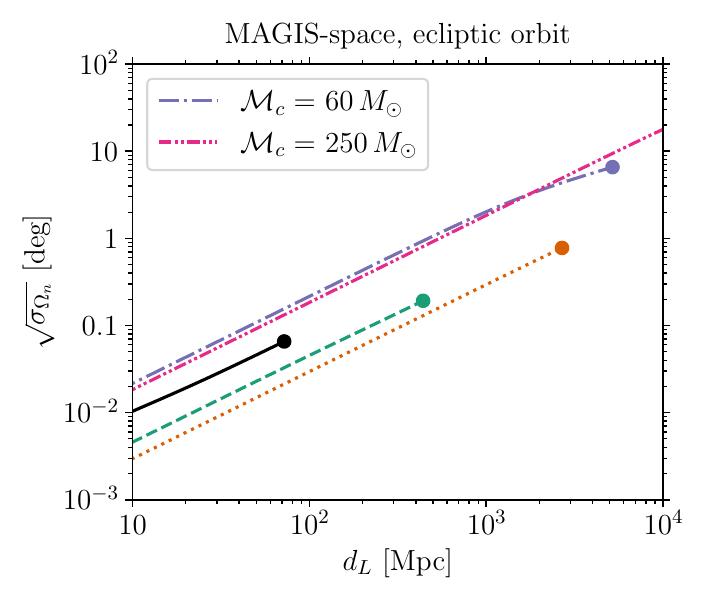}
    \includegraphics[height=7.0cm, trim=2.1cm 0cm 0cm 0cm, clip]{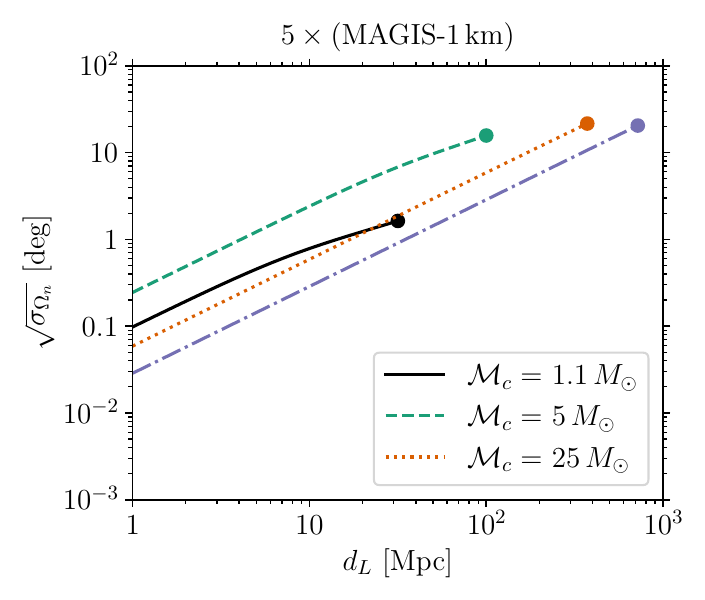}
    \caption{{\bf \boldmath Sky-localization error ($\sqrt{\sigma_{\Omega_n}}$)} as a function of the luminosity distance ($d_L$) for five benchmark choices of the detector frame chirp mass: $\mathcal{M}_c = \{1.1,~5,~25,~60,~250\}\,M_\odot$, as denoted in the legends. All other parameters are set to the benchmark values given in Tab.~\ref{tab:Parameters}. We terminate each line at the luminosity distance where the source would give rise to a signal with SNR $\rho = 5$. {\it Left:} Results for a single ``MAGIS-space'' detector in ecliptic orbit. {\it Right:} Results for a network of five terrestrial ``MAGIS-1\,km'' detectors, located at Homestake, Sudbury, Renstr\"om, Tautona, and Zaoshan; see Sec.~\ref{sec:MidbandGWDetection}. Note that we do not show results for $\mathcal{M}_c = 250\,M_\odot$ in the right panel: such a binary would merge within the sensitivity band, hence, estimating $\sqrt{\sigma_{\Omega_n}}$ would require going beyond the inspiral GW waveforms we use in this work.}
    \label{fig:angRes_dL_BP}
\end{figure}

At small chirp masses, $\mathcal{M}_c \lesssim 3\,M_\odot$, we can see that the sky-localization capabilities of ``MAGIS-1\,km'' detectors improve drastically. For example, for a binary neutron star system with $\mathcal{M}_c \sim 1\,M_\odot$ at a distance $d_L \sim 30\,$Mpc, which would give rise to a $\rho \sim 5$ signal, we find $\sqrt{\sigma_{\Omega_n}} \sim 1^\circ$. This is because for such chirp masses, the lifetime of the signal becomes sufficiently long for the detector to make use of the changing Doppler shift of the signal in the detector as Earth travels around the Sun; for example, the lifetime of the signal from a $\mathcal{M}_c = 1.1\,M_\odot$ source in the ``MAGIS-1\,km'' sensitivity band is ${\sim}\,40\,$days. Thus, terrestrial detectors can still have excellent sky localization for lower mass sources such as binary neutron stars.

In order to further illustrate the ability of GW detectors in the mid-band to localize GW sources in the sky, in Fig.~\ref{fig:angRes_dL_BP} we show the sky-localization error for five benchmark choices of the chirp mass, $\mathcal{M}_c = \{1.1,~5,~25,~60,~250\}\,M_\odot$, as a function of the luminosity distance. As in the previous figures, the left panel is for the ``MAGIS-space'' detector, while the right panel is for the \TMN network. As expected for any parameter uncertainty, for each benchmark value of $\mathcal{M}_c$, the sky-localization error scales as $\sqrt{\sigma_{\Omega_n}} \propto d_L \propto 1/\rho$. Exceptions from this scaling are visible for some of the results where $d_L$ approaches the value where $\rho = 5$ (the $d_L$ where we terminate each line in Fig.~\ref{fig:angRes_dL_BP}): such deviations from $\sqrt{\sigma_{\Omega_n}} \propto d_L$ are due to the effects of the effective priors we include in our Fisher matrix analysis to constrain angular variables to their physical range. The effect visible in Fig.~\ref{fig:angRes_dL_BP} arises mainly from the prior on the inclination angle $\iota$ affecting $\sqrt{\sigma_{\Omega_n}}$ via the partial degeneracy of $\iota$ with the sky-location in the GW amplitude. 

Note that, in Fig.~\ref{fig:angRes_dL_BP}, sources at the same $d_L$ with different choices of $\mathcal{M}_c$ now give rise to signals with vastly different $\rho$. This variation in SNR (see Fig.~\ref{fig:ChirpMass_SNR_Scans}) then leads to changes in sky-localization error, with both effects appearing in Fig.~\ref{fig:angRes_dL_BP}.

Let us summarize two main points demonstrated in this section: First, GW detectors in the mid-band are not particularly well-suited to measuring the luminosity distance to GW sources. This is primarily the result of substantial degeneracies between $d_L$ and other parameters describing the binary that inherent to inspiral waveforms. Note, however, that this could be alleviated by combining GW observations in the mid-band with observations of the same source at higher frequencies, e.g., in LVK or future terrestrial laser-interferometer detectors.

On the other hand, GW detectors in the mid-band are ideally suited to measuring the position of GW sources on the sky, especially for signals that spend at least a few months within their observable frequency band. This capability results from the large displacements of these detectors around the Sun compared to their observed GW wavelengths, an advantage that is unique to the mid-band.

\subsection{Dependence on Sky Location} \label{sub:SkyLocationScaling}

\begin{figure}
   \centering 
   \text{MAGIS-space}
   \vspace{0.2cm}
   
   \includegraphics[width=0.32\linewidth, trim=3.6cm 5.5cm 3.6cm 0cm, clip]{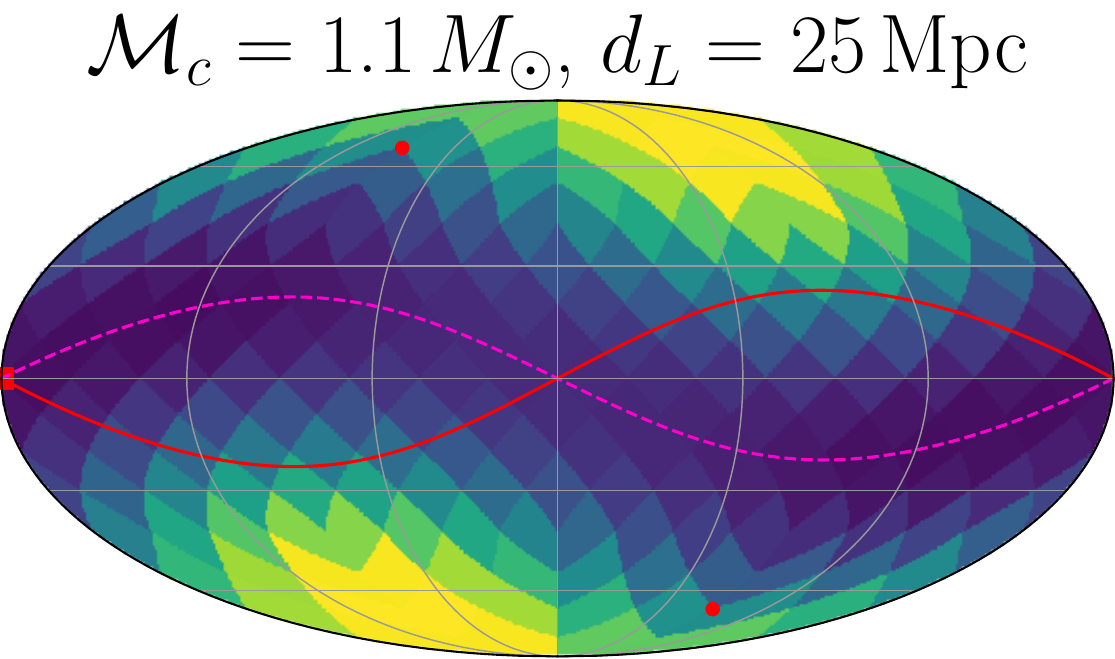}
   \includegraphics[width=0.32\linewidth, trim=3.6cm 5.5cm 3.6cm 0cm, clip]{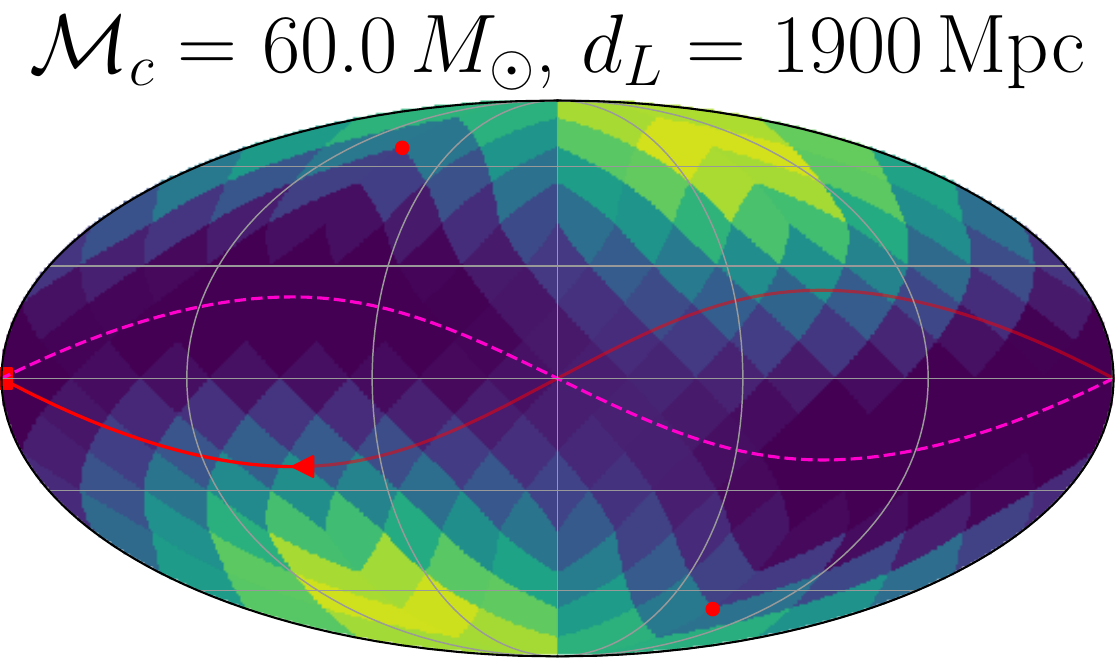}
   \includegraphics[width=0.32\linewidth, trim=3.6cm 5.5cm 3.6cm 0cm, clip]{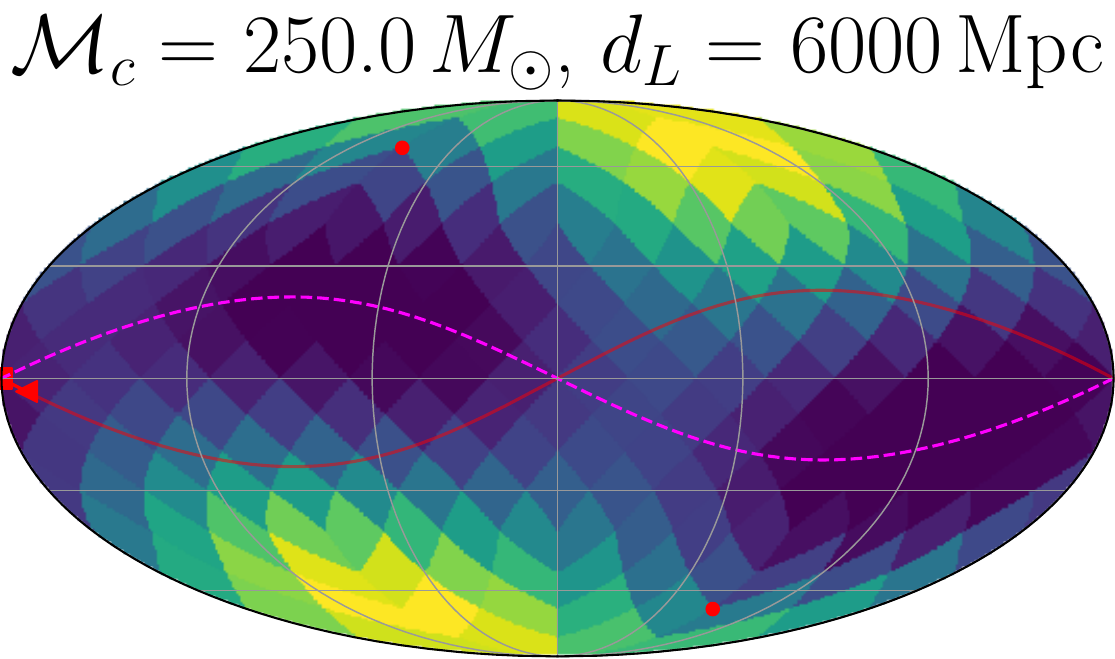}
   \vspace{0.1cm}
   
   \includegraphics[width=0.32\linewidth, trim=3.6cm 5.5cm 3.6cm 0cm, clip]{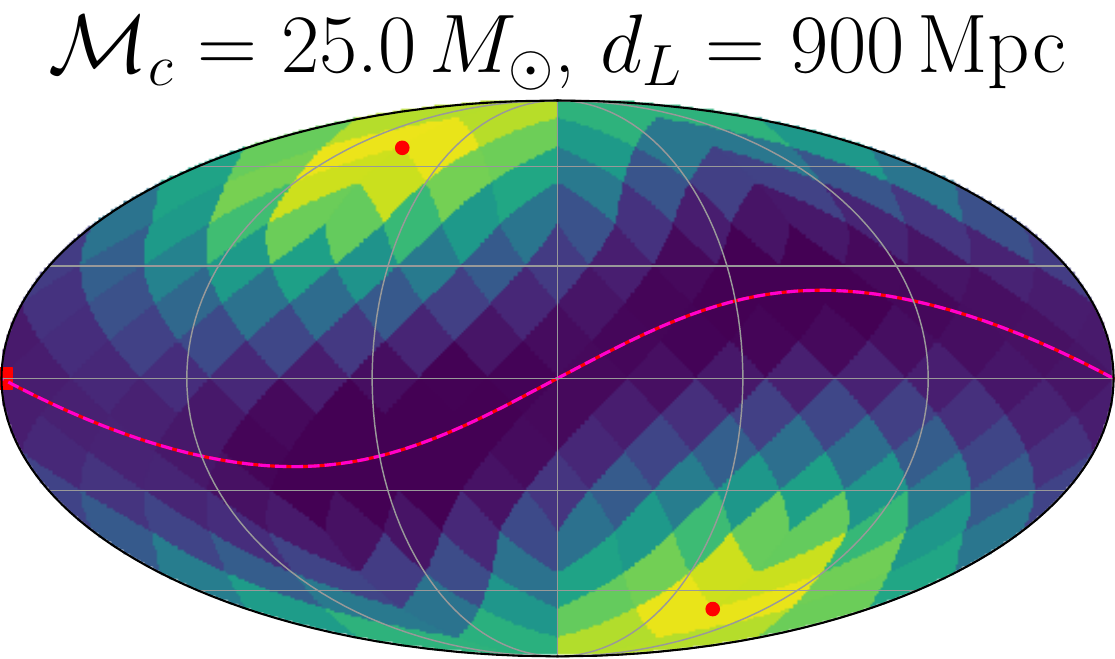}
   \includegraphics[width=0.32\linewidth, trim=3.6cm 5.5cm 3.6cm 0cm, clip]{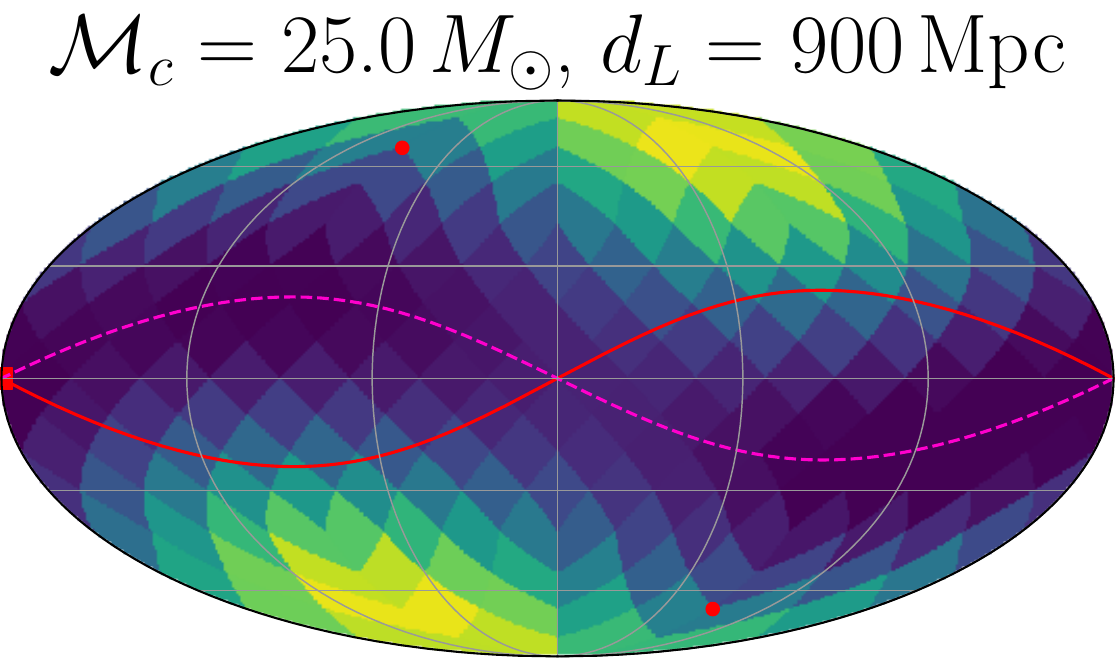}
   \includegraphics[width=0.32\linewidth, trim=3.6cm 5.5cm 3.6cm 0cm, clip]{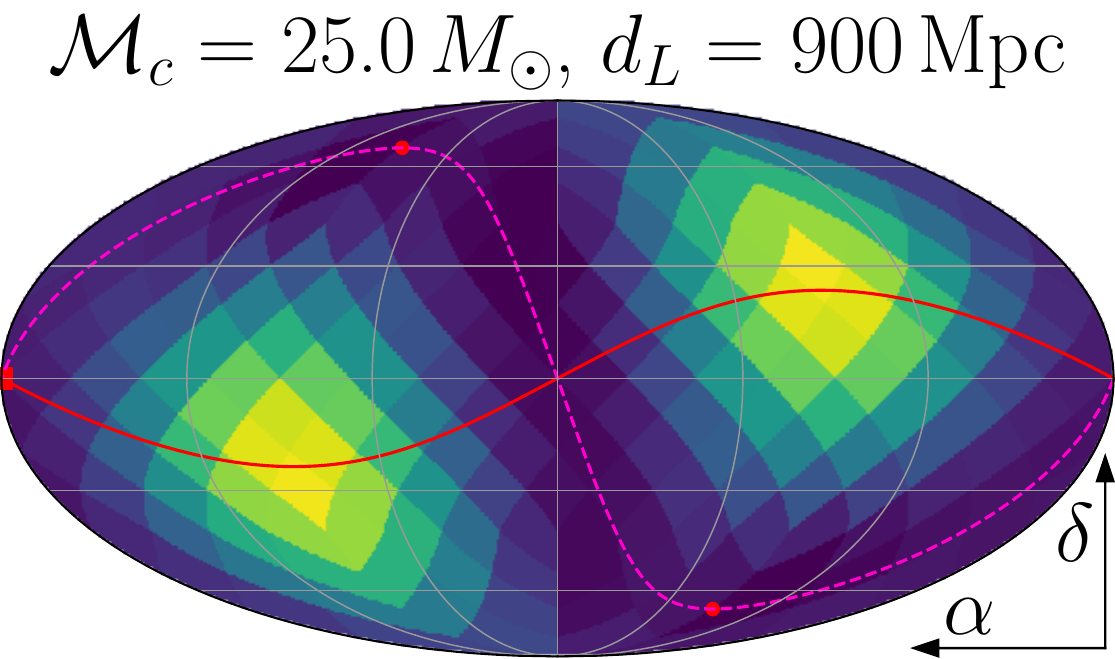}
       
   \includegraphics[width=0.45\linewidth, trim=0cm 0.2cm 0cm 11.5cm, clip]{Figs/skymap_SNR_Space_Mc25_45deg}
   \vspace{-0.3cm}
   \caption{{\bf Signal-to-noise ratio (SNR)} as a function of the source's sky-location for a ``MAGIS-space'' detector. All plots are Mollweide projections of equatorial coordinates, right ascension ($\alpha$) increases from right to left, declination ($\delta$) increases from bottom to top, and the origin ($\alpha = \delta = 0$) is at the center of each map. The solid red line marks the ecliptic plane and the red circles indicate the ecliptic poles. The red triangle and square are the start and end points, respectively, of the detector location as seen from the Sun over the course of the observation; note that these are identical for signals observed for a full year. The dashed magenta line visualizes the detector orbit. {\it Upper row, left-to-right:} Results for $\mathcal{M}_c = \{1.1,~60,~250\}\,M_\odot$, respectively, for a detector in an orbit $45^\circ$ from the ecliptic. {\it Lower row, left-to-right:} Results for $\mathcal{M}_c = 25\,M_\odot$ for a space detector in an orbit $0^\circ,~45^\circ,~90^\circ$ from the ecliptic, respectively. All other parameters are set to the values given in Tab.~\ref{tab:Parameters}.}
   \label{fig:skymap_SNR_Space}
   \vspace{0.4cm}
   
   \includegraphics[width=0.32\linewidth, trim=4.1cm 5.7cm 3.9cm 0cm, clip]{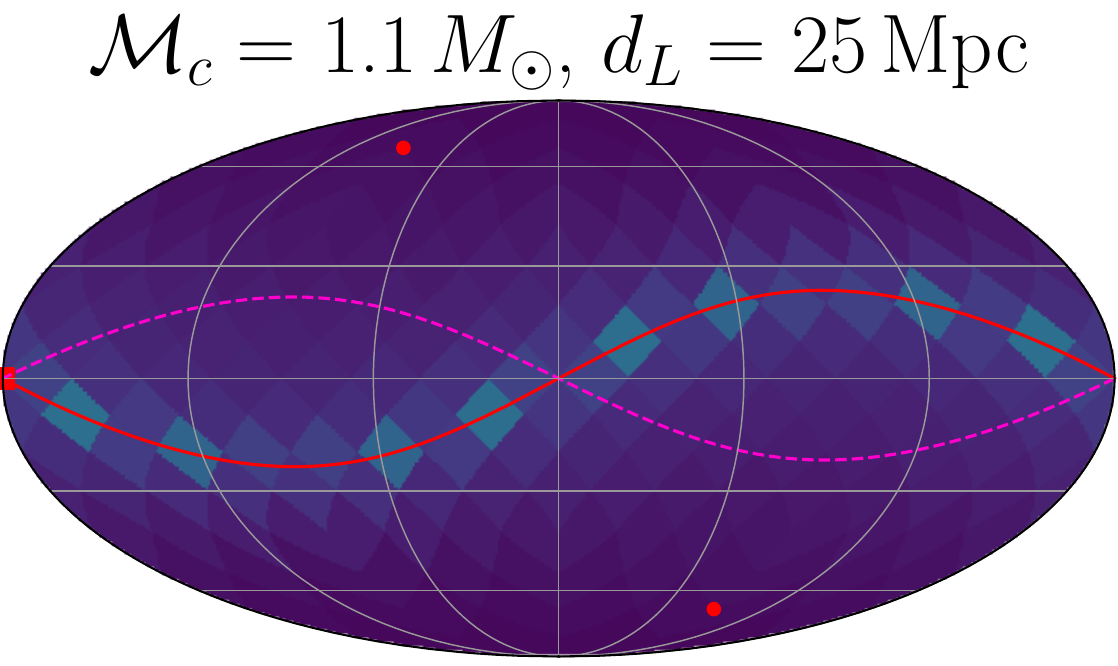}
   \includegraphics[width=0.32\linewidth, trim=4.1cm 5.7cm 3.9cm 0cm, clip]{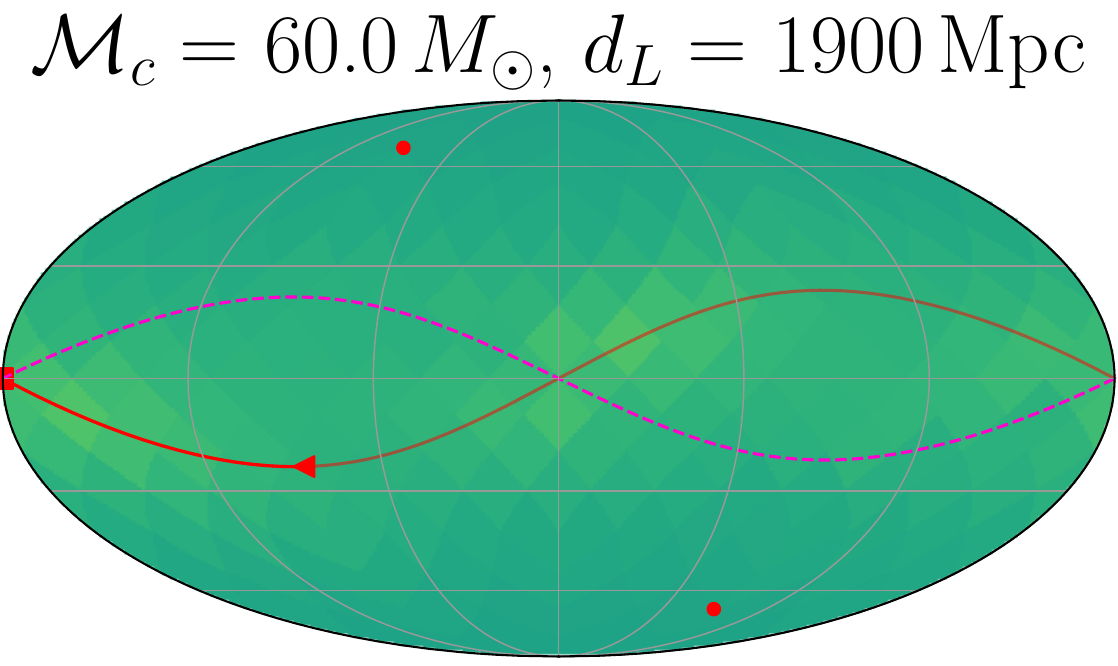}
   \includegraphics[width=0.32\linewidth, trim=4.1cm 5.7cm 3.9cm 0cm, clip]{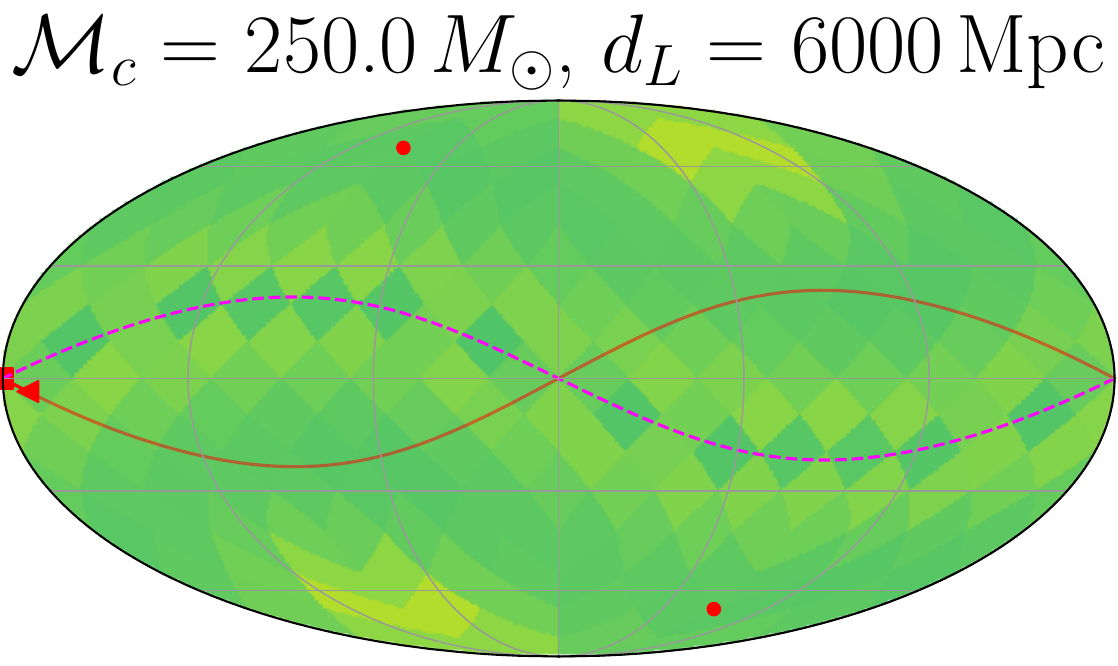}
   \vspace{0.1cm}
   
   \includegraphics[width=0.32\linewidth, trim=4.1cm 5.7cm 3.9cm 0cm, clip]{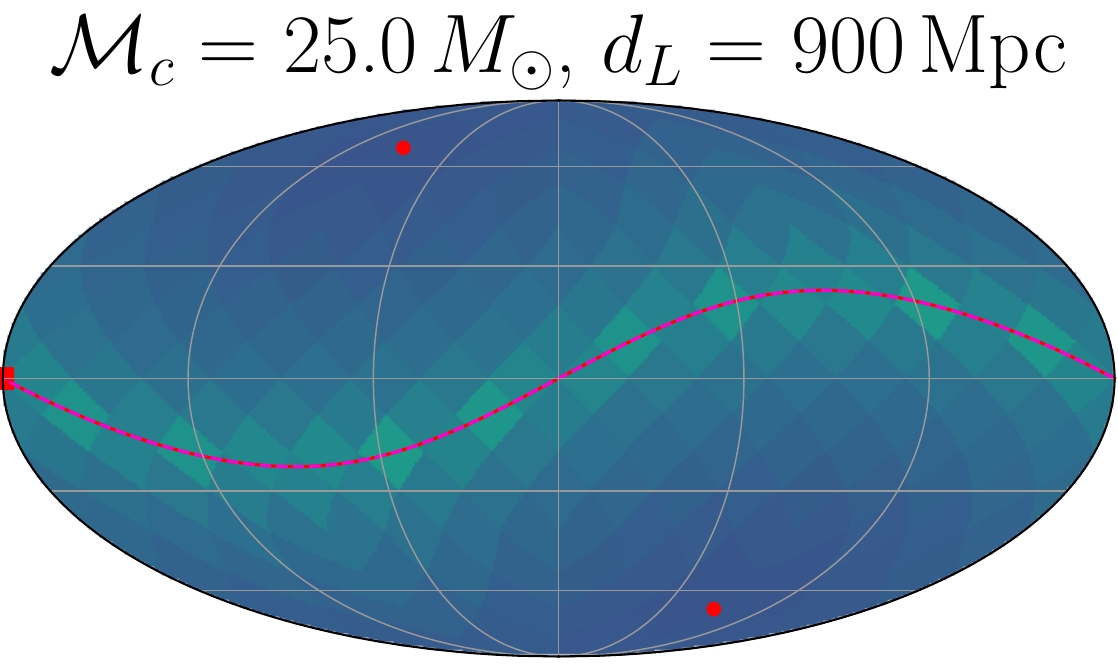}
   \includegraphics[width=0.32\linewidth, trim=4.1cm 5.7cm 3.9cm 0cm, clip]{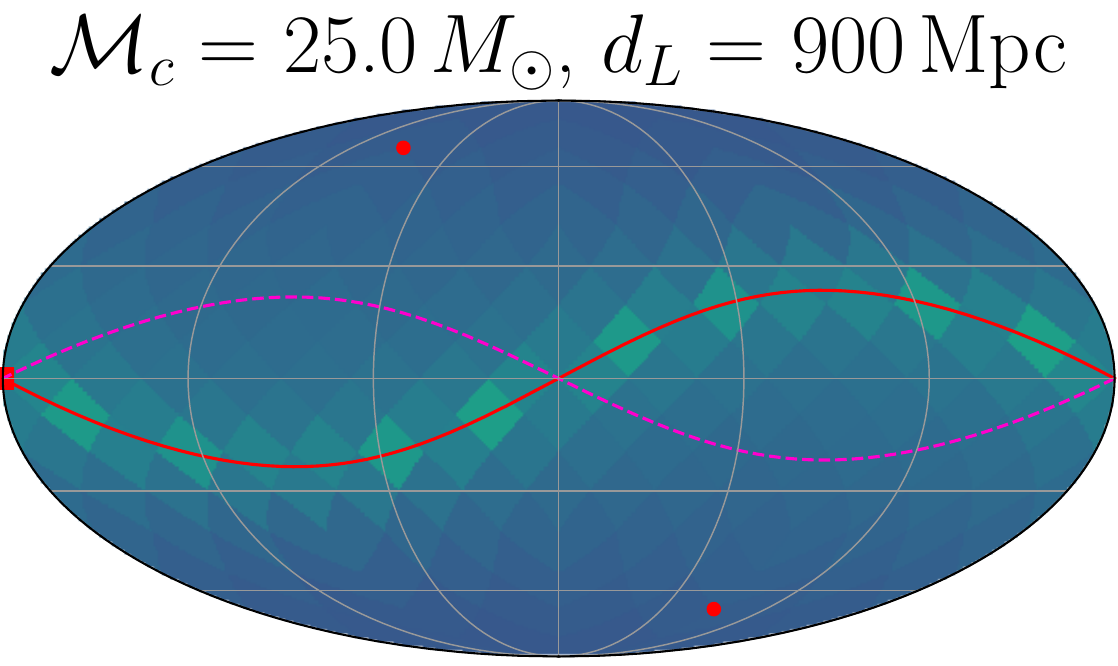}
   \includegraphics[width=0.32\linewidth, trim=4.1cm 5.7cm 3.9cm 0cm, clip]{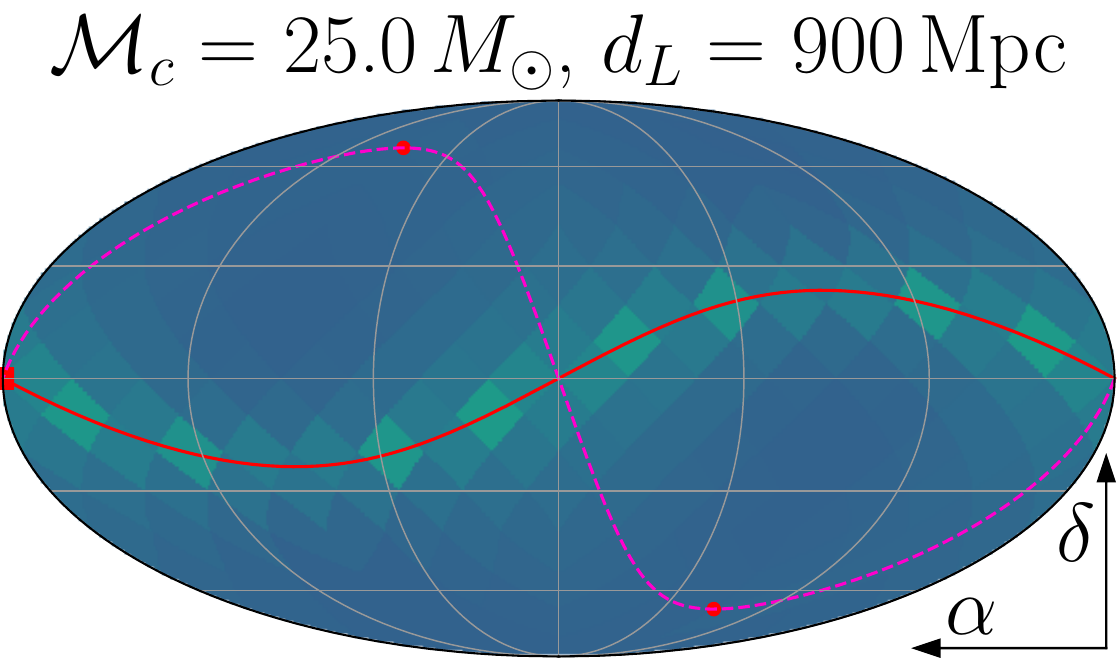}
       
   \includegraphics[width=0.45\linewidth, trim=0cm 0.3cm 0cm 11.6cm, clip]{Figs/skymap_angRes_Space_Mc25_45deg}
   \vspace{-0.3cm}
   \caption{{\bf \boldmath Sky-localization error ($\sqrt{\sigma_{\Omega_n}}$)} [see Eq.~\eqref{eq:OmegaNUncertaintyDefinition}] as a function of the source's sky-location for a ``MAGIS-space'' detector, otherwise the same as Fig.~\ref{fig:skymap_SNR_Space}.}
   \label{fig:skymap_angRes_Space}
\end{figure}

In this section, we discuss how the ability of atom-interferometer GW detectors in the mid-band to measure the signal from compact binaries and to reconstruct the source's sky location changes with the position of the source on the sky. Let us begin by discussing the satellite-borne ``MAGIS-space'' detector: in Fig.~\ref{fig:skymap_SNR_Space} we show sky-maps of the SNR ($\rho$), and, in Fig.~\ref{fig:skymap_angRes_Space}, of the sky-localization error ($\sqrt{\sigma_{\Omega_n}}$). 

First, let us note that the luminosity distances for the various benchmark values of $\mathcal{M}_c$ are chosen such that all sources have comparable SNR; in particular, we set $d_L$ such that the largest SNR (as a function of sky-position) for each source is $\rho \approx 25$. Since all of the signals we consider have lifetimes much longer than the orbital period of the satellites, the detector's antenna function is largely averaged around the orbital plane. The SNR therefore depends primarily on the source direction relative to that plane. Since the only propagating GW degrees of freedom are transverse to the GW's propagation direction, we can expect that the SNR is largest for sources orthogonal to the orbital plane of the detector, and smallest for sources located in the orbital plane. 

Before moving on, let us also note that the variations in SNR with the sky-position for ``MAGIS-space'' are relatively small: all panels in Fig.~\ref{fig:skymap_SNR_Space} show a variation of less than a factor of two in $\rho$. These results demonstrate that a single ``MAGIS-space'' detector in medium Earth orbit would re-orient its baseline sufficiently fast to avoid any pronounced blind-spots in its sky coverage for signals from ${\sim}\,1-100\,M_\odot$ GW sources.

In Fig.~\ref{fig:skymap_angRes_Space} we show sky-maps for how well a ``MAGIS-space'' detector could measure the position of GW sources on the sky as function of their sky position. The choices for the chirp mass of and luminosity distances to the GW sources, as well as for the detector orbits, are the same as in the respective panels of Fig.~\ref{fig:skymap_SNR_Space}. With the exception of the top right panel, to which we return below, we can see that the patterns of the variation of the sky-localization error ($\sqrt{\sigma_{\Omega_n}}$) with the sources position on the sky is largely independent of the SNR (see Fig.~\ref{fig:skymap_SNR_Space}) or the choice of the detector's orbital plane (see the three panels in the lower row of Fig.~\ref{fig:skymap_angRes_Space} ). Instead, we find the smallest $\sqrt{\sigma_{\Omega_n}}$ for sources orthogonal to the ecliptic plane, marked by the red line in Fig.~\ref{fig:skymap_angRes_Space}. As we have stressed throughout this work, the primary way in which a GW detector in the mid-band could localize GW sources in the sky is by measuring the changing Doppler shift of the GW signal as the detector accelerates relative to the GW source. For signals that can be measured for at least a few months in the detector, the dominant source of the acceleration of the detector is its orbit around the Sun. For a detector in geocentric orbit, as we consider here, this orbit is in the ecliptic plane. In order to understand the pattern of $\sqrt{\sigma_{\Omega_n}}$ in Fig.~\ref{fig:skymap_angRes_Space}, note that the inference of the source's sky position is not given by the magnitude of the change of the Doppler shift of the signal the detector can observe, but by the derivative of that magnitude with respect to the sky-position. While the magnitude of the change in Doppler shift is largest for a source in the ecliptic plane, its derivative is largest for a source orthogonal to the ecliptic plane, i.e. a source at the ecliptic poles. Note that for sources with lifetimes ${\gtrsim}\,1\,$yr, i.e., for $\mathcal{M}_c = 1.1\,M_\odot$ and $\mathcal{M}_c = 25\,M_\odot$ in Fig.~\ref{fig:skymap_angRes_Space}, the variations of $\sqrt{\sigma_{\Omega_n}}$ apparent along the ecliptic plane are mainly due to our discretization of the sky: each pixel shows $\sqrt{\sigma_{\Omega_n}}$ for a GW source located at the center of the pixel, and the precise alignment of the pixels with the ecliptic plane varies along the ecliptic plane.

Let us now compare the results for $\sqrt{\sigma_{\Omega_n}}$ we can see in Fig.~\ref{fig:skymap_angRes_Space} for the different choices of the chirp mass. The signals from both the $\mathcal{M}_c = 1.1\,M_\odot$ (top-left panel) and from the $\mathcal{M}_c = 25\,M_\odot$ source can be observed over the full year before the signal leaves the detector's sensitivity band we include in our analysis. For both sources, we find very similar patterns for $\sqrt{\sigma_{\Omega_n}}$, except that, across the sky, $\sqrt{\sigma_{\Omega_n}}$ can be measured with uncertainty roughly one order of magnitude smaller for a $\mathcal{M}_c = 1.1\,M_\odot$ source than for $\mathcal{M}_c = 25\,M_\odot$. This is due to our choice of including only the signal recorded during the last year before the it leaves the detectors sensitivity band in the analysis. For the $\mathcal{M}_c = 25\,M_\odot$ source, this means that most of the recorded signal is at some tens of mHz, while the signal from a $\mathcal{M}_c = 1.1\,M_\odot$ source starts at $f_{\rm GW} \sim 0.2\,$Hz one year before it reaches the upper edge of the ``MAGIS-space'' sensitivity band, see Fig.~\ref{fig:NoiseCurve}. Since we expect $\sqrt{\sigma_{\Omega_n}} \sim 1/(\rho f_{\rm GW} L)$, increasing $f_{\rm GW}$ by one order of magnitude leads to similar improvement in $\sqrt{\sigma_{\Omega_n}}$.

The $\mathcal{M}_c = 60\,M_\odot$ source, for which we show results in the top-middle panel of Fig.~\ref{fig:skymap_angRes_Space}, has a lifetime of approximately three months in the ``MAGIS-space'' sensitivity band; see the red triangle and square along the ecliptic, marking the position of the detector at the beginning and end of the period it observes the signal, respectively. Because the signal cannot be measured long enough for the detector to complete its full orbit around the Sun, we find larger sky-localization errors than for the $\mathcal{M}_c = 25\,M_\odot$ source. Furthermore, $\sqrt{\sigma_{\Omega_n}}$ starts to show variations along the ecliptic plane for $\mathcal{M}_c = 60\,M_\odot$ that no longer stem from the discretization effects: instead, since the detector completes only a segment of its orbit around the Sun while observing the signal, its ability to measure the sky-position from the changing Doppler shift now depends on the orientation of the source in the ecliptic plane relative to the segment of the orbit the detector completes.

The signal from a $\mathcal{M}_c = 250\,M_\odot$ source, shown in the top-right panel of Fig.~\ref{fig:skymap_angRes_Space}, can only be observed for ${\sim}\,8\,$days in ``MAGIS-space''. For such short-lived signals, the ability to measure the sky-location no longer stems from the changing Doppler shift induced by the detector orbiting the Sun, but predominantly from the detector's ability to measure the GW strain in different direction as the detector's baseline rotates. Thus, we can see that the pattern of $\sqrt{\sigma_{\Omega_n}}$ over the sky aligns (inversely) with the SNR pattern we see in Fig.~\ref{fig:skymap_SNR_Space}. The exception for this is that we find an improvement in $\sqrt{\sigma_{\Omega_n}}$ for sources close to the detector's orbital plane, where the SNR is smallest. The reason for this improvement is that, for such a source, there are times where the detector's baseline points right at the GW source. At such times, no strain signal is measured in the detector. Due to this effect, sources located close to the plane of the detector's rotation can be localized particularly well.

\begin{figure}
   \centering
   \text{MAGIS-1\,km}
   \vspace{0.2cm}
   
   \includegraphics[trim=3.2cm 5.5cm 3.6cm 0cm, clip, width=0.32\linewidth]{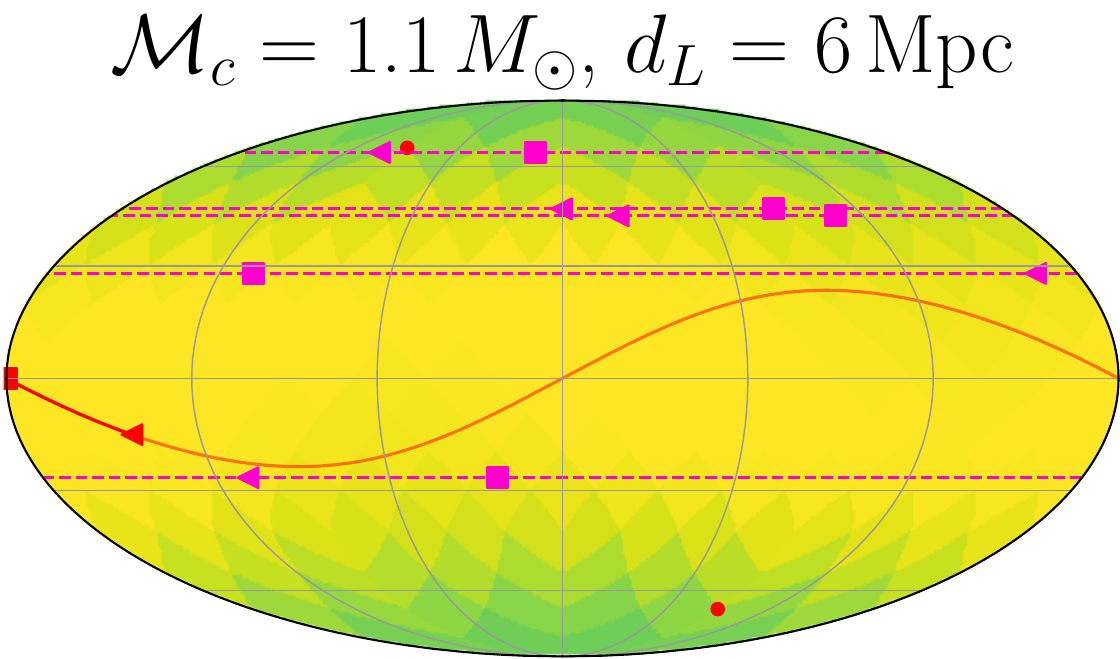}
   \includegraphics[trim=3.2cm 5.5cm 3.6cm 0cm, clip, width=0.32\linewidth]{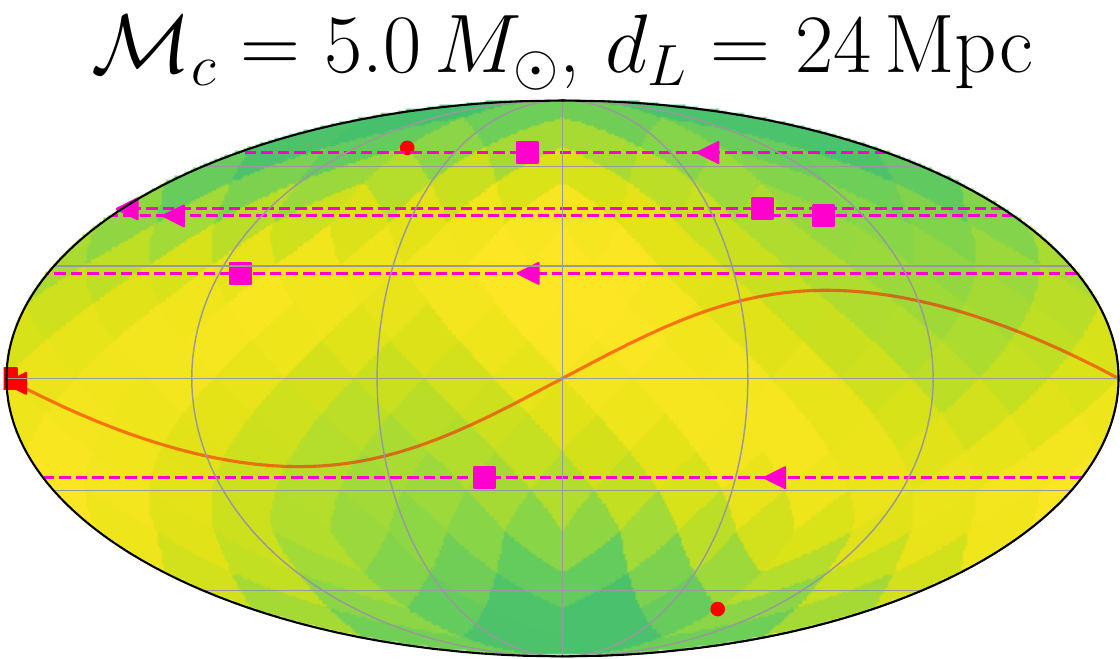}
   \includegraphics[trim=3.2cm 5.5cm 3.6cm 0cm, clip, width=0.32\linewidth]{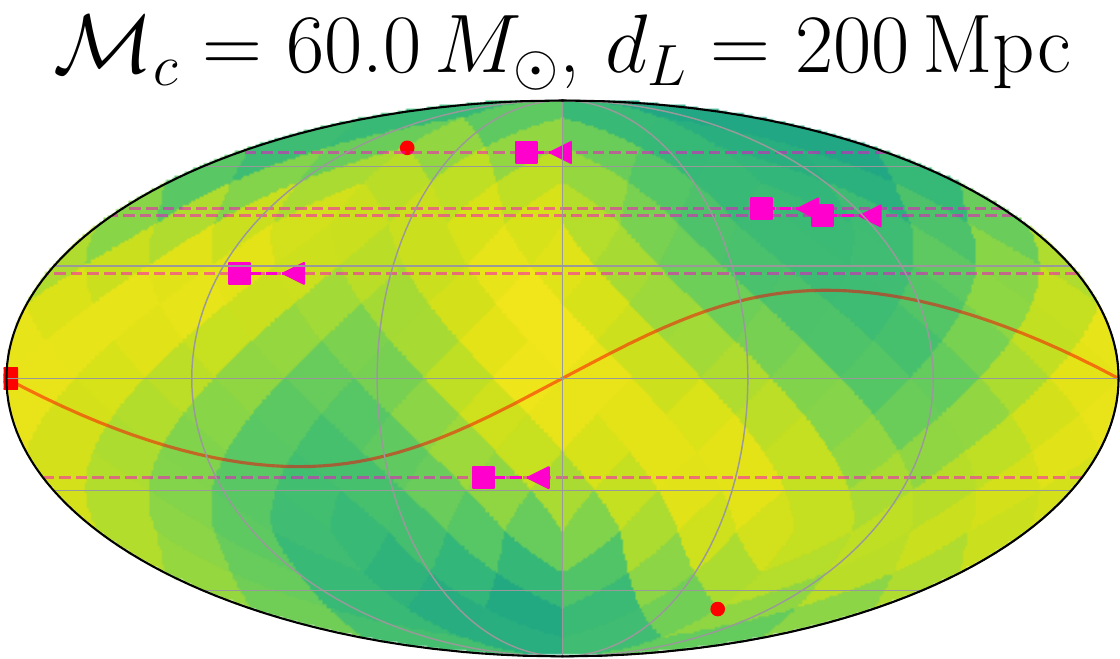}
   \vspace{0.1cm}
   
   \includegraphics[trim=3.2cm 5.5cm 3.6cm 0cm, clip, width=0.32\linewidth]{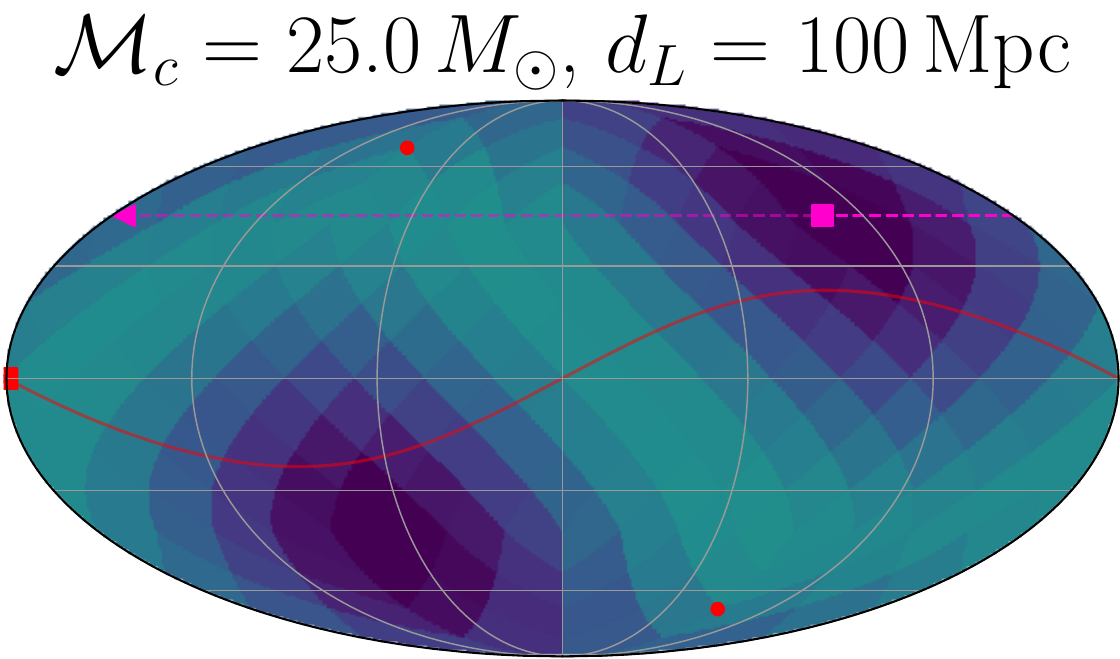}
   \includegraphics[trim=3.2cm 5.5cm 3.6cm 0cm, clip, width=0.32\linewidth]{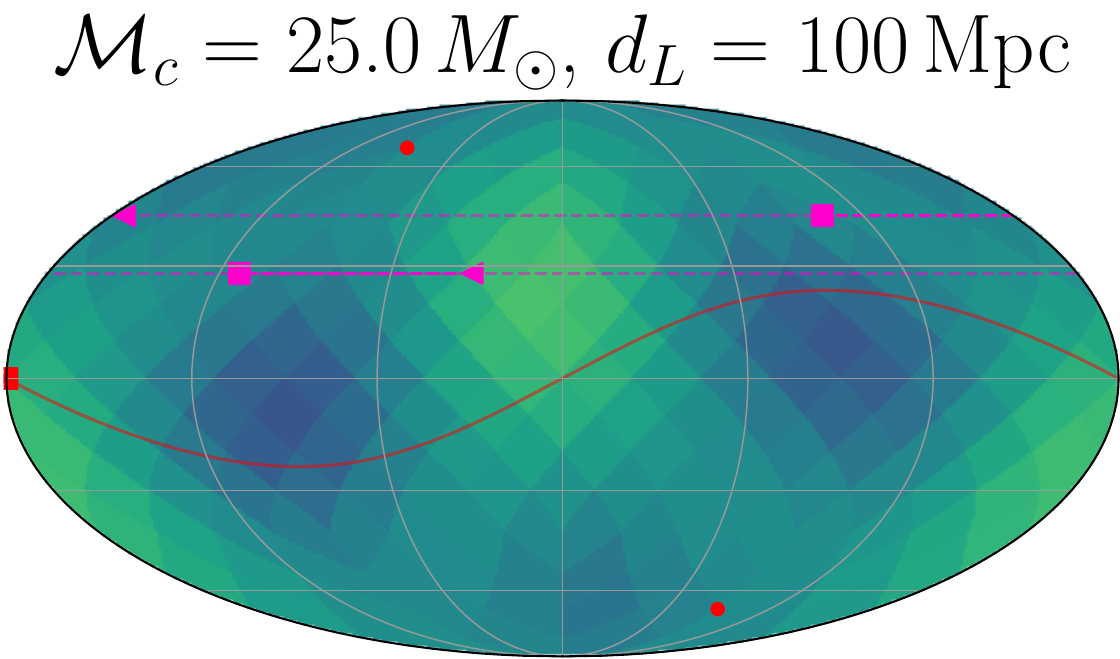}
   \includegraphics[trim=3.2cm 5.5cm 3.6cm 0cm, clip, width=0.32\linewidth]{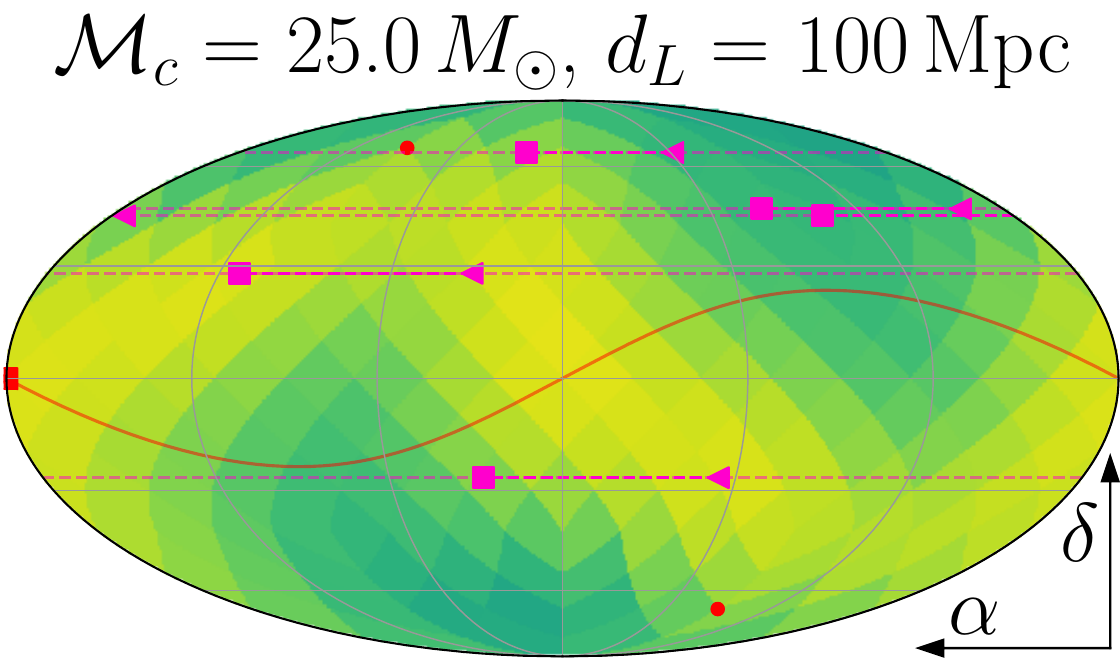}

   \includegraphics[trim=0cm 0.3cm 0cm 11.5cm, clip, width=0.43\linewidth]{Figs/skymap_SNR_Ground_Mc25_c}
   \vspace{-0.3cm}
   \caption{{\bf SNR} as a function of the source's sky-location for ``MAGIS-1\,km''. All plots are Mollweide projections of equatorial coordinates, right ascension ($\alpha$) increases from right to left, declination ($\delta$) increases from bottom to top, and the origin ($\alpha = \delta = 0$) is at the center of each map. The solid red line marks the ecliptic plane and the red circles indicate the ecliptic poles. The red triangle and square are the start and end points, respectively, of Earth's location as seen from the Sun over the course of the observation. The dashed magenta lines show the directions the baselines of each detector are pointing; magenta triangles and squares mark their start and end locations, respectively. {\it Upper row, left-to-right:} Results for $\mathcal{M}_c = \{1.1,~5,~60\}\,M_\odot$, respectively, for the \TMN network (i.e., detectors at Homestake, Sudbury, Renstr\"om, Tautona, and Zaoshan). {\it Lower row, left-to-right:} Results for $\mathcal{M}_c = 25\,M_\odot$ for Homestake alone, Homestake and Zaoshan, and for all five detectors. All other parameters are set to the values given in Tab.~\ref{tab:Parameters}.}
   \label{fig:skymap_SNR_Ground}
   \vspace{0.4cm}
   
   \includegraphics[trim=3.8cm 5.8cm 3.9cm 0cm, clip, width=0.32\linewidth]{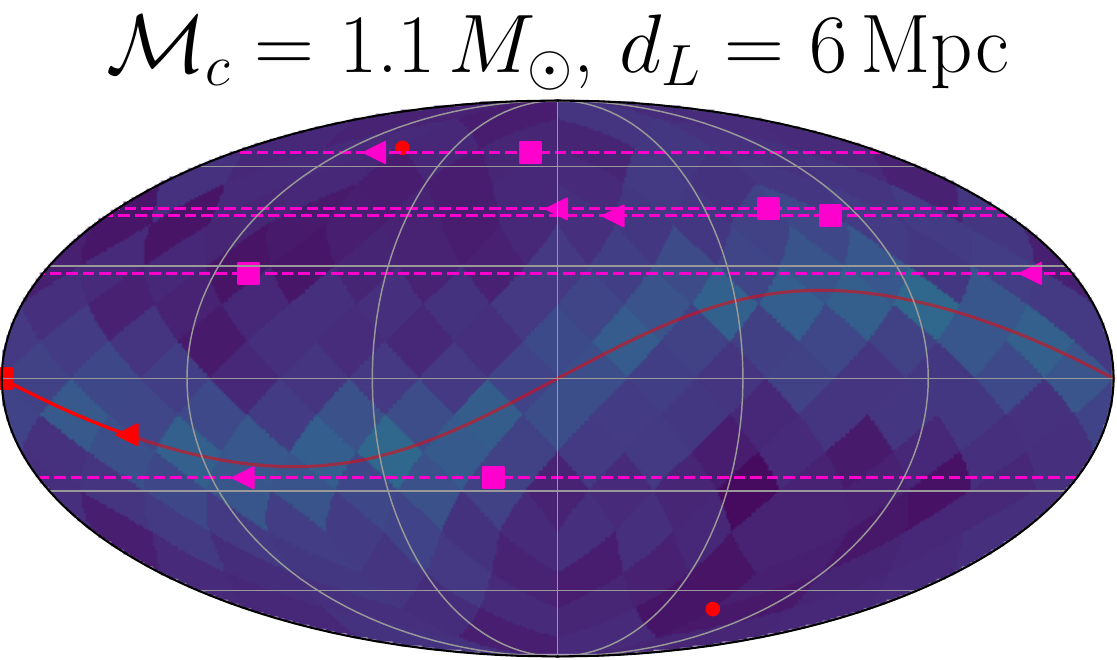}
   \includegraphics[trim=3.8cm 5.8cm 3.9cm 0cm, clip, width=0.32\linewidth]{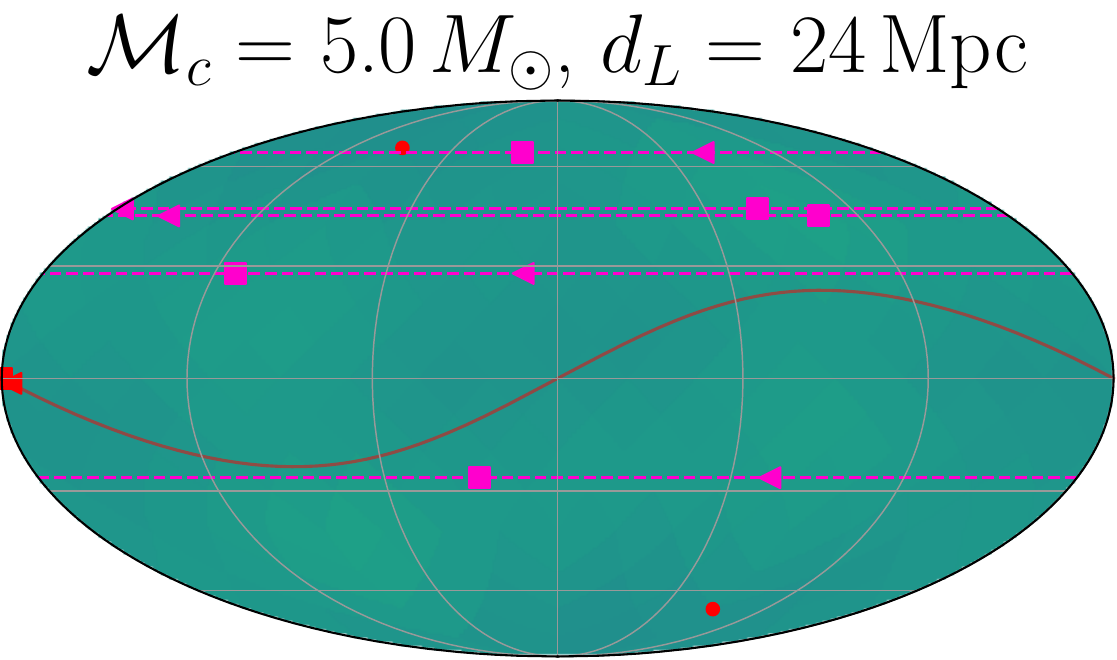}
   \includegraphics[trim=3.8cm 5.8cm 3.9cm 0cm, clip, width=0.32\linewidth]{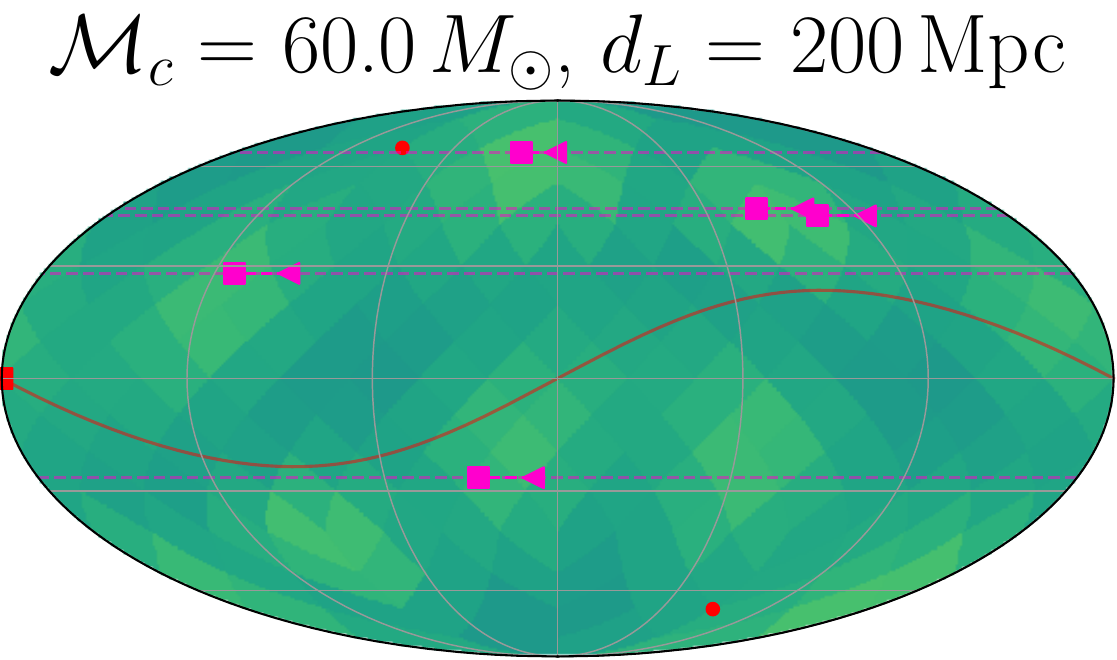}
   \vspace{0.1cm}
   
   \includegraphics[trim=3.8cm 5.8cm 3.9cm 0cm, clip, width=0.32\linewidth]{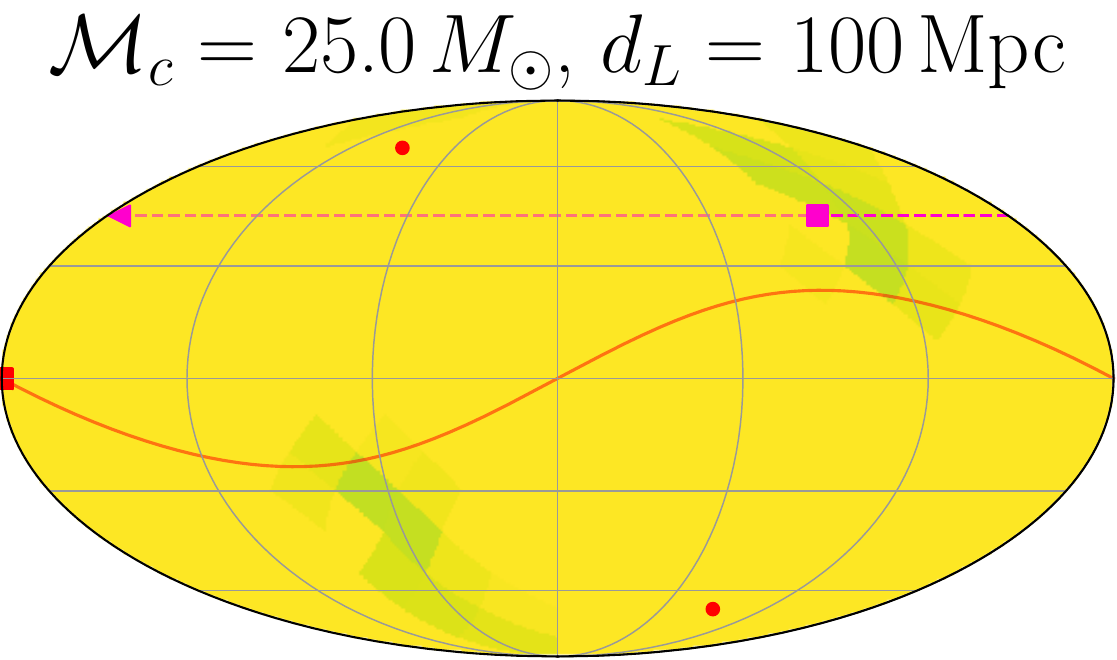}
   \includegraphics[trim=3.8cm 5.8cm 3.9cm 0cm, clip, width=0.32\linewidth]{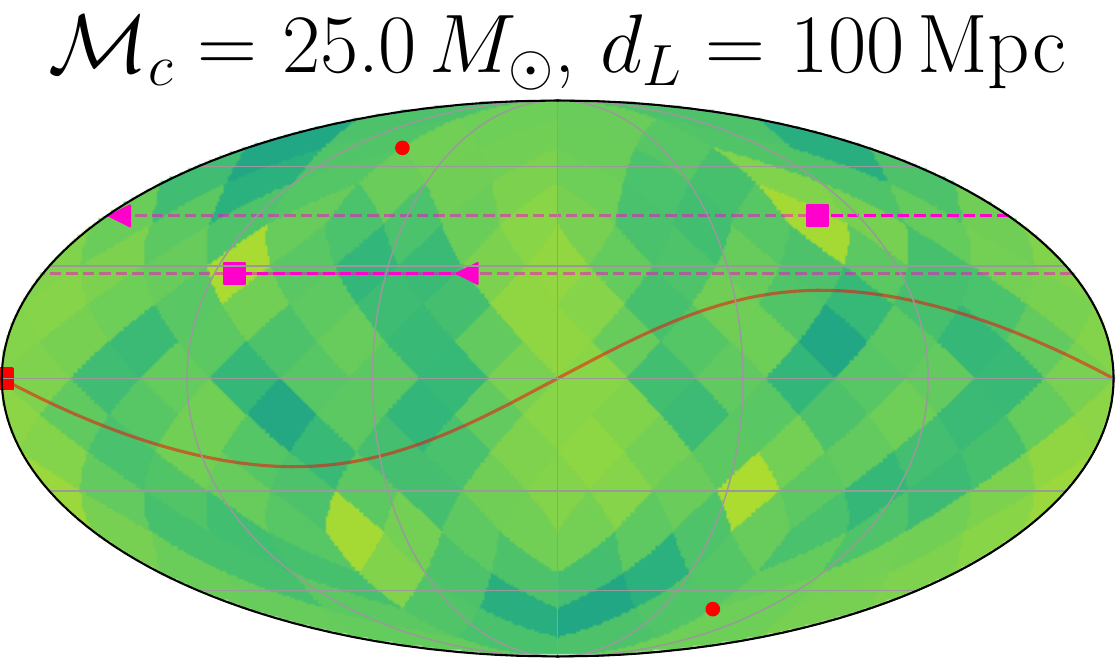}
   \includegraphics[trim=3.8cm 5.8cm 3.9cm 0cm, clip, width=0.32\linewidth]{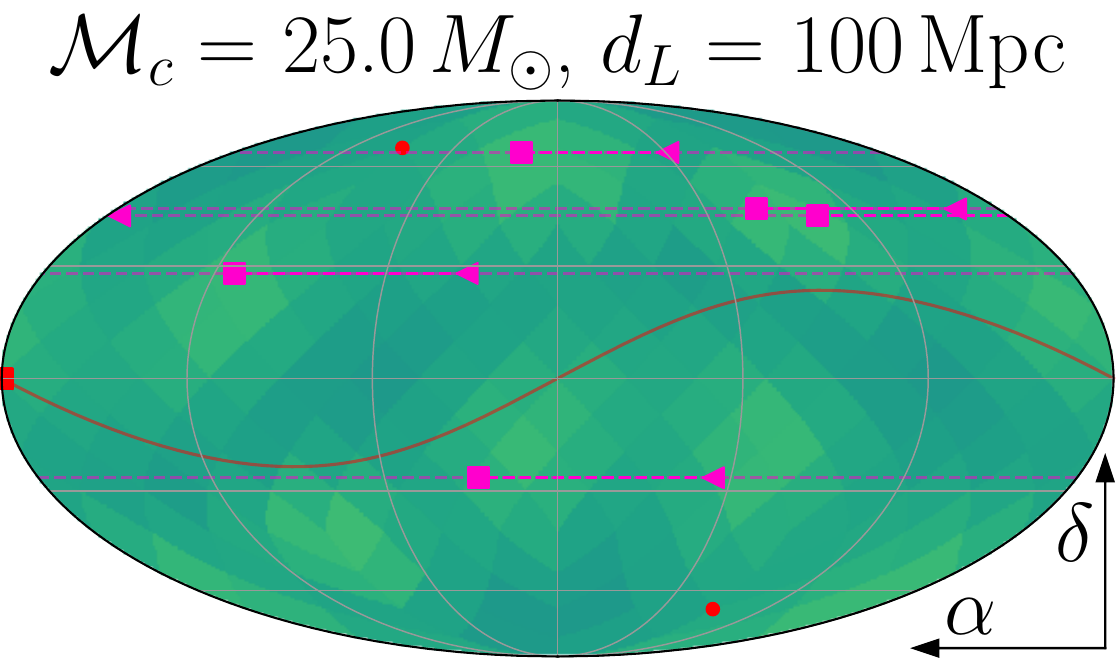}

   \includegraphics[trim=0cm 0.3cm 0cm 11.5cm, clip, width=0.45\linewidth]{Figs/skymap_angRes_Ground_Mc25_c}
   \vspace{-0.3cm}
   \caption{{\bf \boldmath Sky-localization error ($\sqrt{\sigma_{\Omega_n}}$)} [see Eq.~\eqref{eq:OmegaNUncertaintyDefinition}] as a function of the source's sky-location for ``MAGIS-1\,km'', otherwise the same as Fig.~\ref{fig:skymap_SNR_Ground}.}
   \label{fig:skymap_angRes_Ground}
\end{figure}

Let us now turn our attention to terrestrial ``MAGIS-1\,km'' detectors: in Fig.~\ref{fig:skymap_SNR_Ground}, we show sky-maps of the SNR, and, in Fig.~\ref{fig:skymap_angRes_Ground}, of the sky-localization error. We can immediately note that the patterns of SNR and $\sqrt{\sigma_{\Omega_n}}$ are rather different than what we found for ``MAGIS-space'' in Figs.~\ref{fig:skymap_SNR_Space} and~\ref{fig:skymap_angRes_Space}. Many of these qualitative differences are driven by the much shorter lifetime of signals in the ``MAGIS-1\,km'' sensitivity band (see Tab.~\ref{tab:LifetimeHorizon}).

In order to understand the patterns of SNR we see in Fig.~\ref{fig:skymap_SNR_Ground}, the relevant time-scale to compare the signals' lifetime to is again the time it takes the detectors to re-orient the direction of their baselines. Since we are considering terrestrial detectors, this timescale is set by Earth's rotational period of $24\,$h. While for ``MAGIS-space'', all signals could be observed for time-scales long compared to the time it takes a satellite-borne detector in medium Earth orbit to reorient, this is no longer true for the shorter lifetime of signals in the sensitivity band of terrestrial ``MAGIS-1\,km'' detectors. Hence, we see increasing variations of the SNR along the direction in which the baselines rotate with increasing $\mathcal{M}_c$ in Fig.~\ref{fig:skymap_SNR_Ground}; see especially the three panels in the top row. For the particular choice of the detector locations in the \TMN network, note that the detectors at Homestake, Sudbury and Tautona all have baselines pointing in rather similar directions (up to the orientation ${\bm \ell} \to -{\bm \ell}$). Thus, despite considering a network of five detectors, we find that for sources with lifetimes shorter than the timescale over which the detectors reorient [see especially the right-most panels in Fig.~\ref{fig:skymap_SNR_Ground} for $\mathcal{M}_c = 60\,M_\odot$ (top) and $\mathcal{M}_c = 25\,M_\odot$ (bottom)], the SNR for sources with sky-location approximately aligned with the direction of the baselines of these three detectors is still a factor of almost two smaller than for sources orthogonal to the directions these three detectors' baselines point in. For longer-lived sources (see especially $\mathcal{M}_c = 1.1\,M_\odot$ shown in the top-left panel), the rotation of the detectors' baselines smoothes out the coverage over the sky and we see only much smaller variations in the SNR. 

Conversely, the variation of the SNR across the sky does not increase substantially for the two-detector Homestake$+$Zaoshan network compared to the \TMN network, although the signal from the $\mathcal{M}_c = 25\,M_\odot$ source shown in the bottom row of Fig.~\ref{fig:skymap_SNR_Ground} has a lifetime of only $\sim 5\,$h in the detectors' sensitivity band. This is because the direction of these two specific baselines\footnote{Recall that, throughout this work, we assume a vertical layout of the detectors.} differs by approximately $90^\circ$, the optimal choice in order to realize the most uniform SNR coverage to sources across the sky for a network of two single-baseline detectors. Indeed, the variation in SNR we see across the sky for the two-detector network in Fig.~\ref{fig:skymap_SNR_Ground} is approximately $\sqrt{2} \sim 1.4$, as one would obtain for a short-lived signal observed in a network of two single-baseline detectors with baselines differing by $90^\circ$. 

In the bottom-left panel of Fig.~\ref{fig:skymap_SNR_Ground}, besides the overall reduction of the SNR due to the choice of a single ``MAGIS-1\,km'' detector, we find a much larger variation of the SNR across the sky: the smallest and largest values of the SNR we observe differ by more than one order of magnitude. Note that the smallest SNR is found for sources whose sky-location is approximately aligned with the direction of the detector's baseline when the signal leaves the detectors frequency band (marked by the black square). This can be understood from comparing the shape of the ``MAGIS-1\,km'' sensitivity curve to the frequency-dependence of the strain signal, see Fig.~\ref{fig:NoiseCurve}. As discussed in App.~\ref{app:Detectors}, the sensitivity of terrestrial atom-interferometer detectors to GW signals is limited by seismic Newtonian Gravity Gradient Noise at $f_{\rm GW} \lesssim 1\,$Hz. Thus, for a ``MAGIS-1\,km'' detector, the signal's GW strain is largest compared to the detector's noise at $f_{\rm GW} \approx 1\,$Hz. Since the chirp of the signals is a steep function of frequency [$d f_{\rm GW}/dt \propto f_{\rm GW}^{11/3}$; see Eq.~\eqref{eq:FrequencyEvolution}], the time at which the GW signal is produced at $1\,$Hz is much closer to the time when the signal leaves the detector's frequency band (when $f_{\rm GW} = 10\,$Hz) than to the time the signal enters the frequency band (when $f_{\rm GW} = 0.1\,$Hz). However, as discussed above, a network of two, or more, terrestrial detectors placed in different locations leads to much more homogeneous sky-coverage.

Let us now discuss the variation of the sky-localization uncertainty, $\sqrt{\sigma_{\Omega_n}}$, as a function of sky-position we show in Fig.~\ref{fig:skymap_angRes_Ground}. Recall that only the signal from a $\mathcal{M}_c = 1.1\,M_\odot$ source, with a lifetime of ${\sim}\,40\,$days, can be measured for long enough for terrestrial detectors to complete a significant fraction of Earth's orbit around the Sun. Accordingly, we see in the top-left panel of Fig.~\ref{fig:skymap_angRes_Ground} that the sky-localization error depends on the relative orientation of the source with respect to the ecliptic plane, with the smallest $\sqrt{\sigma_{\Omega_n}}$ found for sources close to the ecliptic poles. 

The signals from all other sources shown in Fig.~\ref{fig:skymap_angRes_Ground} have lifetimes in the ``MAGIS-1\,km'' sensitivity band too short for the detectors to make use of the changing Doppler shift as Earth orbits the Sun. The resulting patterns in $\sqrt{\sigma_{\Omega_n}}$ we can observe in the top-middle, top-right, and bottom-right panel of Fig.~\ref{fig:skymap_angRes_Ground} thus have small variations that arise from a combination of SNR (see Fig.~\ref{fig:skymap_SNR_Ground}) and the relative orientation of the baselines with respect to the sources: as we discussed for the ``MAGIS-space'' results above, if a GW source is oriented such that one of the baselines can be pointed directly at the source, the ability to localize such a source on the sky improves. Otherwise, the ability of a \TMN network to localize sources with signals with lifetimes $\ll 1\,$month in the sensitivity band is almost independent of $\mathcal{M}_c$ (when adjusting $d_L$ to keep the SNR constant). Across the top-middle, top-right, and bottom-right panels Fig.~\ref{fig:skymap_angRes_Ground} we find $\sqrt{\sigma_{\Omega_n}} \sim 10^\circ$. 

Comparing the ability of different networks of ``MAGIS-1\,km'' detectors to localize sources on the sky, we can see from the panels in the bottom row of Fig.~\ref{fig:skymap_angRes_Ground} that, while a two-detector network (shown in the bottom-middle panel) loses some of the ability to localize sources on the sky compared to the full \TMN network, a single ``MAGIS-1\,km'' detector has almost no ability to constrain the sky-position of a GW source with signal lifetime shorter than a month. This motivates building multiple terrestrial detectors in different locations on Earth.

Although we leave an exhaustive discussion of the optimal combination of terrestrial detectors for future work, let us here give a broad overview based on preliminary studies we have performed. As discussed in the previous paragraphs, there are clear advantages of a network of multiple terrestrial detectors over a single ``MAGIS-1\,km'' detector. Many GW sources that could be observed in MAGIS-1\,km-like detectors give rise to signals with short lifetimes compared to the rotational period of the Earth. Operating a network of multiple detectors thus brings two advantages: the GW strain could be measured along multiple different directions, and the difference in the arrival time of the signal at detectors in different locations allows one to localize the source on the sky, which in turn also leads to better constrains on any other parameter that is (partially) degenerate with the source's sky-location. For two detectors, a separation of ${\sim}\,90^\circ$ is optimal from the viewpoint of achieving the most homogeneous SNR coverage and parameter reconstruction abilities for sources across the sky. On the other hand, a network of two detectors with opening angle of ${\sim}\,45^\circ$ would lead to the smallest errors in the binary's parameters for sources located orthogonally to the plane spanned by the detector's  baselines -- then, such a configuration would be maximally sensitive to the strain in both the $h_+$ and the $h_\times$ polarization modes. Furthermore, for sources that can be measured on time-scales longer than $\mathcal{O}(1)\,$h, only north-south separation matters, as east-west separation is washed out by Earth's rotation. Generally, it would of course be advantageous to build as many detectors in as many different places on Earth as possible. Taking into account practicalities, it is interesting to note that a network of three detectors in China, Northern Europe, and the north-east of the US/south-west of Canada would be a promising network configuration maximizing differences in arrival time, achieving approximately homogeneous sky-coverage by featuring baselines separated by ${\sim}\,90^\circ$, and allowing one to break degeneracies between parameters by featuring pairs of baselines differing by ${\sim}\,45^\circ$.

\subsection{Dependence on Time to Merger} \label{sub:ObservationTimeScaling}

Gravitational wave detectors in the mid-band are well-suited to pinpointing the position of sources on the sky. Furthermore, due to their frequency coverage, they can observe the signals from ${\sim}\,1-100\,M_\odot$ compact binaries well before these binaries merge. This combination gives mid-band detectors a unique role to play in multi-messenger astronomy: by predicting the sky-location and time of merger of compact binaries in advance, they could enable electromagnetic telescopes to plan targeted observations. In this section, we investigate the ability of a mid-band GW detector to predict the sky-location and time of merger as a function of the time until merger; see, e.g., Refs.~\cite{Cannon:2011vi,Adams:2015ulm,Chan:2018csa,Sachdev:2020lfd,Magee:2021xdx,Dax:2021tsq,Nitz:2021pbr,Kovalam:2021bgg,Hu:2023hos,Chen:2023qga} for work on early-warning capabilities in other frequency bands.

\begin{figure}
   \centering
   \text{MAGIS-space, ecliptic orbit}
   \vspace{0.3cm}
   
   \includegraphics[trim=0.3cm 0.3cm 0.3cm 0.3cm, clip, width=0.32\linewidth]{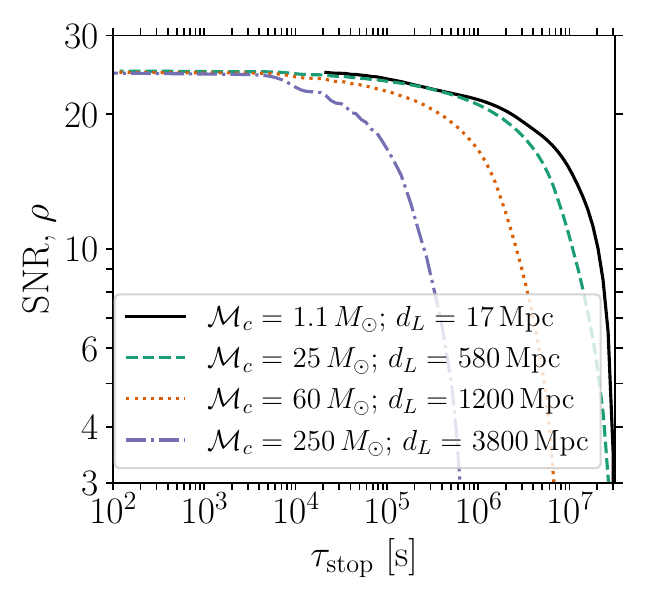}
   \includegraphics[trim=0.3cm 0.3cm 0.3cm 0.3cm, clip, width=0.32\linewidth]{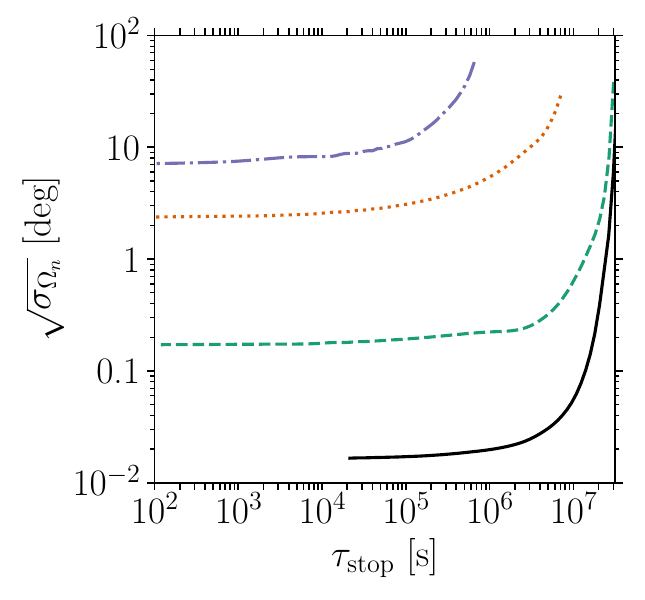}
   \includegraphics[trim=0.3cm 0.3cm 0.3cm 0.3cm, clip, width=0.32\linewidth]{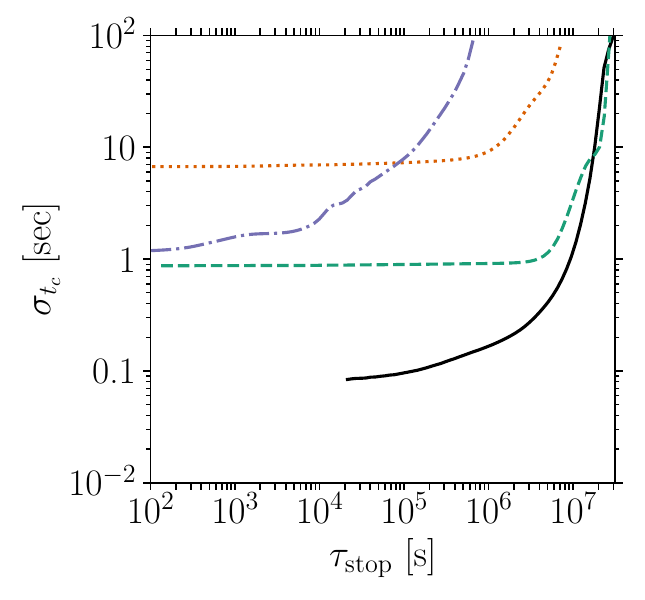}
   \caption{{\bf ``MAGIS-space'': early warning capability.} From left to right, the three panels show the SNR ($\rho$), the sky-localization uncertainty ($\sqrt{\sigma_{\Omega_n}}$), and the uncertainty with which the time of merger can be measured ($\sigma_{t_c}$) as a function of how early before the (true) merger time the observation is stopped in a ``MAGIS-space'' detector in geocentric orbit parallel to the ecliptic plane. In particular, we include the data a detector would record between either the time when the signal enters the detector's frequency band or one year before it leaves the detector's frequency band (which ever time is later) and the time $\tau_{\rm stop}$ before the merger-time in the analysis. The different lines are for different choices of the chirp mass ($\mathcal{M}_c$) and luminosity distance ($d_L$) as denoted in the legend in the left panel; note that we have chosen $d_L$ for each source such that $\rho = 25$ for $\tau_{\rm stop} \to 0$. All other parameters are set to the benchmark values given in Tab.~\ref{tab:Parameters}. The value of $\tau_{\rm stop}$ where $\rho \to 0$ for each source corresponds to the time before merger when the signal enters the detectors sensitivity band [$\tau(f_{\rm GW} \approx 30\,{\rm mHz})$]. Towards small values of $\tau_{\rm stop}$, we end each line at the time before merger when the signal leaves the detector's sensitivity band [$\tau_{\rm stop}(f_{\rm GW} \approx 3\,{\rm Hz})$].}
   \label{fig:advanceTime_Space}
   
   \vspace{0.5cm}
   \text{$5 \times \left(\text{MAGIS-1\,km}\right)$}
   \vspace{0.3cm}
   
   \includegraphics[trim=0.3cm 0.3cm 0.3cm 0.3cm, clip, width=0.32\linewidth]{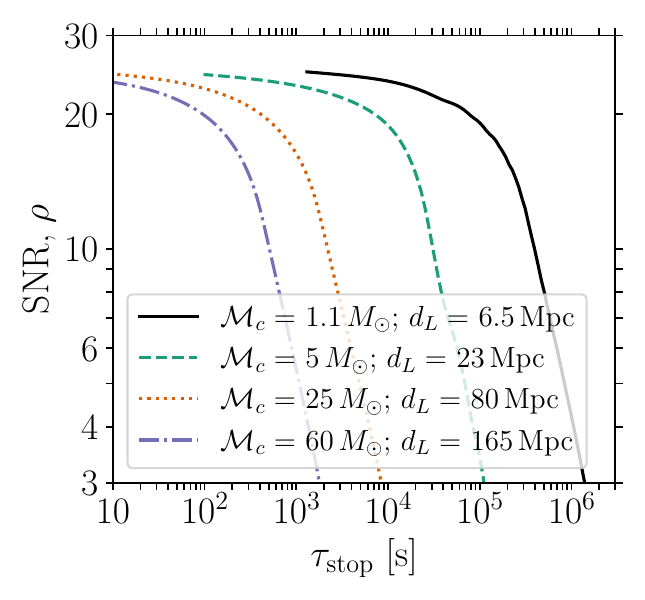}
   \includegraphics[trim=0.3cm 0.3cm 0.3cm 0.3cm, clip, width=0.32\linewidth]{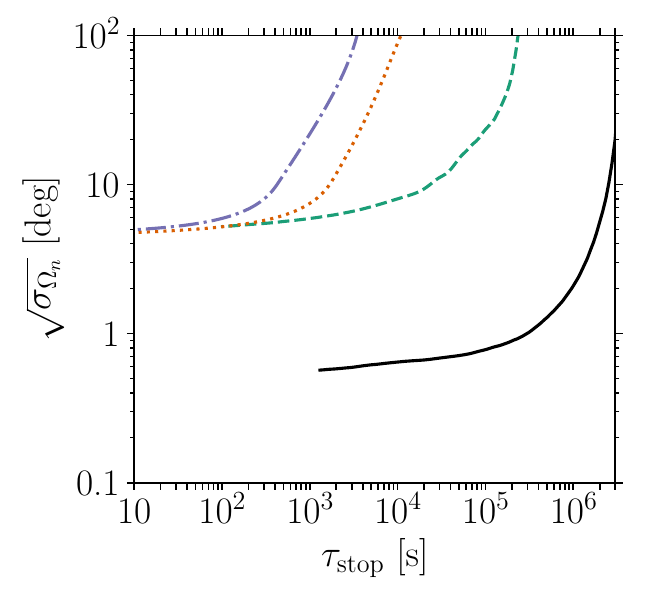}
   \includegraphics[trim=0.3cm 0.3cm 0.3cm 0.3cm, clip, width=0.32\linewidth]{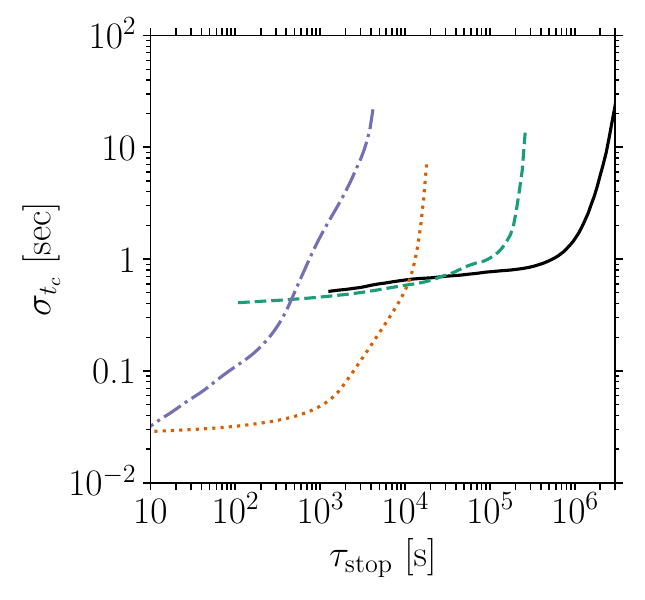}
   \caption{{\bf ``MAGIS-1\,km'': early warning capability.} Same as Fig.~\ref{fig:advanceTime_Space}, but for a terrestrial \TMN network. Note that the choices of $\mathcal{M}_c$ and $d_L$ for the respective lines differ from those in Fig.~\ref{fig:advanceTime_Space}. The value of $\tau_{\rm stop}$ where $\rho \to 0$ for each source corresponds to the time before merger when the signal enters the detector's sensitivity band [$\tau(f_{\rm GW} \approx 0.5\,{\rm Hz})$]. Towards small values of $\tau_{\rm stop}$, we end each line at the time before merger when the signal leaves the detectors sensitivity band [$\tau_{\rm stop}(f_{\rm GW} \approx 10\,{\rm Hz})$].}
   \label{fig:advanceTime_Ground}
\end{figure}

In Figs.~\ref{fig:advanceTime_Space} and~\ref{fig:advanceTime_Ground} we show the SNR ($\rho$), the sky-localization uncertainty ($\sqrt{\sigma_{\Omega_n}}$), and the uncertainty of the time-of-merger ($\sigma_{t_c}$) as a function of how long before merger of the respective source the observation is stopped for ``MAGIS-space'' and for the terrestrial \TMN network. Concretely, let us define the time-before-merger as $\tau \equiv t_c - t$, where $t_c$ is the binary's time of merger. Furthermore, let us denote the smallest and largest GW frequencies a detector is sensitive to as $f_{\rm GW}^{\rm min}$ and $f_{\rm GW}^{\rm max}$, respectively (for ``MAGIS-space'', $f_{\rm GW}^{\rm min} \approx 30\,$mHz and $f_{\rm GW}^{\rm max} \approx 3\,$Hz, while for ``MAGIS-1\,km'', $f_{\rm GW}^{\rm min} \approx 0.5\,$Hz and $f_{\rm GW}^{\rm max} = 10\,$Hz). In Figs.~\ref{fig:advanceTime_Space} and~\ref{fig:advanceTime_Ground}, we show how $\rho$, $\sqrt{\sigma_{\Omega_n}}$, and $\sigma_{t_c}$ change if we include only the data the detector would record between $\tau_{\rm start} \equiv \min\{ \tau(f_{\rm GW}^{\rm min}), \tau(f_{\rm GW}^{\rm max}) + 1\,{\rm yr} \}$ and $\tau_{\rm stop}$ with different choices of $\tau_{\rm stop}$. The smallest value of $\tau_{\rm stop}$ we show for each GW source in Figs.~\ref{fig:advanceTime_Space} and~\ref{fig:advanceTime_Ground} is $\tau_{\rm stop} = \tau(f_{\rm GW}^{\rm max})$, and as $\tau_{\rm stop} \to \tau(f_{\rm GW}^{\rm min})$, we of course find $\rho \to 0$ and $\sqrt{\sigma_{\Omega_n}}, \sigma_{t_c} \to \infty$. 

Let us first discuss the results for a ``MAGIS-space'' detector, shown in Fig.~\ref{fig:advanceTime_Space}. We show results for the same choices of the chirp mass as before, albeit with different values of $d_L$ here set such that for each source, $\rho \approx 25$ as $\tau_{\rm stop} = \tau(f_{\rm GW}^{\rm max})$. 

The main result from Fig.~\ref{fig:advanceTime_Space} is that the time and location of the merger can be predicted well in advance. Neither the SNR nor the ability to determine when and where a merger will occur are worsened by more than order-unity factors unless an order-unity fraction of the signal is excluded from the analysis. Even for the shortest-lived source we consider here, $\mathcal{M}_c = 250\,M_\odot$, the observation can be stopped $\tau_{\rm stop} \sim 10^4\,{\rm s} \approx 3\,$hours before the time of merger without any appreciable loss in SNR or relevant increase in $\sqrt{\sigma_{\Omega_n}}$ or $\sigma_{t_c}$ compared to observing the signal until it leaves the ``MAGIS-space'' sensitivity band. In fact, in almost all cases the merger can be predicted accurately many days in advance. In the context of multi-messenger astronomy, an early-warning time of at least a few hours is sufficient for any telescope to set its sights on the sky-location where the merger event is expected to occur -- even space-based telescope such as Hubble Space Telescope or James Webb Space Telescope need less than one hour to slew their telescope by $180^\circ$. 

In the right panel of Fig.~\ref{fig:advanceTime_Space} we show how well the merger time, $t_c$, could be constrained. As we discussed in Sec.~\ref{sec:Code}, in our computation, $t_c$ is defined in an inertial frame (co-moving with the solar system) located at the position of the center of Earth at the mid-point in time between the signal entering and leaving the sensitivity band of the detector to compute the GW signal. The detector measures a quantity related to $t_c$ but defined in its accelerated frame; the relation of that quantity to $t_c$ involves a frame transformation that requires knowing the sky-position of the source. Parametrically, for a source whose signal is measured over the time-span of a year in a detector orbiting Sun with the Earth, this induces an uncertainty in $t_c$ of order $1\,{\rm AU}/\sqrt{\sigma_{\Omega_n}}$, e.g., for $\sqrt{\sigma_{\Omega_n}} \sim 1^\circ$, this amounts to an uncertainty on $t_c$ of ${\sim}\,1\,$s from the frame transformation. Taking the lifetime of sources in the detector band as well as the respective values we find for $\sqrt{\sigma_{\Omega_n}}$ for the choices of $\mathcal{M}_c$ made in Fig.~\ref{fig:advanceTime_Space}, we can understand the results for $\sigma_{t_c}$. In the context of multi-messenger astronomy, let us note that the merger time of all sources considered in Fig.~\ref{fig:advanceTime_Space} could be forecasted with $\sigma_{t_c} < 10\,$s at least a day before the merger event, which is more than sufficient timing forecast to plan targeted observations with electromagnetic observatories.

In Fig.~\ref{fig:advanceTime_Ground}, we show equivalent results for a terrestrial \TMN network. Note that the choices of $\mathcal{M}_c$ for which we show results differ from those in Fig.~\ref{fig:advanceTime_Space}, and we have again adjusted the luminosity distance of each source such that $\rho \approx 25$ as $\tau_{\rm stop} = \tau(f_{\rm GW}^{\rm max})$. The behavior of the SNR (left panel), sky-localization uncertainty (middle panel), and the uncertainty with which the merger time can be predicted (right panel) with $\tau_{\rm stop}$ is analogous to what we discussed above for ``MAGIS-space.'' Again, sky localization and merger time prediction ability are not significantly affected as long as $\tau_{\rm stop} \lesssim 10\,$min. For many sources, the merger time and location could be predicted hours or days in advance with the terrestrial detector network.

In summary, the results shown in this section demonstrate that GW detectors in the mid-band are well-suited for multi-messenger astronomy. For essentially any event that can be well-localized from observations in a mid-band GW detector, both the sky-position and the time of merger can be predicted well in advance of the merger event, giving telescopes a chance to observe any potential electromagnetic signal generated in association with the merger. 

\section{Conclusions} \label{sec:Conclusion}

In this work, we have presented a detailed discussion of the capabilities of GW detectors in the $f_{\rm GW} \sim 30\,{\rm mHz} - 10\,$Hz ``mid-band'' to measure the signals from and constrain the properties of ${\sim}1-100\,M_\odot$ compact binaries. We have focused on atom-interferometer detectors, a promising technology for GW detectors in the mid-band, as we have discussed in Sec.~\ref{sec:MidbandGWDetection}. In order to estimate the SNR and forecast the uncertainty with which the GW source's parameters can be reconstructed, we have performed a Fisher information matrix analysis; we have made the numerical code used to produce all results in this work publicly available at this link\footnote{\href{\linkCode}{\url{\linkCode}}}: \href{\linkCode}{\faGithubSquare}. This is both the first public Fisher analysis code designed to work in the mid-band, and an improvement over some previous codes that ignored certain Doppler corrections to the GW signal \cite{Graham:2017lmg, Cai:2021ooo, Yang:2021xox, Yang:2022iwn}.

We have presented numerical results for two particular detector benchmarks: a (network) of terrestrial km-size detectors with strain sensitivity given by the ``MAGIS-1\,km'' curve in Fig.~\ref{fig:NoiseCurve}, and a satellite-born detector in medium Earth orbit with sensitivity given by the ``MAGIS-space'' curve in Fig.~\ref{fig:NoiseCurve}. While the precise numerical results are of course tied to these particular detector choices, the qualitative results we obtain here are applicable to any GW detector in the mid-band. In Sec.~\ref{sec:ResultsDiscussion}, we have presented a number of numerical results and discussed them in detail. Let us summarize some of the key findings:
\begin{itemize}
    \item A network of five terrestrial ``MAGIS-1\,km'' detectors has a horizon at which it could observe a source with SNR $\rho=5$ of ${\sim}\,30\,$Mpc for a system with a (detector frame) chirp mass of $\mathcal{M}_c = 1.1\,M_\odot$, while the horizon for a $\mathcal{M}_c = 25\,M_\odot$ source is ${\sim}\,400\,$Mpc. A satellite-born ``MAGIS-space'' detector, on the other hand, has a horizon of ${\sim}\,100\,$Mpc for $\mathcal{M}_c = 1.1\,M_\odot$ sources, the horizon for $\mathcal{M}_c = 25\,M_\odot$ sources is ${\sim}\,3\,$Gpc, and the signal from a black hole binary with $\mathcal{M}_c = 250\,M_\odot$ could be observed out to a distance of ${\sim}\,20\,$Gpc corresponding to a cosmological redshift of $z \sim 2$. (See Fig.~\ref{fig:ChirpMass_SNR_Scans})

    \item A network of multiple ``MAGIS-1\,km'' detectors has far superior ability to measure the signals from GW sources across the sky and to reconstruct their parameters compared to a single detector, beyond the trivial improvement in SNR from increasing the number of detectors. When operating a network of only two detectors, placing them ${\sim}\,90^\circ$ apart on the surface of Earth appears ideal from a standpoint of achieving homogeneous sensitivity across the sky and parameter reconstruction abilities. For example, the SNR of signals in such a two-detector network would vary by a factor of less than ${\sim}\,1.5$ across the sky. When considering a network of three detectors, one detector in the north-east of the US or the south-east of Canada, one detector in Northern Europe, and one detector in China seems like a strong compromise between ideal detector placements for homogeneous sky coverage, parameter reconstruction capabilities, and practicality.

    \item Considering a satellite-borne detector, the SNR for ${\sim}\,1-100\,M_\odot$ sources would vary less than a factor of two across the sky for a single ``MAGIS-space'' detector in medium Earth orbit. Operating a satellite born detector in geocentric rather than heliocentric orbit has multiple advantages: not only would such a detector change the direction in which its baseline points on hour-timescales, but furthermore, its orbit around Earth is important for localizing sources on the sky that have signals with lifetimes shorter than $\mathcal{O}(1)\,$month in the detector. (See Appendix~\ref{app:SupplementalPlots} and Fig.~\ref{fig:spaceOrbits}.)

    \item Gravitational wave detectors in the mid-band are poorly placed to measure the luminosity distance of GW sources. For ${\sim}\,1-100\,M_\odot$ sources, even if the source is close enough to give rise to a $\rho = 25$ signal, we find a typical uncertainty of $\sigma_{d_L}/d_L \sim 10\,\%$, and for a ``MAGIS-space'' detector, the ability to measure the luminosity distance to sources with $\mathcal{M}_c \lesssim 10\,M_\odot$ is even poorer. This is because GW detectors in the mid-band would measure the GW signal produced during the inspiral phase of ${\sim}\,1-100\,M_\odot$ binaries, leading to degeneracies between the luminosity distance and other parameters describing the binary (most importantly, its inclination angle). (See Fig.~\ref{fig:ChirpMass_Omegan_Scans})

    \item Gravitational wave detectors in the mid-band have a unique capability to pinpoint the location of sources on the sky. Many GW sources can give rise to signals that have lifetimes of at least ${\sim}\,1\,$month in the sensitivity band of mid-band detectors: for terrestrial ``MAGIS-1\,km'' detectors, this is true for $\mathcal{M}_c \lesssim 3\,M_\odot$ sources, while for a ``MAGIS-space'' detector, $\mathcal{M}_c \lesssim 50\,M_\odot$ sources give rise to signals with lifetimes in excess of a few months. Detectors in the mid-band can localize any such GW source that gives rise to a measurable signal (with SNR $\rho > 5$) with sub-degree angular precision. (See Figs.~\ref{fig:ChirpMass_Omegan_Scans} and~\ref{fig:angRes_dL_BP})

    \item The ability of detectors in the mid-band to pinpoint the location of GW sources on the sky is of particular relevance for multi-messenger astronomy. As we have shown, both the sky-position and the time of merger can be predicted well in advance: more than a few hours before merger for ``MAGIS-space'', with a precision of $\mathcal{O}(1)\,{\rm sec}$ for most sources.  Such an early-warning capability of a mid-band detector would give electromagnetic telescopes ample time to set their sights on the merger event. (See Figs.~\ref{fig:advanceTime_Space} and~\ref{fig:advanceTime_Ground})
\end{itemize}

There are many ways in which this study could be extended in future work. Let us list a few of these directions:
\begin{itemize}
    \item Throughout this work, we neglected effects induced by the spins of the binary's constituents. Depending on the magnitude and orientation of the spins, as well as the values of the other parameters, the effects of spins can either break degeneracies between parameters, or add new degeneracies. It would thus be worthwhile to study how the ability of GW detectors in the mid-band to constrain the property of GW sources changes with their constituents' spins. Similarly, the effects of eccentric orbits should be considered (see Ref.~\cite{Yang:2022tig} for work in this direction). 

    \item In this work, we studied signals from GW sources in the ${\sim}\,1-100\,M_\odot$ mass range, for which mid-band detectors would observe the GW signal generated during the inspiral phase of the binaries. Of course, such detectors could potentially also observe GW signals from sources that merge in the mid-band, for example, white dwarf binaries or black hole binaries with masses $\gtrsim 100\,M_\odot$. We leave considerations of such signals for future work.

    \item As we discussed above, mid-band detectors are poorly placed to measure the luminosity distance to ${\sim}\,1-100\,M_\odot$ sources. Detectors operating at higher frequencies, for example, terrestrial laser interferometer detectors, could observe the GW signal produced during the merger and ringdown phase of such binaries. It would be interesting to study how multi-band observations, combining mid-band detectors with detectors aimed at higher frequencies, could be used to further constrain the properties of GW sources, combining the strengths of detectors in different GW frequency bands.

    \item Throughout this work, we have assumed that only one GW signal is present at the detector at any given time. Given the lifetime of GW signals in the mid-band as well as the horizon of proposed detectors, this assumption may not be well-justified. One tool for atom-interferometer detectors to mitigate problems arising from overlapping signals is their ability to operate in a resonant readout mode, shaping the response of the detector to a narrow frequency range (see the discussion in App.~\ref{subapp:DetectorBenchmarks}). It would be worthwhile to study the effects of overlapping signals in mid-band detectors, and how well particular capabilities of atom-interferometer detectors, such as their resonant readout mode, can be used to disentangle overlapping signals and reduce the confusion noise. 
\end{itemize}
In the upcoming decades, a major aim of physicists and astronomers is to extend the frequency coverage of GW detectors to as much of the range from the nHz to the kHz (and beyond) with the greatest sensitivity possible. In this endeavour, it is crucial to evaluate what detector concepts aimed at particular regions of this vast frequency space excel at, what their weaknesses are, and how their designs can be optimized. In this work, we took a detailed look at the capabilities of atom-interferometer detectors aimed at the mid-band to measure signals from ${\sim}\,1-100\,M_\odot$ binaries and to constrain their properties. While only a first step, we hope that this work is a useful resource, and perhaps a starting point for future studies, in the community's effort to understand the capabilities of proposed detector concepts and to design the gravitational wave detectors of the future.

\acknowledgments
We are indebted to Thomas~Wilkason for important discussions during the preliminary stages of this work. 
We thank the authors of Refs.~\cite{Cai:2021ooo,Yang:2021xox,Yang:2022iwn}, especially Sunghoon~Jung and Tao~Yang, for providing results from their parameter estimation code for comparison with our results. 
The authors acknowledge support by NSF Grants PHY-2310429 and PHY-2014215, Simons Investigator Award No. 824870, DOE HEP QuantISED award \#100495, the Gordon and Betty Moore Foundation Grant GBMF7946, and the U.S. Department of Energy (DOE), Office of Science, National Quantum Information Science Research Centers, Superconducting Quantum Materials and Systems Center (SQMS) under contract No. DEAC02-07CH11359.
The work of SB was in part performed at the Aspen Center for Physics, which is supported by National Science Foundation grant PHY-2210452. 
ZB is supported by the National Science Foundation Graduate Research Fellowship under Grant No. DGE-1656518 and by the Robert and Marvel Kirby Fellowship and the Dr. HaiPing and Jianmei Jin Fellowship from the Stanford Graduate Fellowship Program. 
Some of the computing for this project was performed on the Sherlock cluster. We would like to thank Stanford University and the Stanford Research Computing Center for providing computational resources and support that contributed to these research results. 
We acknowledge the use of the Python scientific computing packages NumPy~\cite{Harris:2020xlr} and SciPy~\cite{Virtanen:2019joe}, as well as the graphics environment Matplotlib~\cite{Hunter:2007ouj}. Some of the results in this paper have made use of the healpy~\cite{Zonca:2019vzt,healpy} and HEALPix packages~\cite{Gorski:2004by}. 


\appendix

\section{Gravitational Waves from Compact Binaries} \label{app:Signal}

In this and the following appendix, we describe the GW signal from compact binaries and the response of a GW detector to this signal. For the purposes of this work, we neglect the possibilities of non-zero spins of the binary's components and of eccentric orbits. The GW signal and the response of the detector are then controlled by nine parameters: the chirp mass, $\mathcal{M}_c$, and the mass ratio, $q$, which are intrinsic to the binary, and seven parameters that describe the binary's position and orientation in space-time relative to the detector: the luminosity distance, $d_L$; a reference phase of the binary, $\Phi_0$; the time of merger of the binary, $t_c$; two angles describing the orientation of the binary's orbital momentum, which we described by the inclination angle, $\iota$, and the polarization angle, $\psi$; and the position of the binary on the sky, which we describe in heliocentric equatorial coordinates parameterized by right ascension and declination, $\alpha$ and $\delta$, respectively. These parameters are summarized in Tab.~\ref{tab:Parameters}.

We will focus on compact binaries with component masses in the ${\sim}\,1 - 100\,M_\odot$ mass range. A GW detector in the mid-band -- which, for the purposes of this work, we define as spanning the $f_{\rm GW} = 30\,{\rm mHz} - 10\,$Hz frequency range -- would observe the GW signal produced in the {\it inspiral} phase of the binary, i.e., when the binary is still approximately described by a pair of orbiting point particles. In this regime, the GW signal is well-described by the leading terms in the post-Newtonian (PN) expansion of General Relativity. For our numerical computations, we will employ 3.5/3.0 PN order (frequency evolution/amplitude correction) waveforms; see, for example, Refs.~\cite{Maggiore:2007ulw,Buonanno:2009zt,Blanchet:2013haa,Mishra:2016whh,Isoyama:2020lls} for the corresponding expressions. Here, we discuss the leading-order expression for the waveforms, which suffice to illustrate the qualitative effects the parameters have on the GW signal. 

We begin with the time evolution of the GW frequency, $f_{\rm GW}$, described by (see, for example, Refs.~\cite{Cutler:1994ys, Poisson:1995ef, Holz:2005df})
\begin{align}
    \frac{d f_{\rm GW}}{dt} &= \frac{96}{5}\pi^{8/3}\mathcal{M}_c^{5/3}f_{\rm GW}^{11/3} \;, \label{eq:FrequencyEvolutionApp} 
\end{align}
controlled by the chirp mass, which is related to the component masses $m_{1,2}$ as
\begin{equation}
    \mathcal{M}_c \equiv \frac{(m_1 m_2)^{3/5}}{(m_1 + m_2)^{1/5}} \,.
\end{equation}
Throughout this work, we order the component masses as $m_1 > m_2$, and define the mass ratio $q \equiv m_1/m_2$, which enters the frequency evolution at higher PN order. In General Relativity, the GW signal from a source at redshift $z'$ with component masses $m_{1,2}'$ is perfectly degenerate with the signal from a local ($z=0$) source with masses $m_{1,2} = (1+z') m_{1,2}'$ (up to a change in the amplitude proportional to the change in $d_L$). In other words, GW signals display a perfect mass-redshift degeneracy. Thus, throughout this paper, all masses are {\it detector frame} masses\footnote{Strictly speaking, for a detector that moves on an accelerated trajectory, the ``detector frame masses'' must be defined in some inertial frame. We will return to this subtlety later.} unless noted explicitly. 

There are only two propagating polarization degrees of freedom of a GW signal. As is common in the field, we write them in transverse traceless gauge, such that the GW perturbation of the metric tensor can be written as
\begin{equation} \label{eq:GWh}
    h_{\mu\nu}(t) = h_+ e_{\mu\nu}^+ + h_{\times} e_{\mu\nu}^\times \;,
\end{equation}
where the $e^{+/\times}$ are the polarization tensors. This polarization basis is fixed by the direction of the GW source on the sky (as seen from the GW detector), ${\bm n}$, and the angular momentum of the binary, ${\bm L}$, see, for example, Refs.~\cite{Cutler:1994ys,Nissanke:2009kt}. Choosing a Cartesian coordinate system where $\hat{\bm n}$ defines the $\hat{\bm z}$ direction, one can construct the remaining axes as
\begin{equation}
    \hat{\bm x} = \frac{ \hat{\bm n} \times \hat{\bm L} }{ | \hat{\bm n} \times \hat{\bm L} | } \;, \quad \hat{\bm y} = - \frac{\hat{\bm n} \times \hat{\bm x}}{| \hat{\bm n} \times \hat{\bm x} |} \;.
\end{equation}
The spatial components of the polarization tensors are then 
\begin{equation} \label{eq:PolarizationTensors}
    e_{ij}^+ = \hat{x}_i \hat{x}_j - \hat{y}_i \hat{y}_j \;, \quad e_{ij}^\times = \hat{x}_i \hat{y}_j + \hat{y}_i \hat{x}_j\;.
\end{equation}

The amplitude of the GW {\it strain} signal along the two polarization directions are given by
\begin{equation} \begin{split}
    \label{eq:AmplitudeApp}
    h_+(t) &= \frac{2 \mathcal{M}_c^{5/3} \pi^{2/3} f_{\rm GW}(t)^{2/3}}{d_L} \left[ 1 + \left(\hat{\bm L} \cdot \hat{\bm n}\right)^2 \right] \cos\Phi(t) \;, \\
    h_\times(t) &= \frac{4\mathcal{M}_c^{5/3} \pi^{2/3} f_{\rm GW}(t)^{2/3}}{d_L} \left( \hat{\bm L} \cdot \hat{\bm n} \right) \sin\Phi(t) \;,
\end{split} \end{equation}
where $\Phi(t) = \int_{t_{\rm ref}}^t dt'~2\pi f_{\rm GW}(t') + \Phi_0$ is the GW phase with $\Phi_0 = \Phi(t_{\rm ref})$ the phase at some reference time $t_{\rm ref}$. As mentioned above, we use a heliocentric equatorial coordinate system\footnote{Concretely, we use the International Celestial Reference System (ICRS).}, thus, the sky-position ${\bm n}$ of a GW source is parameterized by the luminosity distance, $d_L$, the right ascension, $\alpha$, and the declination, $\delta$. 

The orientation of the binary's angular momentum is parameterized by the inclination angle, $\iota$, defined via $\cos\iota = \hat{\bm L} \cdot \hat{\bm n}$, and the polarization angle, $\psi$, measuring the (counter-clockwise) angle of ${\bm L}$ with respect to a line parallel to the celestial equator around the axis ${\bm n}$ (see, for example, Ref.~\cite{Wheelan13Notes}). From Eqs.~\eqref{eq:PolarizationTensors} and~\eqref{eq:AmplitudeApp} we can immediately see the usefulness of this parameterization: while the inclination angle of the binary, $\iota$, affects the amplitudes in the two different polarization directions, $h_+$ and $h_\times$, changes in the polarization angle, $\psi$, correspond to a rotation of the polarization tensors in the plane orthogonal to ${\bm n}$ on the sky.  

From Eq.~\ref{eq:AmplitudeApp} we can also already notice the well-known $\iota$--$d_L$ degeneracy. For $\iota = 0$ or $\pi$, a change in $\iota$ is perfectly degenerate with a change in $d_L$ at leading order in the waveform (and $\psi$ becomes ill-defined in this limit since $\hat{\bm n} \times \hat{\bm L} = 0$).
The $\iota$--$d_L$ degeneracy is partially broken for other values of $\iota$. In addition, the $\iota$--$d_L$ degeneracy is broken at higher PN order (as implemented in our numerical results). Since PN effects become less and less prominent the deeper the GW signal is generated during the binary's inspiral phase (in other words, the longer before the merger one observes a GW signal), we can already anticipate that the ability of GW detectors in the mid-band to measure the luminosity distance (and the inclination angle) of GW sources will depreciate the lower the chirp mass of the binary is.

Close to the merger of the binary, the (analytic) PN waveforms we use in this work become inaccurate. In that regime, numerical General Relativity must be employed to accurately calculate the waveforms\footnote{Another technique of interest for calculating waveforms close to the merger is the effective one-body approach~\cite{Buonanno:1998gg,Damour:2009zoi}.}. We will restrict ourselves to considering signals for which the PN waveforms are good approximations of the full solution. For the mid-band, this corresponds to restricting ourselves to considering binaries with $\mathcal{M}_c \lesssim 300\,M_\odot$ for the space-based atom-interferometer GW detector benchmark we consider (for which we consider observations up to a frequency of $f_{\rm GW} \approx 3\,$Hz) and binaries with $\mathcal{M}_c \lesssim 50\,M_\odot$ for the terrestrial detector benchmark we consider (with a largest frequency of $f_{\rm GW} = 10\,$Hz). As a rough justification of these bounds, note that the orbital frequency of a test particle around a BH with mass $M$ is $\nu_{\rm ISCO} = \frac{1}{6\sqrt{6}} \frac{1}{2 \pi} \frac{c^3}{G M}$ at the innermost stable circular orbit (ISCO); our choices ensure that the largest frequency observed by our detectors is a factor of a few below the merger frequency.

As stated in the beginning of this section, we neglect the effects of spins of the binary constituents in this work. In general, these spins would appear in the higher-PN corrections to inspiral waveforms~\cite{Cutler:1994ys, Poisson:1995ef}, we leave consideration of the effects of spin to future work. We note that, depending on the values of other parameters, including non-zero spins can either improve or worsen the precision with which parameters can be reconstructed from a GW measurement because spins can both break degeneracies and introduce new ones. 

The nature of the objects in a compact binary has only a small effect on their inspiral waveform, such that our discussion will generally be agnostic to binary composition, except insofar as it affects the companions' masses. A variety of binaries could be observed by mid-band GW detectors, including most sources observable by LVK~\cite{KAGRA:2013rdx, KAGRA:2021duu, Graham:2017lmg}, as well as some sources so massive that they merge at frequencies below LVK's sensitivity. In this work, we will focus on neutron stars and stellar- and intermediate-mass black holes (chirp masses of roughly $1-300\,M_\odot$), leaving consideration of lighter objects (e.g. white dwarfs) as well as heavier ones (e.g. supermassive black holes) to future work. We will similarly restrict to mass ratios of the two binary companions near unity, typically taking that ratio to be $q \equiv m_1 / m_2 = 1.15$ or a symmetric mass ratio $\eta \equiv m_1 m_2/(m_1+m_2)^2 \approx 0.249$.

\section{Atom Interferometer Detectors} \label{app:Detectors}

In this appendix, we discuss the response of a detector to the GW signal from compact binaries. We focus on atom-interferometer detectors as a promising technology to probe GW signals in the mid-band, $f_{\rm GW} \sim 30\,{\rm mHz} - 10\,$Hz. In App.~\ref{subapp:AIreview}, we give an expanded review of atom-interferometers as GW detectors. In App.~\ref{subapp:AntennaFuns}, we discuss the antenna functions of a single-baseline atom-interferometer detector, and in App.~\ref{subapp:DetectorBenchmarks}, we discuss the concrete detector benchmark scenarios and their associated sensitivity curves that we use in the numerical results in this paper. 

\subsection{Review of Atom Interferometers} \label{subapp:AIreview}

Conceptually, an atom-interferometer GW detector is similar to using two atomic clocks to measure the light-travel time of laser pulses sent in between the clocks, and hence the distance between the clocks~\cite{Kolkowitz:2016wyg}. By tracking changes to this distance, one could search for GW signals. However, such a setup would be subject to a number of noise sources. Perhaps most importantly for our discussion, atom clouds held in an optical lattice (as is common in atomic clocks) would be subject to mechanical vibrations in the experimental apparatus, including seismic noise. Furthermore, such a setup would also be subject to the noise in the phase of the laser pulses sent in between the clocks. Although there may be ways to mediate such noise sources, we focus on a different experimental approach~\cite{Dimopoulos:2008sv}.

By using two (or more) atom-interferometers driven by common laser pulses, one can mitigate noise from mechanical vibrations and laser phase noise. In an atom-interferometer, one prepares a cloud of ultra-cold atoms (typically in a magneto-optical trap), and then releases this cloud into free-fall. Using a sequence of laser pulses with the appropriate frequency and temporal intensity profile, the atom cloud is ``split'' into a superposition of states with different momenta and excitation states of the atoms and subsequently recombined. Finally, one measures the spatial distribution of the atom cloud after the laser pulse sequence has been completed. Such a setup can be understood as a Mach-Zehnder interferometer: the interference pattern observed in the final atom cloud measures changes in the relative physical length of the paths the atoms can take, as well as changes in the difference in time spent in the excited state along each path; see Ref.~\cite{Dimopoulos:2008hx} for a detailed discussion of atom interferometry and, in particular, the general computation of the output of the interferometer.
In analogy with a Mach-Zehnder interferometer, we will refer to the two paths (in physical space as well as in the space of excitation of the atoms) as the ``arms'' of the interferometer.

We can immediately appreciate one of the advantages of this setup: since atom-interferometers use clouds of atoms that are in free-fall during the duration of the measurement, they are decoupled from mechanical vibrations of the experimental apparatus due to machines, humans, seismic waves, and other sources. 

Second, the free-falling atom clouds used in atom interferometers are electrically neutral. This makes atom-interferometers much more robust against electromagnetic background fields, which plague laser-interferometer based GW detectors or any other technique that uses macroscopic test masses which, invariably, are subject to electrostatic charging. 

One can then realize a GW detector by using common laser pulses to drive two physically separated atom interferometers. Correctly adding up the momentum kicks delivered to the two arms of each atom interferometer by the laser pulses, such that the arms eventually recombine, requires a sequence of counter-propagating laser pulses to excite and de-excite the atoms clouds in each atom interferometer. In order to drive two atom-interferometers separated by hundreds of meters to tens of thousands of kilometers with the same laser pulses, as envisaged for space-based GW detectors, it is advantageous to use single-photon-excited ``clock'' transitions rather than two-photon ``Raman'' transitions (which, in order to transfer substantial momentum, would require two counter-propagating pulses to interact with the atom clouds at the same time). Using clock transitions, it is possible to find a pulse sequence that allows one to drive two macroscopically separated atom interferometers with common laser pulses~\cite{Graham:2012sy}. Such a setup has multiple advantages: first, by comparing the response of the two atom-interferometers, each of which measure the difference in phase accumulated in their respective arms driven by counter-propagating pulses, one can choose to measure the second (time) derivative of the light-travel time between the atom interferometers averaged over the interrogation time (i.e. the total duration of the laser pulse sequence from the first beam-splitter pulse to the final laser pulse that recombines the wave packets). In other words, such a setup directly measures the relative {\it acceleration} of the atom clouds in the separate interferometers. This accelerometer configuration of two atom-interferometers is insensitive to the (precise) distance between or relative velocity of the atom clouds when released from their respective magneto-optical traps, making the setup robust to mechanical vibrations of the experimental apparatus that affect the launch conditions of the atom cloud from their opto-mechanical traps.

Second, since the atom interferometers are driven by {\it common} laser pulses, and each interferometer measures the differences in phase between the various laser pulses driving it, the comparison of the interferometers' outputs is insensitive to laser phase noise~\cite{Graham:2012sy}. Thus, atom-interferometers can eliminate laser phase noise even in a single-baseline configuration formed by two atom-interferometers (with their atom clouds acting as test masses) driven by a single set of counter-propagating laser pulses. In contrast, laser-interferometer detectors such as LVK or LISA must use Michelson-interferometer-type setups, interfering a beam-split laser propagating in two different spatial directions to suppress laser phase noise, requiring at least three mirrors acting as test masses.

Let us briefly summarize the discussion of atom interferometers above: using two (or more) atom interferometers, one can realize a GW detector with electrically neutral free-falling test masses -- making the test masses insensitive to electromagnetic background fields and mechanical vibrations of the experimental apparatus. Driving the transitions in both interferometers with common laser pulses renders laser phase noise irrelevant in a single-baseline detector: only changes to the time it takes the pulses to travel between the two atom clouds affect the difference in the output measured by the two interferometers.

\subsection{Antenna Functions for Atom-Interferometer GW Detectors} \label{subapp:AntennaFuns}

Atom interferometer GW detectors with a single baseline have a simple but unique antenna function: only the component of a GW's strain along the direction of laser propagation contributes to the sensitivity. The antenna functions in the polarization basis are obtained by contracting the spatial components of the polarization tensors defined in Eq.~\eqref{eq:PolarizationTensors} with a unit vector in the relative direction of one interferometer to the other (i.e., along the direction that the laser pulses between the interferometers propagate), $\hat{\bm \ell}$,
\begin{align} \label{eq:AntennaFun}
    F_{+,\times}(t) = \sum_{i,j} \hat{\ell}_i(t) ~ e^{+,\times}_{ij} ~ \hat{\ell}_j(t) \;.
\end{align}
As we emphasize by explicitly denoting their time-dependence, $F_i = F_i(t)$, the antenna functions inherit the time-dependence of $\hat{\bm{\ell}}$. The response of a single-baseline detector to a GW signal parameterized in the polarization basis by $h_{+,\times}(t)$, see Eq.~\eqref{eq:Amplitude}, is then
\begin{equation}
    h(t) = h_+(t) F_+(t) + h_\times(t) F_\times(t)\;.
\end{equation}

A terrestrial detector would change the direction its baseline is pointing on hour timescales due to Earth's rotation. Likewise, the direction $\hat{\bm \ell}$ for a satellite-borne detector in medium Earth orbit would rotate by $360^\circ$ over the few-hours orbital period of such a detector. Thus, for GW signals that can be measured for hours, or longer, even a single terrestrial or satellite-borne detector could measure the GW strain along many different spatial directions. For signals that can be measured for times shorter than the timescale over which the detector reorients, e.g., signals from sources with chirp masses $\mathcal{M}_c \gtrsim 20\,M_\odot$ in terrestrial detectors (see Tab.~\ref{tab:LifetimeHorizon}), using multiple detectors with different $\hat{\bm \ell}$ remains advantageous.

The motion of the detector during the time it measures a GW signal has a second important effect: as the detector follows its accelerated trajectory around Earth's center, and together with Earth around the Sun, the Doppler-shift of the GW signal seen by the detector changes. This change in the Doppler shift depends on the location of the GW source relative to the detector's motion, thus, it can be used to measure the location of GW sources on the sky. As we will see, using this effect, a single detector can achieve excellent directional resolution for sufficiently long-lived GW signals. 

\subsection{Detector Benchmarks} \label{subapp:DetectorBenchmarks}

Since the first proposal of atom-interferometer based GW detectors~\cite{Dimopoulos:2007cj} more than 15\,years ago, a number of collaborations have formed with the aim of building such detectors, including AEDGE~\cite{AEDGE:2019nxb, Bertoldi:2021rqk}, AION~\cite{Badurina:2019hst}, MAGIS~\cite{Graham:2017pmn,MAGIS-100:2021etm}, MIGA~\cite{Canuel:2017rrp}, SAGE~\cite{Tino:2019tkb}, VLBAI~\cite{Schlippert:2019hzx}, and ZAIGA~\cite{Zhan:2019quq}. Although different collaborations of course follow different design choices and plans, a typical strategy is to first build a terrestrial detector with $\mathcal{O}(100)\,$m separation between the atom interferometers, and then a terrestrial detector with $\mathcal{O}(1)\,$km separation of the atom sources. Most collaborations envisage a vertical layout of the detectors -- since Earth's gravity accelerates the atom clouds vertically once released from their traps, a vertical layout has some technical advantages -- limiting terrestrial detectors to the size of available shafts on Earth. The deepest existing mine-shaft is the ${\sim}\,4\,$km shaft of the Mponeng Gold Mine in South Africa. Ultimately, the collaborations are aiming to operate one (or multiple) atom-interferometer GW detectors in space, linking two satellites carrying atom-interferometers placed $\mathcal{O}(10,000)\,$km or more apart via lasers driving the interferometers.

In this work, we will use two benchmark scenarios for the sensitivity of an atom-interferometer detector to GW signals: a terrestrial detector with atom sources separated by 1\,km (which we call ``MAGIS-1\,km'', adapted from Ref.~\cite{MAGIS-100:2021etm}) and a satellite-based detector in medium Earth orbit (``MAGIS-space'', adapted from Ref.~\cite{Graham:2017pmn}). Note that while these sensitivity curves are adapted from forecasts for the detectors of the MAGIS collaboration\footnote{Some of us are members of the MAGIS collaboration.}, our results apply to any single-baseline detector in the mid-band. In Fig.~\ref{fig:NoiseCurve}, we show the sensitivity curves -- specifically, the square root of the one-sided power spectral density of the detector's strain noise -- for these detector benchmarks. 

Let us continue with a heuristic explanation of the frequency range in which atom-interferometer detectors could measure the GW {\it strain}, the fractional change in the distance between the test masses. At the low-frequency end, the sensitivity is limited by the stability of the test masses. For a satellite-borne atom-interferometer based detector, the most important constraint arises from the time the atom clouds can be kept in free fall. The free-fall time sets the upper limit for the {\it interrogation time} of the atom interferometer sequence, $T$, the time between the first beam-splitter pulse and the final pulse of the sequence re-combining the interferometer's arms. As we discussed above, an atom-interferometer GW detector measures the acceleration between the two test masses averaged over $T$, thus, the sensitivity to the GW strain signal is suppressed for $f_{\rm GW} \lesssim 1/T$. A combination of vacuum quality and atomic drift in the satellite limit the free-fall time to, at most, a few hundred seconds~\cite{Graham:2016plp, Hogan:2011tsw, Dimopoulos:2008sv, Dimopoulos:2007cj}.  Thus, space-based detectors are well suited to searching for signals at frequencies above some tens of mHz.

For terrestrial detectors, technically feasible free-fall times of the atoms are limited to a few seconds. More importantly, the stability of the test masses in a terrestrial detector is affected by Newtonian Gravity Gradient Noise (GGN). While the free-falling atom clouds functioning as atom-interferometers' test masses are decoupled from mechanical vibrations, they are subject to gravitational forces produced by variations of Earth's (or any other nearby) mass-distribution on timescales comparable to the (inverse) frequency of the signal. Below frequencies of $\mathcal{O}(1)\,$Hz, a terrestrial detector will be affected by GGN from seismic (surface) waves and possibly other~\cite{Carlton:2023ffl} sources. Various strategies for reducing GGN have been proposed~\cite{Hughes:1998pe, Harms:2013raa, Chaibi:2016dze, Harms:2019dqi}. For our ``MAGIS-1\,km'' benchmark, we do not take any such improvements into account. However, we do assume a level of GGN at the optimistic end of the range of the seismic GGN estimates at the site of the Homestake mine Ref.~\cite{MAGIS-100:2021etm} made based on seismic measurements in Ref.~\cite{Harms:2010mp}; see also Ref.~\cite{Harms:2019dqi} for an in-depth discussion of GGN. 

At high frequencies, the strain sensitivity of an atom-interferometer GW detector operated using a series of counter-propagating laser pulses is limited by the finite light-travel time between the atom interferometers. If the test masses' acceleration induced by the GW oscillates faster than the light travel time, the signal is averaged out. Thus, the strain sensitivity depreciates at frequencies $f_{\rm GW} \gtrsim L^{-1}$, where $L$ is the separation of the atom interferometers. 

The discussion so far suffices to understand the features of the ``MAGIS-1\,km'' sensitivity curve in Fig.~\ref{fig:NoiseCurve}. At frequencies $f_{\rm GW} \lesssim 1\,$Hz, the strain sensitivity is limited by GGN. For frequencies $f_{\rm GW} \gtrsim 1\,$Hz, the strain sensitivity is frequency independent, which is the characteristic behavior of an atom-interferometer's sensitivity for $1/T \lesssim f_{\rm GW} \lesssim L^{-1}$ if the limiting noise source is atom shot noise. We terminate the sensitivity curve at $f_{\rm GW} = 10\,$Hz. While this is far below the frequency set by the light-travel time of a detector with atom interferometers separated by 1\,km, signals at frequencies $f_{\rm GW} \gtrsim 10\,$Hz are likely more effectively observed in terrestrial laser interferometer detectors such as LVK, Einstein Telescope~\cite{Hild:2009ns, Hild:2010id} or Cosmic Explorer~\cite{Evans:2021gyd, Srivastava:2022slt}.

For the ``MAGIS-space'' benchmark, the shape of the sensitivity curve is complicated by a number of additional factors. One important ingredient in optimizing an atom interferometer GW detector's sensitivity is maximizing the physical separation between the two arms in each interferometer\footnote{Technically, large separations of the arms are achieved by using a series of counter-propagating laser pulses that repeatedly excite and de-excite the atoms instead of a single excitations to split, redirect, and re-combine the arms of the atom interferometer. Note that Ref.~\cite{Rudolph:2019vcv} has demonstrated a ``Mach-Zehnder'' strontium atom interferometer with momentum transfers of $141\,\hbar k$ using single-photon ``clock''-transitions.}. However, for satellite-born detectors, the possible separation of the arms is limited by the feasible dimensions of the space craft/shield the atom clouds must be contained in to protect them from the effects of photons, cosmic rays, or other interference~(see e.g. Refs.~\cite{Dimopoulos:2008sv, Du:2023eae}). This limitation can be overcome by using a resonantly-enhanced mode to operate the atom-interferometers~\cite{Graham:2016plp}: by repeating identical pulse sequences that periodically interleave the ``arms'' of the interferometer, a phase difference in the two arms from a GW signal with frequency matching the ``repetition rate'' of the pulse sequence can be resonantly added up. The cost of this increase of the sensitivity to signals at a chosen $f_{\rm GW}$ is, of course, that the sensitivity to signals at other frequencies is suppressed. The high-frequency limit to which this strategy can be extended is still set by the light-travel time of the laser between the atom-interferometers. For a space-based detector in medium Earth orbit, such as our ``MAGIS-space'' benchmark, the separation between the test masses is at most a few times ten thousand kilometers. Thus, the largest frequency at which such a detector maintains it's peak sensitivity is limited to a few Hz by the light-travel time. We include the sensitivity up to $f_{\rm GW} \approx 3\,$Hz for our ``MAGIS-space'' detector benchmark.\footnote{We note that the Nyquist frequency associated with the chosen pulse-sequence is not a fundamental limit on the frequencies at which mid-band detectors can be operated -- aliasing at higher frequencies is of little concern when resonantly tracking signals. However, at $f_{\rm GW} \gtrsim 10\,$Hz, terrestrial laser-interferometers appear better placed to measure GW signals than atom-interferometers.}

The ``MAGIS-space'' sensitivity curve shown in Fig.~\ref{fig:NoiseCurve} is the envelope of the sensitivity curves obtained by operating the detector with different resonantly-enhanced pulse sequences~\cite{Graham:2016plp,Graham:2017pmn}. Let us note that operating a GW detector in a frequency-selective resonant readout mode need not be a disadvantage when searching for GW signals in the mid-band. Gravitational wave signals from compact binaries monotonically develop from smaller to larger GW frequencies during the evolution of the binary. Thus, an atom-interferometer detector can be initially operated with a pulse sequence making it most sensitive to GW frequencies at the lower end of the detector's sensitivity band. When a GW signal is detected, one can change the pulse sequence (and hence resonant frequency) in real-time to follow that GW signal's monotonic evolution through the frequency band, maintaining optimal sensitivity via the resonant enhancement of the signal. 

Note that GW signals from compact binaries with chirp masses ranging from ${\sim}\,1\,M_\odot$ to hundreds of solar masses have lifetimes ranging from days to hundreds of years in the mid-band. Thus, it seems quite likely that a mid-band detector with sufficient sensitivity could observe the signal from multiple GW sources at the same time. By operating the detector in a resonant mode and switching between the different frequencies of the multiple GW signals, one could trace multiple sources simultaneously while mitigating confusion noise from signals at different frequencies. Since changing the resonant frequency only requires changing the laser pulse sequence, it can be changed as desired every second or so, set by the frequency with which atom clouds are launched in the interferometer.

\section{Additional Sensitivity Plots} \label{app:SupplementalPlots}

In this appendix, we extend the discussion of Sec.~\ref{sec:ResultsDiscussion}.

\begin{figure}
   \centering
   \text{MAGIS-space}
   \vspace{0.3cm}
   
   \includegraphics[trim=0.3cm 0.3cm 0.3cm 0.3cm, clip, width=0.49\linewidth]{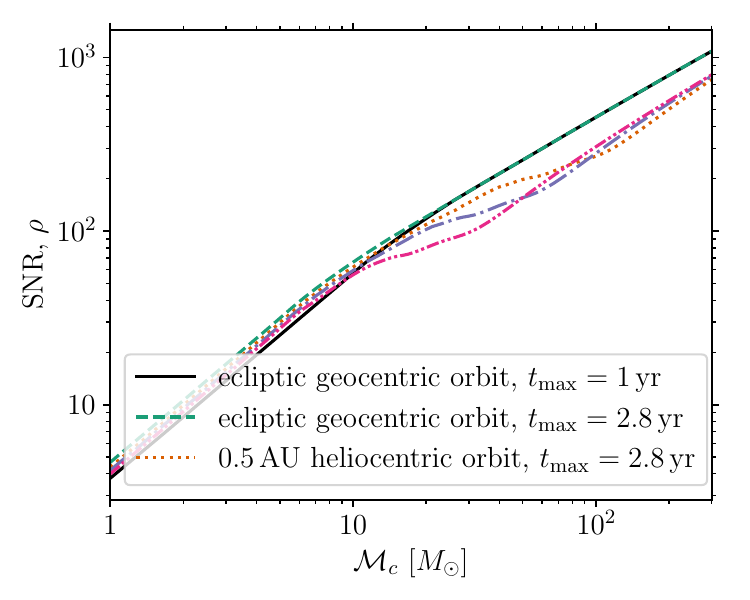}
   \includegraphics[trim=0.3cm 0.3cm 0.3cm 0.3cm, clip, width=0.49\linewidth]{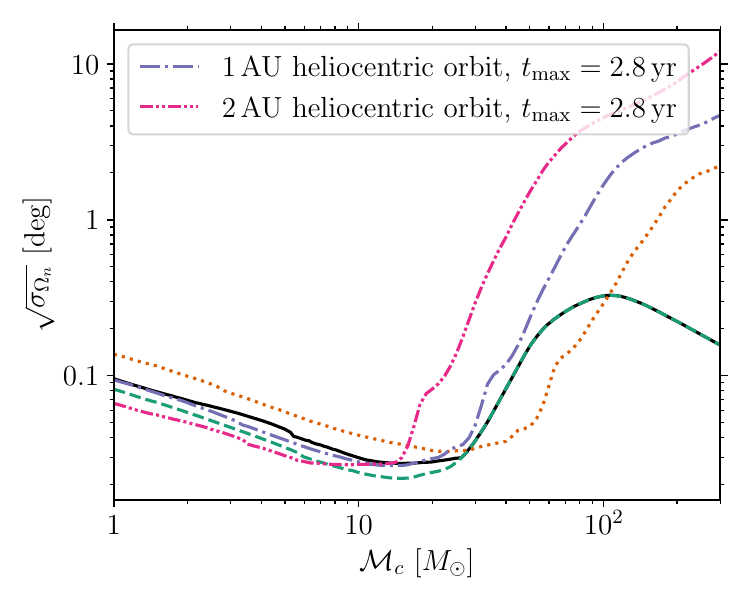}
   \caption{{\bf ``MAGIS-space'': geocentric vs heliocentric orbits.} Plots of SNR (left) and sky-localization uncertainty of the source (right) as a function of chirp mass for different choices for the orbit of a ``MAGIS-space'' detector and maximal observation time ($t_{\rm max}$) as denoted in the legend. Throughout, we have set the luminosity distance to $d_L = 100\,$Mpc. All other parameters are set to the benchmark values given by Tab.~\ref{tab:Parameters}. For all lines except the solid black, we have chosen $t_{\rm max} = 2.8\,$yr, matching the orbital period of a geocentric orbit with $2\,$AU radius. For comparison, the solid black line is for $t_{\rm max} = 1\,$yr, the value used in all other figures in this work.} 
   \label{fig:spaceOrbits}
\end{figure}

First, it is instructive to consider how the ability of a ``MAGIS-space'' detector to measure signals from and reconstruct the parameters of compact binary GW sources changes with different choices of the orbits of the satellites carrying the atom-interferometers. So far we have considered a detector comprised of satellites in medium Earth orbit: such a detector would orbit the Earth with a period of $7.8\,$h, most importantly leading to a reorientation of the detector's ``baseline'' along which it measures the GW strain on hour timescales. Furthermore, Earth, and a detector in geocentric orbit, orbit the Sun with a period of 1\,year in an orbit with a radius of 1\,AU. In Fig.~\ref{fig:spaceOrbits}, we illustrate the effect these motions have on the sensitivity of a ``MAGIS-space'' detector by comparing the SNR and the sky-localization error, as a function of chirp mass, for a ``MAGIS-space'' detector in medium Earth (geocentric) orbit with detectors in heliocentric orbits (i.e., not orbiting Earth) of different radii.

For all orbits, we assume that the detector is comprised of two satellites trailing each other in the same orbit. Thus, for a detector in heliocentric orbit, the time over which the detector would re-orient its baseline is the same as its orbital period. In order to avoid effects stemming from incomplete detector re-orientation for long-lived sources, we have included the signal recorded in the last $t_{\rm max} = 2.8\,$yr before the signal leaves the ``MAGIS-space'' sensitivity band in Fig.~\ref{fig:spaceOrbits}; this choice matches the orbital period of the largest heliocentric orbit ($R = 2\,$AU), we consider in Fig.~\ref{fig:spaceOrbits}. To illustrate the effect of changing $t_{\rm max}$, we show results for the geocentric orbit for both $t_{\rm max} = 2.8\,$yr and $t_{\rm max} = 1\,$yr, the value we use in all other results in this paper. 

In order to understand the results shown in Fig.~\ref{fig:spaceOrbits}, it is important to compare the lifetime of the GW signals to the orbital period of the various heliocentric orbits we consider here. The largest heliocentric orbit we consider has a radius of $R=2\,$AU, corresponding to an orbital period of $2.8\,$yr. The GW signal from a $\mathcal{M}_c \approx 15\,M_\odot$ source has a lifetime matching this period in the ``MAGIS-space'' sensitivity band. For a $R=1\,$AU orbit, the orbital period is $1\,$yr, matching the lifetime of a $\mathcal{M}_c \approx 25\,M_\odot$ source. Finally, for a $R=0.5\,$AU orbit, the lifetime of signals from $\mathcal{M}_c \approx 50\,M_\odot$ sources matches the 4-month period.

In the left panel of Fig.~\ref{fig:spaceOrbits} we compare the SNR as a function of $\mathcal{M}_c$ for the various choices of the orbits we make here. Since we chose a fixed luminosity distance of $d_L = 100\,$Mpc for all sources, the SNR grows with increasing $\mathcal{M}_c$. Comparing the results for the different orbits, the most important feature to note is that for detectors in heliocentric orbit, the SNR of sources with lifetime shorter than the detector's re-orientation time ($\mathcal{M}_c \gtrsim 15\,M_\odot$ for $R = 2\,$AU, $\mathcal{M}_c \gtrsim 25\,M_\odot$ for $R = 1\,$AU, and $\mathcal{M}_c \gtrsim 50\,M_\odot$ for $R = 0.5\,$AU) is smaller than for detectors in geocentric orbit. This change in SNR is due to the relative orientation of the detector's baseline to the source at the time when the signal is recorded, and is thus dependent on the sky-position of the source. More importantly than the suppression of the SNR we can see in the left panel of Fig.~\ref{fig:spaceOrbits} for the particular sky location chosen here, note that a detector in heliocentric orbit would have much less homogeneous sky-coverage for signals with lifetime short compared to its orbital period than a detector in geocentric orbit.

In the right panel of Fig.~\ref{fig:spaceOrbits}, we show the projected sky-localization uncertainty, $\sqrt{\Omega_n}$. To start, let us focus only on the three examples of heliocentric orbits. For small $\mathcal{M}_c$, i.e., signals with long lifetimes, we find that the larger the radius of the heliocentric orbit, the smaller $\sqrt{\Omega_n}$. As discussed extensively above, we expect $\sqrt{\sigma_{\Omega_n}} \propto 1/(\rho f_{\rm GW} L)$, where for signals that have a lifetime comparable to or longer than the orbital period, the effective aperture $L$ is given by the size of the heliocentric orbit $R$. Note that we find the scaling of $\sqrt{\sigma_{\Omega_n}}$ with $R$ to be somewhat slower in Fig.~\ref{fig:spaceOrbits} than $\sqrt{\sigma_{\Omega_n}} \propto 1/R$, this is because of partial degeneracies of the sky-location with other parameters exacerbated by the increasingly slow rotation of the detector. For signals with lifetimes shorter than the orbital period of the detector ($\mathcal{M}_c \gtrsim 15\,M_\odot$ for $R = 2\,$AU, $\mathcal{M}_c \gtrsim 25\,M_\odot$ for $R = 1\,$AU, and $\mathcal{M}_c \gtrsim 50\,M_\odot$ for $R = 0.5\,$AU), we can see that the ability to localize sources on the sky rapidly depreciates the larger $\mathcal{M}_c$ and thus, the shorter the lifetime of the signal. In this regime, we see that $\sqrt{\Omega_n}$ is smaller the smaller the heliocentric orbit, because the acceleration the detector experiences during the sources lifetime is greater the smaller $R$.

Let us now compare the sky-localization ability of a detector in geocentric orbit with that of a detector in heliocentric orbit. From the right panel of Fig.~\ref{fig:spaceOrbits}, we can see that a detector in geocentric orbit has better sky-localization capability than a detector in $R=1\,$AU heliocentric orbit for all values of $\mathcal{M}_c$. Not surprisingly, we find the most striking difference for sources with short-lived signals, recall that for sources with $\mathcal{M}_c \gtrsim 50\,M_\odot$, the lifetime of signals in the ``MAGIS-space'' sensitivity band becomes shorter than a few months. For such sources we see that a detector in geocentric orbit has drastically better source-localization capabilities than a detector in $R=1\,$AU heliocentric orbit. A detector in medium Earth orbit re-orients its baseline on hour time-scales, allowing the detector to measure the strain signal along many spatial directions. Thus, even a single detector in geocentric orbit can (partially) break the degeneracies between different parameters influencing the amplitude of the GW strain signal, allowing it to localize GW sources on the sky. A detector in heliocentric orbit could only measure the GW strain in one spatial direction for sources with lifetimes short compared to the time it takes the detector to orbit the Sun, thus, it would lose its ability to reconstruct many of the source's parameters due to degeneracies.   

These results demonstrate that the option of operating a satellite-borne mid-band GW detector in geocentric orbit not only significantly reduces the launch-cost of such a detector compared to heliocentric orbits, but that the hour-timescale over which such a detector would re-orient allows a single detector to measure the GW strain in many different spatial directions and hence break degeneracies between various parameters controlling the GW signal from compact binaries. Of course, a measurement of the GW strain in multiple spatial directions could also be achieved by operating multiple detectors, but that would further increase the detector's cost and, depending on the technical details of the detectors, may also increase the technical requirements for the mission substantially if more than two satellites have to be flown in a controlled formation.

\begin{figure}[b]
   \includegraphics[height=6.5cm, trim=0cm 0cm 4.1cm 0cm, clip]{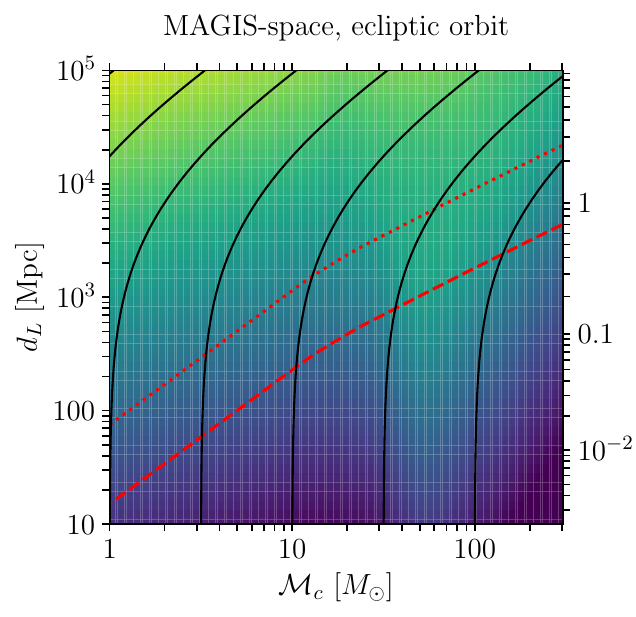}
   \hspace{0.3cm}
   \includegraphics[height=6.5cm, trim=0.9cm 0cm 0cm 0cm, clip]{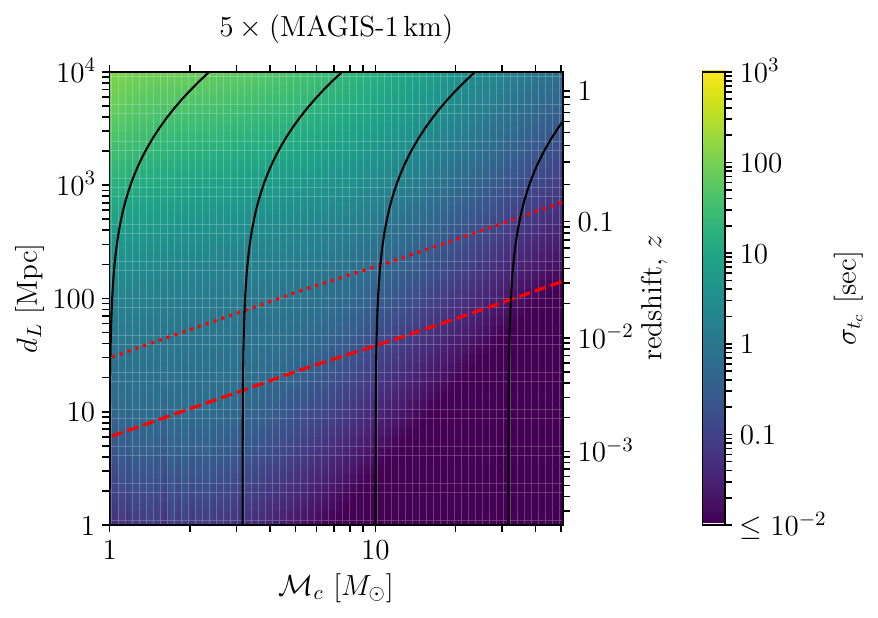}
   \caption{{\bf \boldmath Time-of-merger uncertainty ($t_c$)} in the plane of the (detector frame) chirp mass and luminosity distance of the GW source, all other parameters are set to the benchmark values given in Tab.~\ref{tab:Parameters}. The dotted (dashed) red lines are $\rho=5$ ($\rho=25$) contours of SNR. The right $y$-axes show the corresponding cosmological redshift, $z$, assuming a reference cosmology with a Hubble constant of $H_0 = 67.4\,$km/Mpc/s and a matter density of $\Omega_m = 0.315$. The solid black lines show lines of constant source-frame chirp-masses corresponding to the detector-frame chirp-masses shown on the $x$-axis. {\it Left:} Results for a single ``MAGIS-space'' detector in an ecliptic orbit. {\it Right:} Results for a network of five terrestrial ``MAGIS-1\,km'' detectors, located at Homestake, Sudbury, Renstr\"om, Tautona, and Zaoshan; see Sec.~\ref{sec:MidbandGWDetection}. Note that the range of $d_L$ shown on the vertical axes differs between the two panels.}
   \label{fig:ChirpMass_tc_Scans}
\end{figure}

\begin{figure}
   \centering
   \text{MAGIS-space}
   \vspace{0.2cm}
   
   \includegraphics[trim=3.7cm 5.7cm 3.2cm 0cm, clip, width=0.32\linewidth]{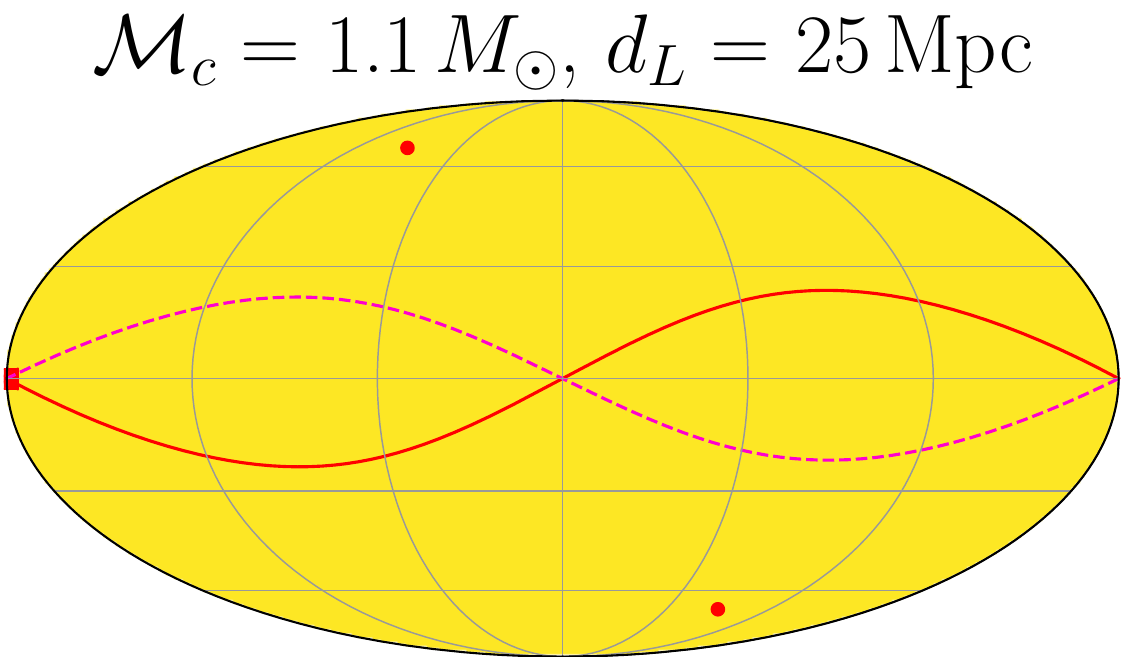}
   \includegraphics[trim=3.7cm 5.7cm 3.2cm 0cm, clip, width=0.32\linewidth]{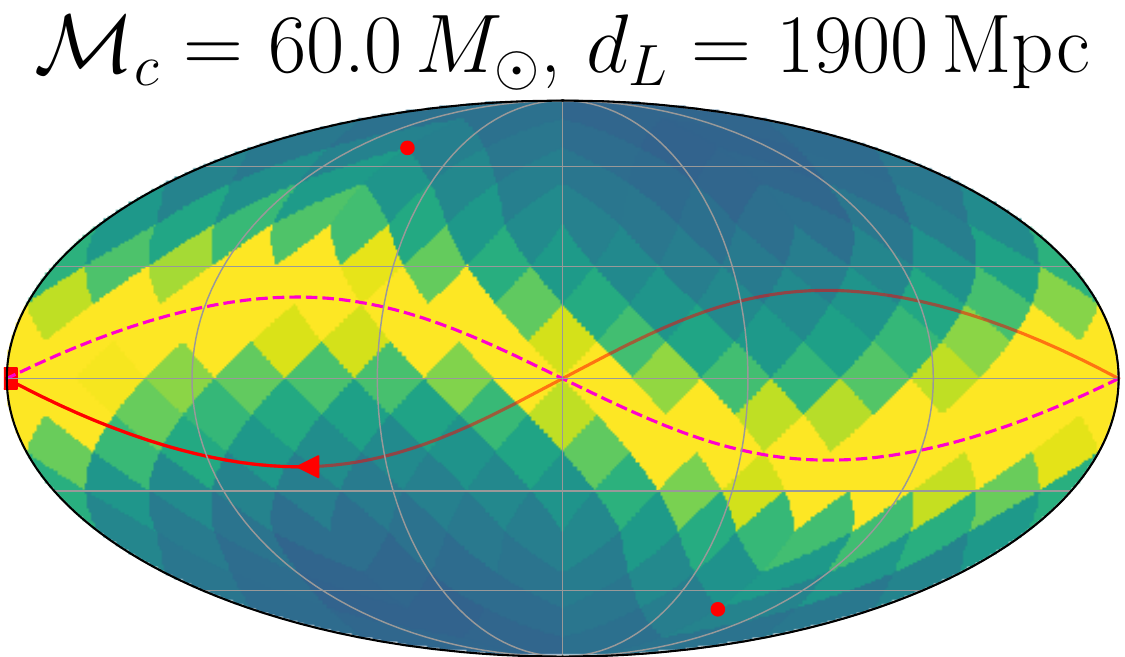}
   \includegraphics[trim=3.7cm 5.7cm 3.2cm 0cm, clip, width=0.32\linewidth]{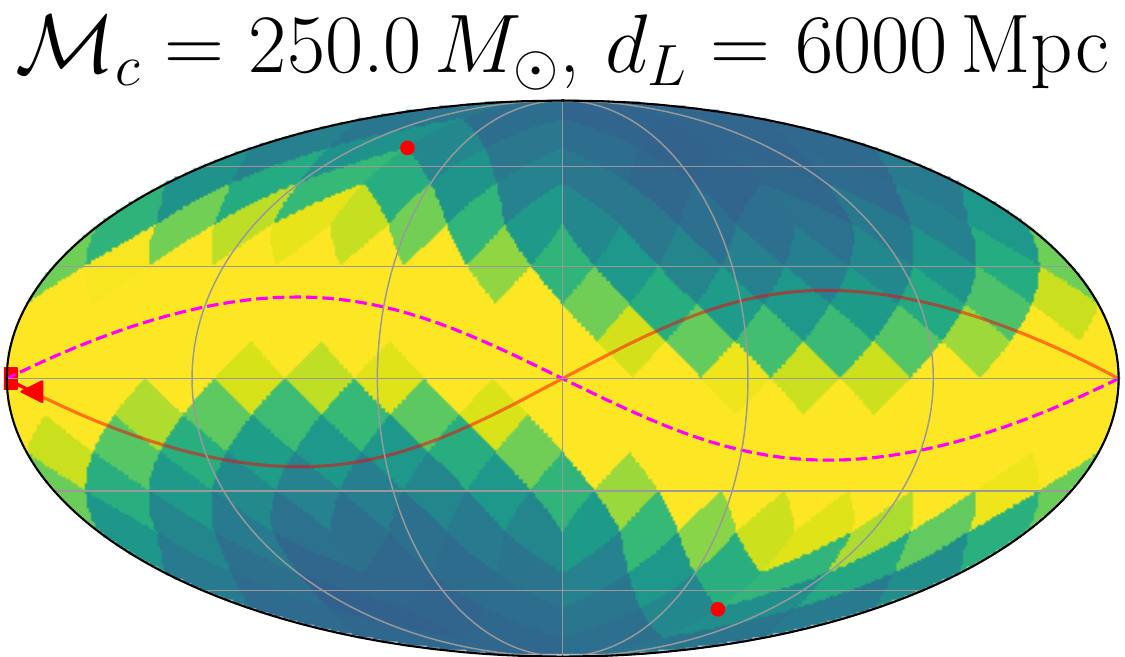}
   \vspace{0.1cm}
   
   \includegraphics[trim=3.7cm 5.7cm 3.2cm 0cm, clip, width=0.32\linewidth]{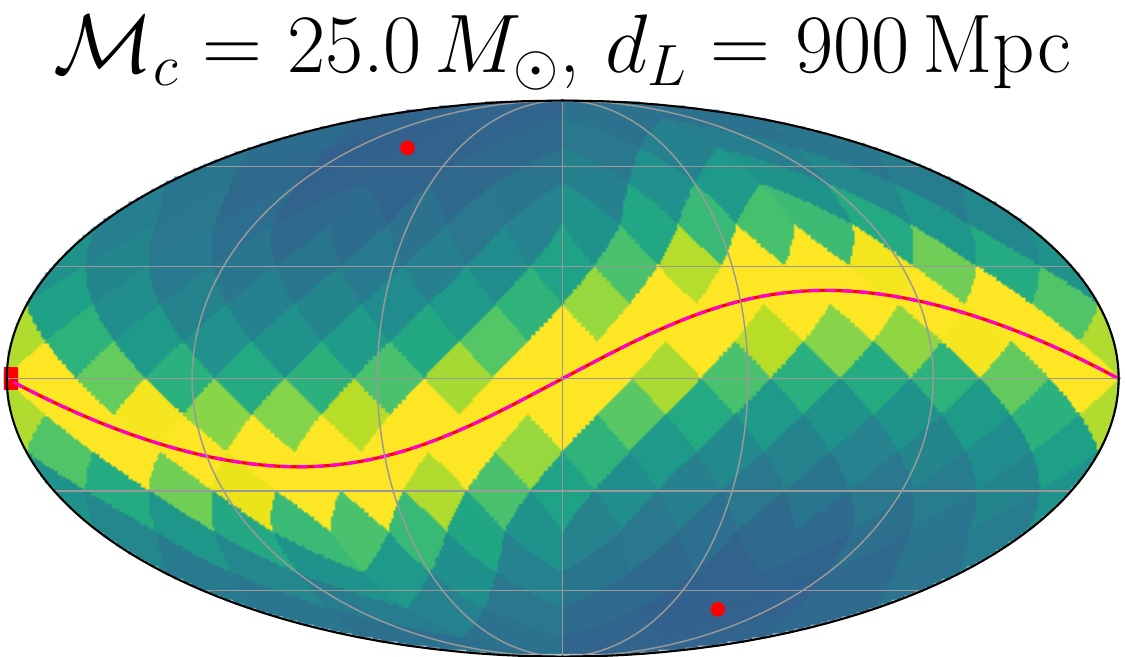}
   \includegraphics[trim=3.7cm 5.7cm 3.2cm 0cm, clip, width=0.32\linewidth]{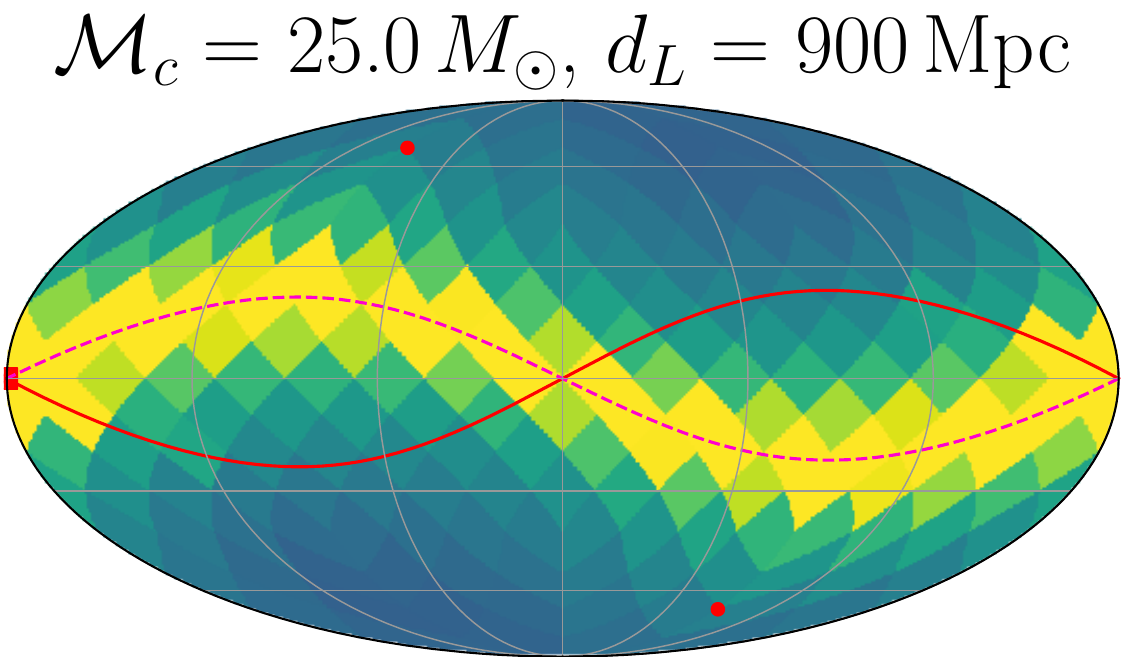}
   \includegraphics[trim=3.7cm 5.7cm 3.2cm 0cm, clip, width=0.32\linewidth]{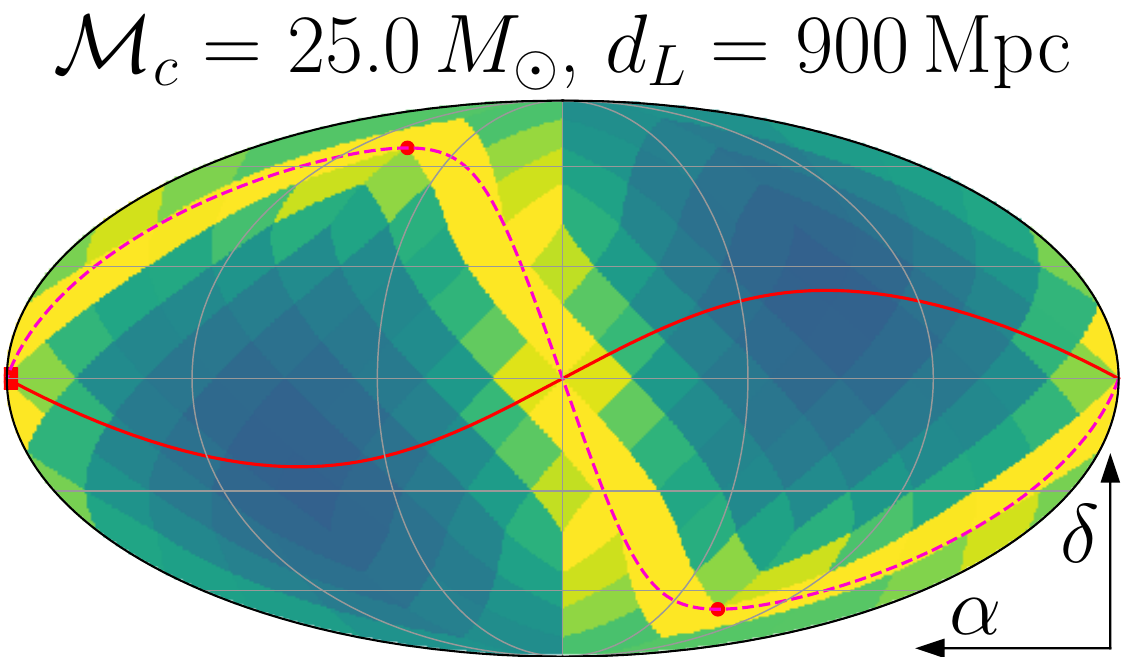}
   
   \includegraphics[trim=0cm 0cm 0cm 11.3cm, clip, width=0.44\linewidth]{Figs/skymap_dL_Space_Mc25_45deg}
   \vspace{-0.3cm}
   \caption{{\bf \boldmath Relative luminosity distance error [$\sigma(d_L)/d_L$]} as a function of the source's sky-location for a ``MAGIS-space'' detector. All plots are Mollweide projections of equatorial coordinates, right ascension ($\alpha$) increases from right to left, declination ($\delta$) increases from bottom to top, and the origin ($\alpha = \delta = 0$) is at the center of each map. The solid red line marks the ecliptic plane and the red circles indicate the ecliptic poles. The red triangle and square are the start and end points, respectively, of the detector location as seen from the Sun over the course of the observation; note that these are identical for signals observed for a full year. The dashed magenta line visualizes the detector orbit. {\it Upper row, left-to-right:} Results for $\mathcal{M}_c = \{1.1,~60,~250\}\,M_\odot$, respectively, for a detector in an orbit $45^\circ$ from the ecliptic. {\it Lower row, left-to-right:} Results for $\mathcal{M}_c = 25\,M_\odot$ for a space detector in an orbit $0^\circ,~45^\circ,~90^\circ$ from the ecliptic, respectively. All other parameters are set to the values given in Tab.~\ref{tab:Parameters}.}
   \label{fig:skymap_dL_Space}
   \vspace{0.3cm}

   \includegraphics[trim=3.5cm 5.8cm 3.9cm 0cm, clip, width=0.32\linewidth]{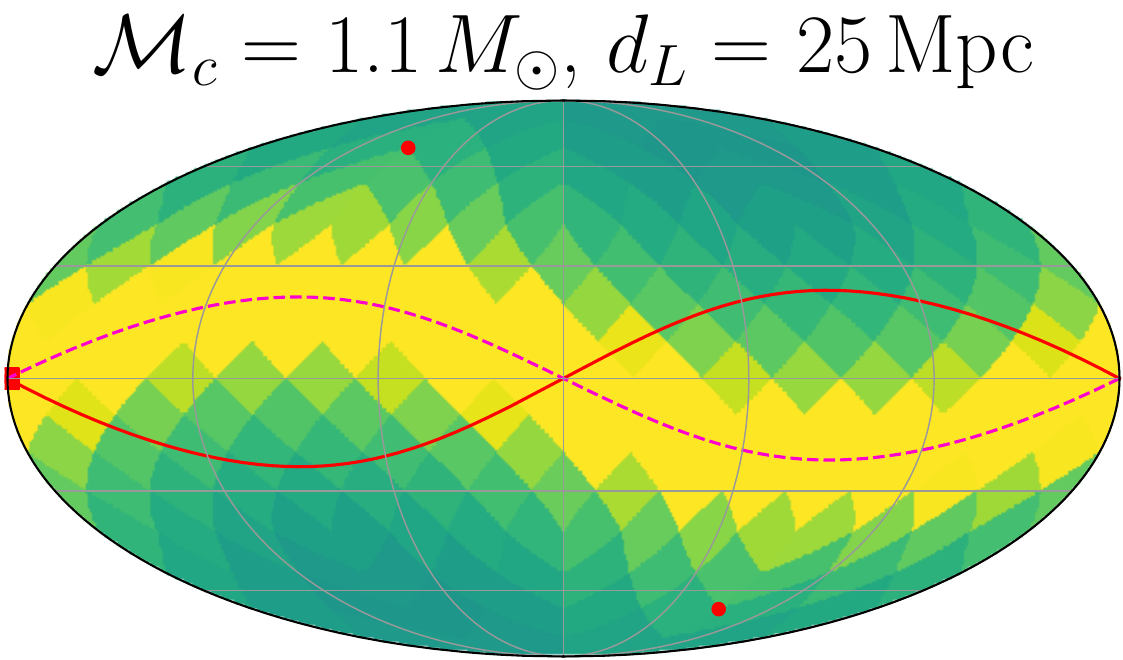}
   \includegraphics[trim=3.5cm 5.8cm 3.9cm 0cm, clip, width=0.32\linewidth]{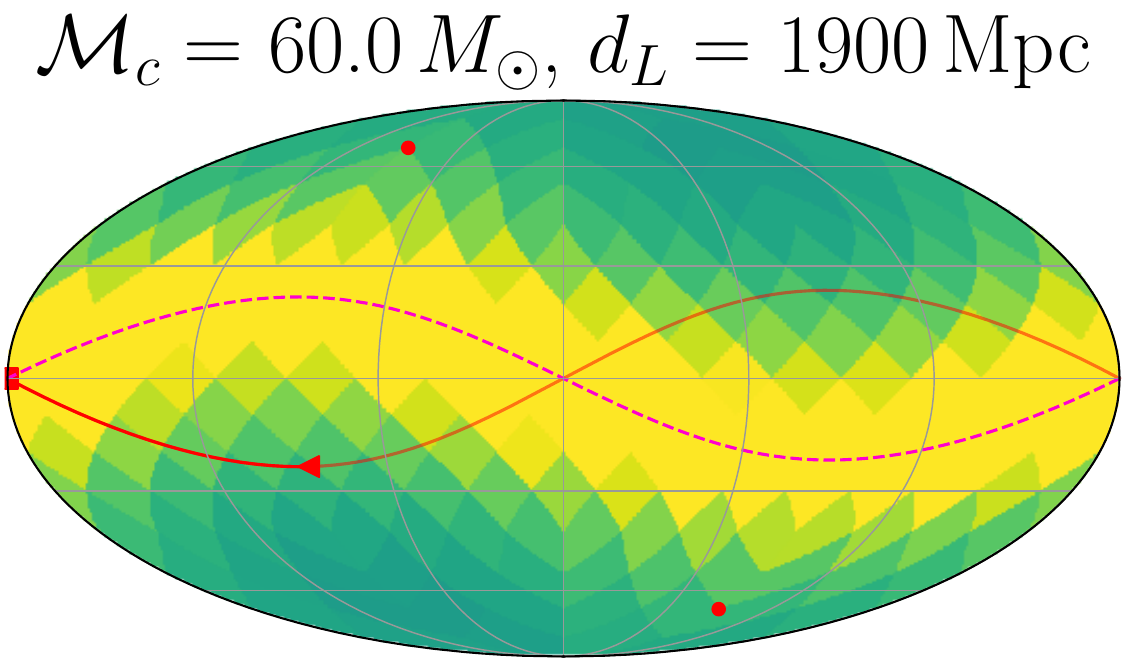}
   \includegraphics[trim=3.5cm 5.8cm 3.9cm 0cm, clip, width=0.32\linewidth]{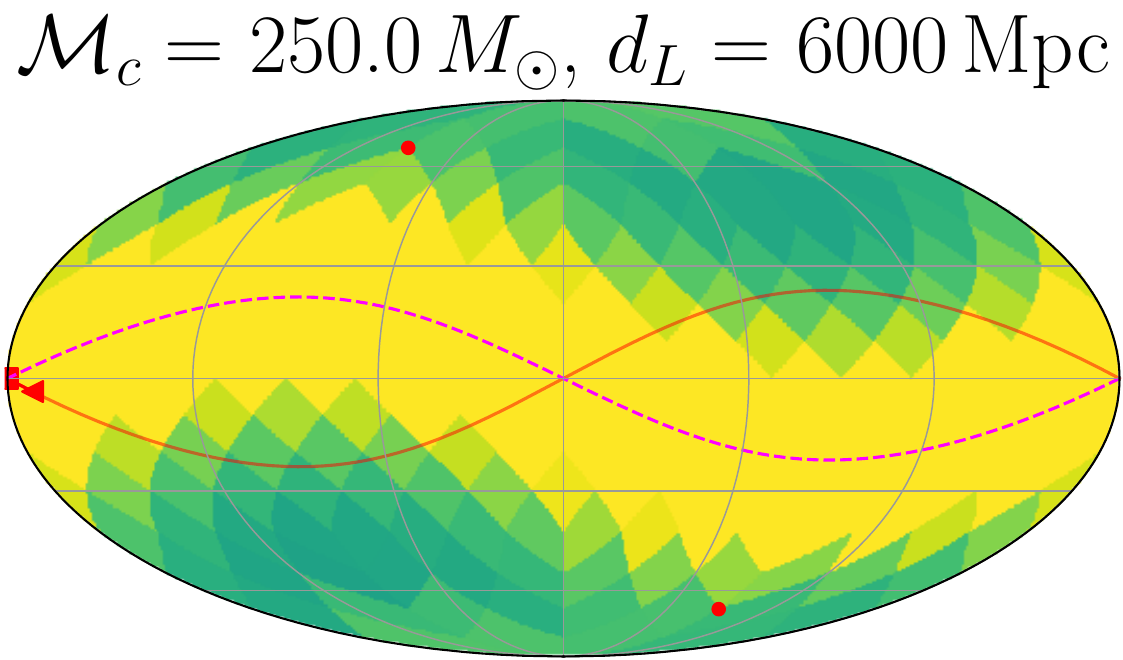}
   \vspace{0.1cm}
   
   \includegraphics[trim=3.5cm 5.8cm 3.9cm 0cm, clip, width=0.32\linewidth]{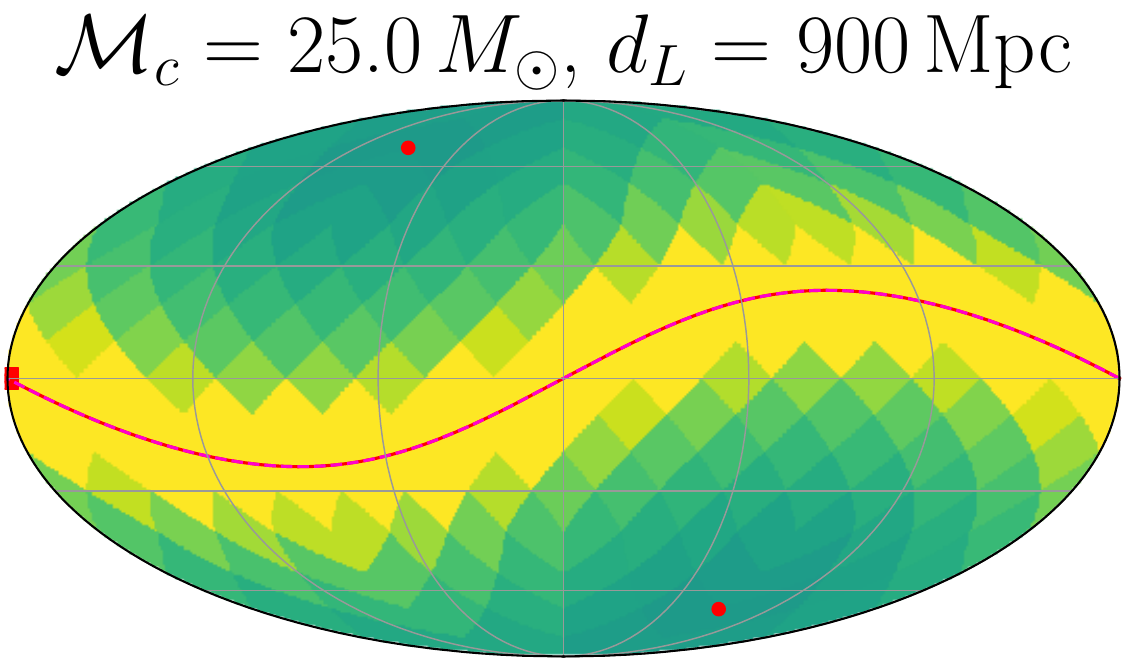}
   \includegraphics[trim=3.5cm 5.8cm 3.9cm 0cm, clip, width=0.32\linewidth]{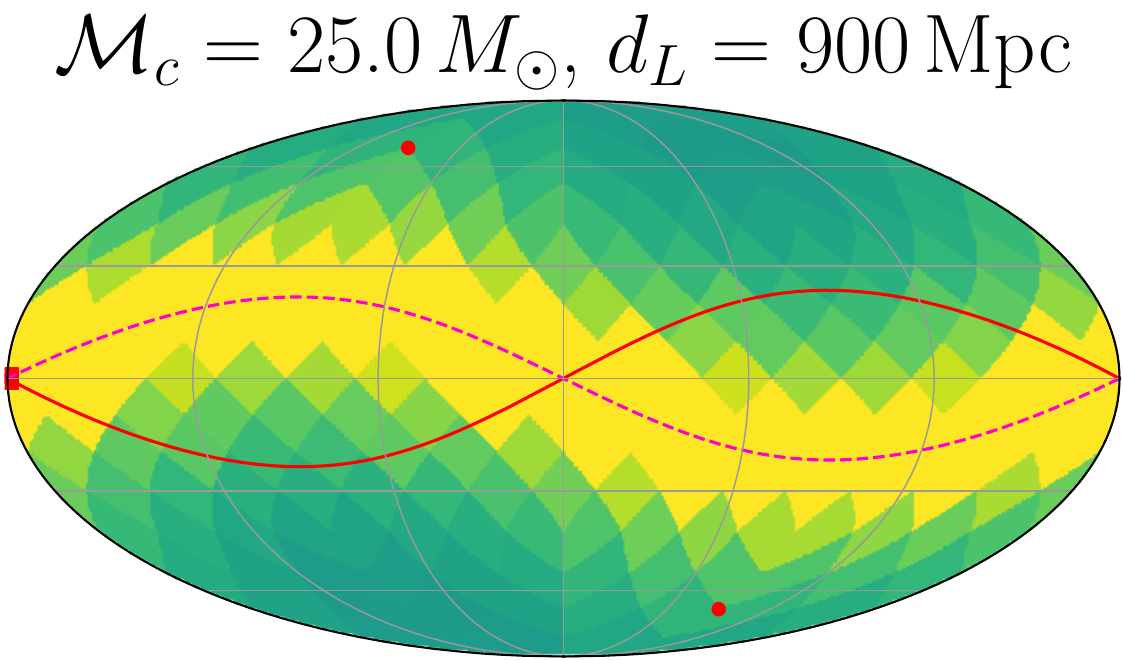}
   \includegraphics[trim=3.5cm 5.8cm 3.9cm 0cm, clip, width=0.32\linewidth]{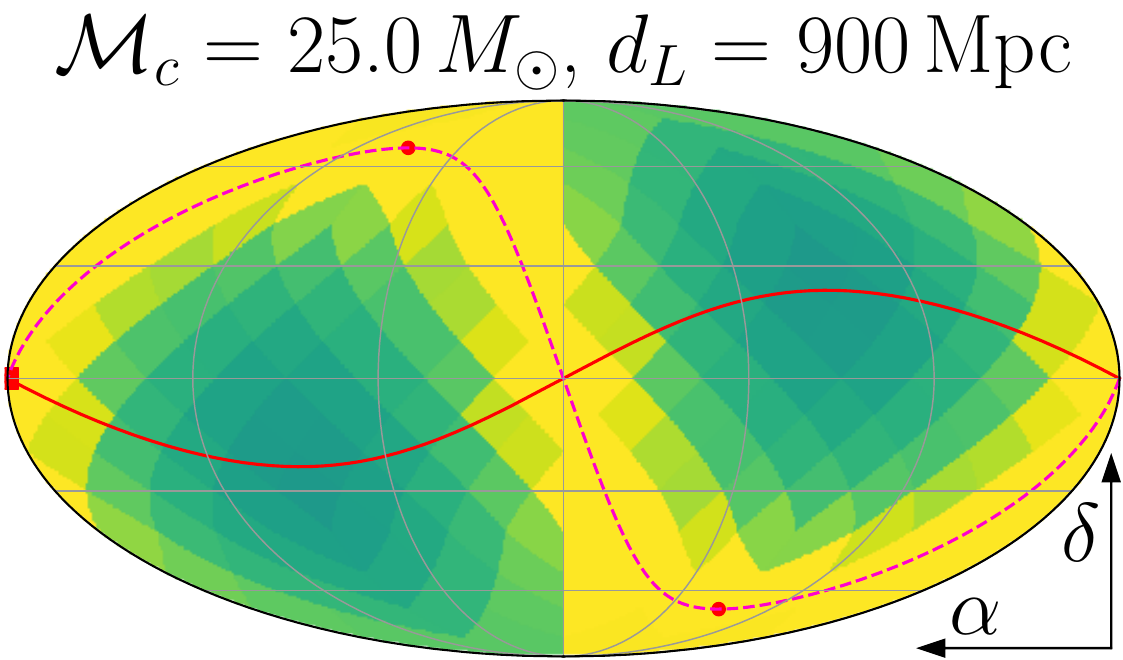}

   \includegraphics[trim=0cm 0cm 0cm 11.3cm, clip, width=0.45\linewidth]{Figs/skymap_Lres_Space_Mc25_45deg}
   \vspace{-0.3cm}
   \caption{{\bf \boldmath Uncertainty in the direction of the binary's angular momentum ($\sqrt{\sigma_{\Omega_L}}$)} [see Eq.~\eqref{eq:OmegaLUncertaintyDefinition}] for a ``MAGIS-space'' detector, otherwise the same as Fig.~\ref{fig:skymap_dL_Space}.}
   \label{fig:skymap_Lres_Space}
\end{figure}

\begin{figure}
   \centering
   \text{MAGIS-1\,km}
   \vspace{0.2cm}
   
   \includegraphics[trim=3.7cm 5.6cm 3.3cm 0cm, clip, width=0.32\linewidth]{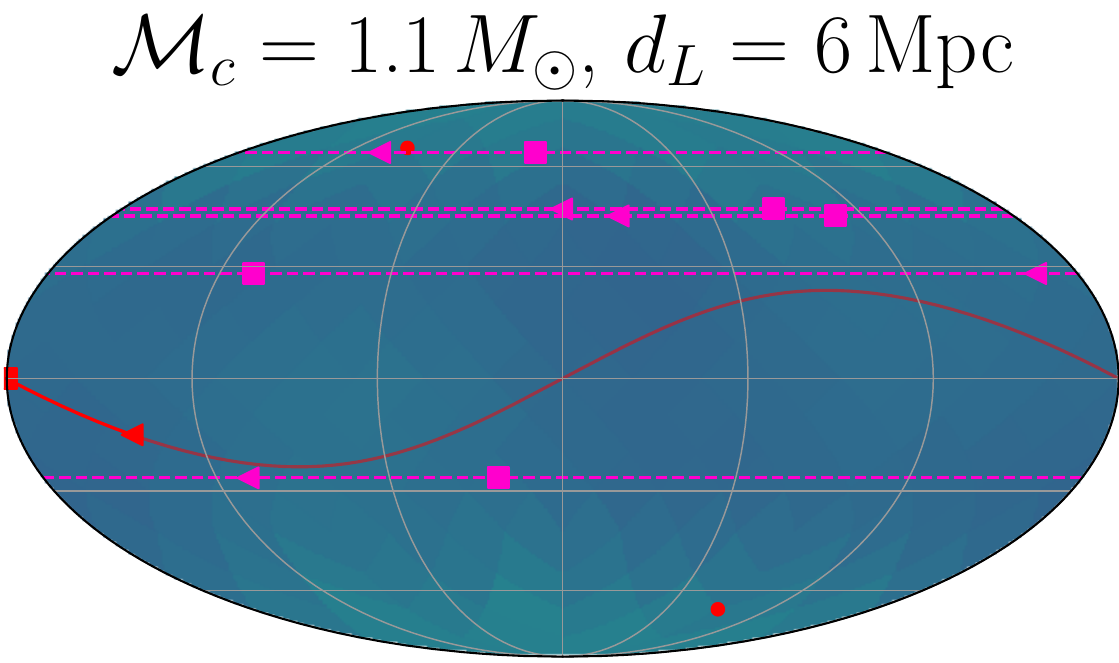}
   \includegraphics[trim=3.7cm 5.6cm 3.3cm 0cm, clip, width=0.32\linewidth]{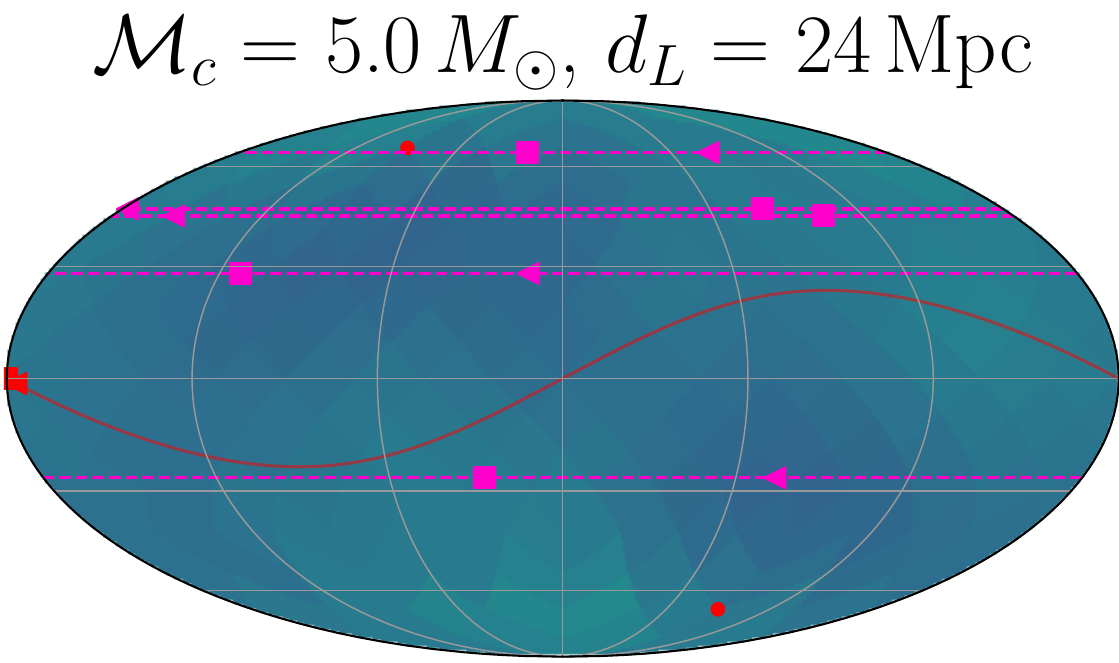}
   \includegraphics[trim=3.7cm 5.6cm 3.3cm 0cm, clip, width=0.32\linewidth]{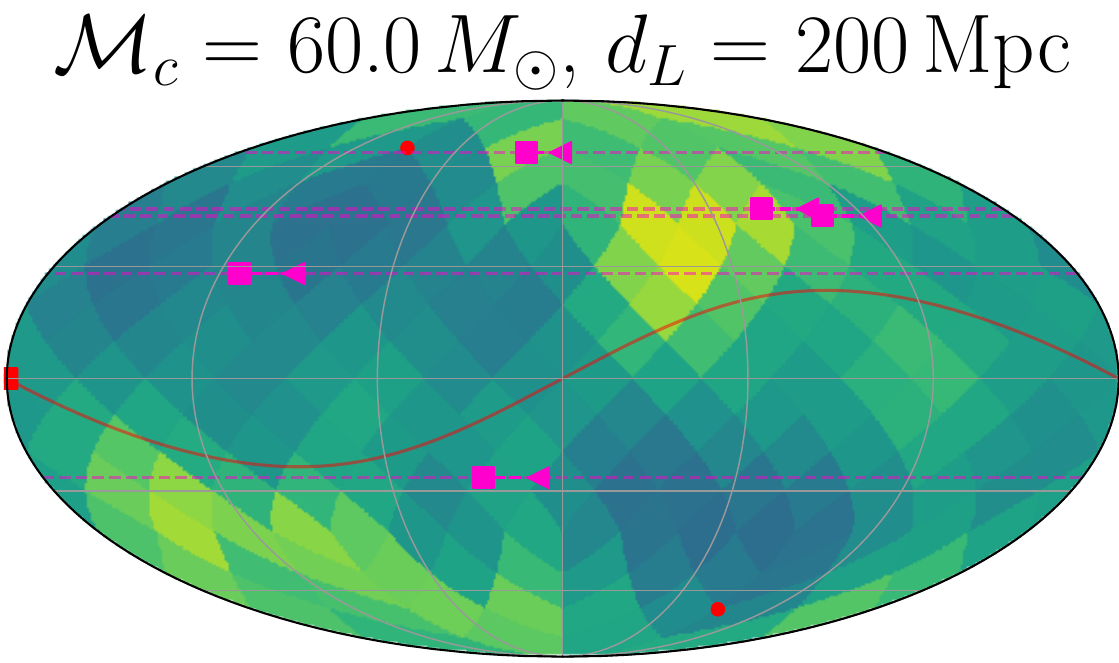}
   \vspace{0.1cm}
   
   \includegraphics[trim=3.7cm 5.6cm 3.3cm 0cm, clip, width=0.32\linewidth]{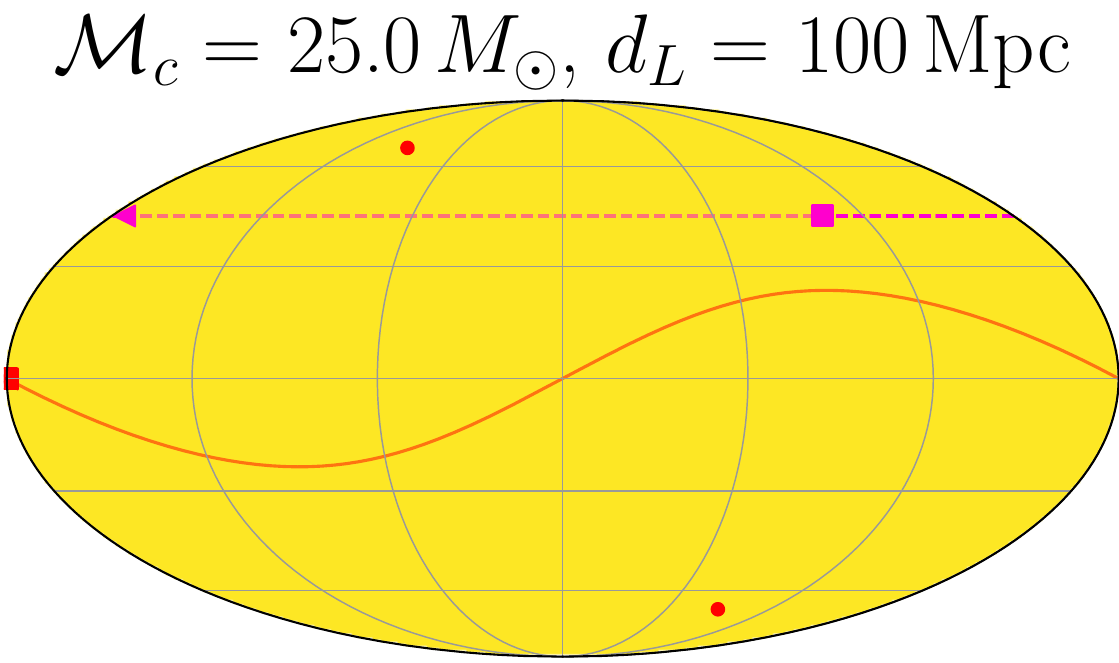}
   \includegraphics[trim=3.7cm 5.6cm 3.3cm 0cm, clip, width=0.32\linewidth]{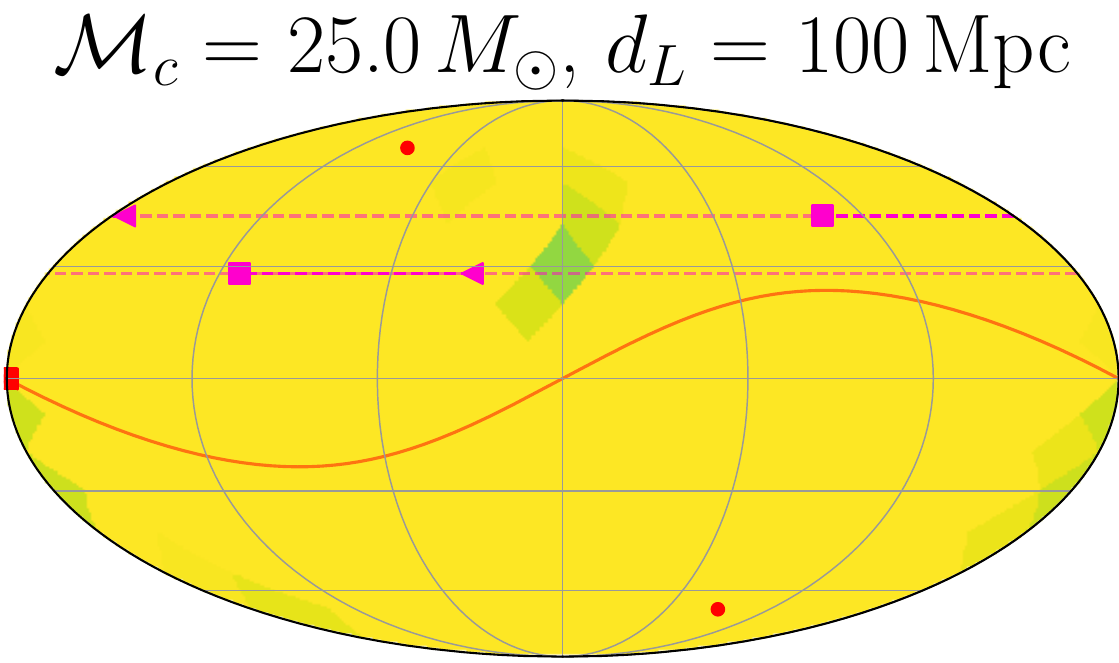}
   \includegraphics[trim=3.7cm 5.6cm 3.3cm 0cm, clip, width=0.32\linewidth]{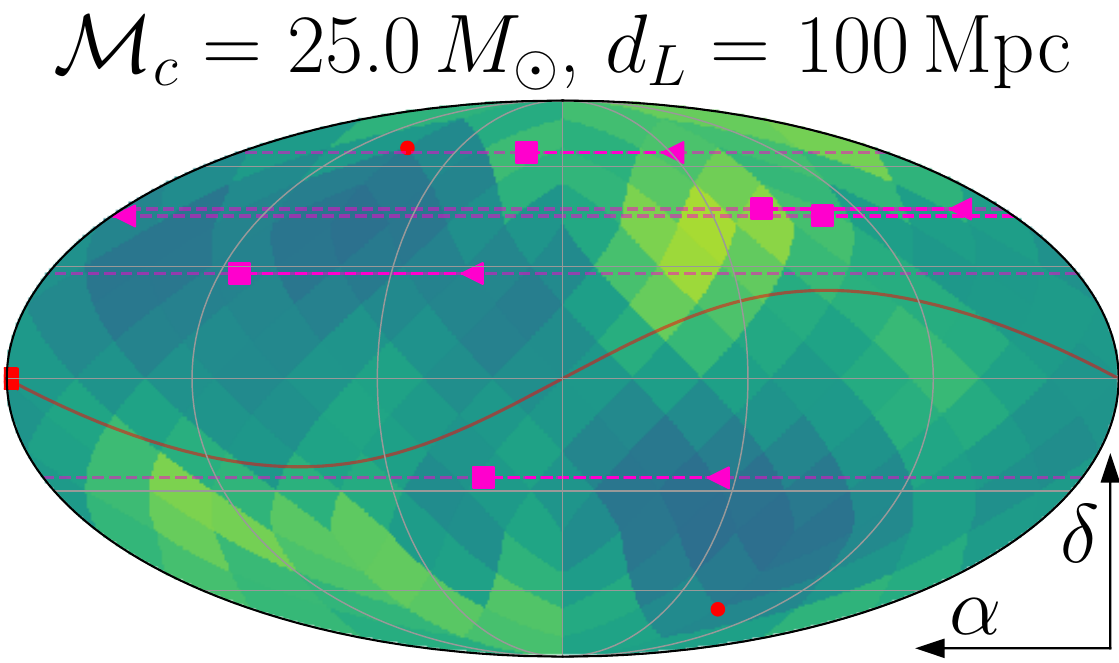}

   \includegraphics[trim=0cm 0cm 0cm 11.5cm, clip, width=0.45\linewidth]{Figs/skymap_dL_Ground_Mc25_c}
   \vspace{-0.5cm}
   \caption{{\bf \boldmath Relative luminosity distance error [$\sigma(d_L)/d_L$]} as a function of the source's sky-location for ``MAGIS-1\,km''. All plots are Mollweide projections of equatorial coordinates, right ascension ($\alpha$) increases from right to left, declination ($\delta$) increases from bottom to top, and the origin ($\alpha = \delta = 0$) is at the center of each map. The solid red line marks the ecliptic plane and the red circles indicate the ecliptic poles. The red triangle and square are the start and end points, respectively, of Earth's location as seen from the Sun over the course of the observation. The dashed magenta lines show the directions the baselines of each detector are pointing; magenta triangles and squares mark their start and end locations, respectively. {\it Upper row, left-to-right:} Results for $\mathcal{M}_c = \{1.1,~5,~60\}\,M_\odot$, respectively, for the \TMN network (i.e., detectors at Homestake, Sudbury, Renstr\"om, Tautona, and Zaoshan). {\it Lower row, left-to-right:} Results for $\mathcal{M}_c = 25\,M_\odot$ for Homestake alone, Homestake and Zaoshan, and for all five detectors. All other parameters are set to the values given in Tab.~\ref{tab:Parameters}.}
    \label{fig:skymap_dL_Ground}
   \vspace{0.2cm}
   
   \includegraphics[trim=3.6cm 5.8cm 3.9cm 0cm, clip, width=0.32\linewidth]{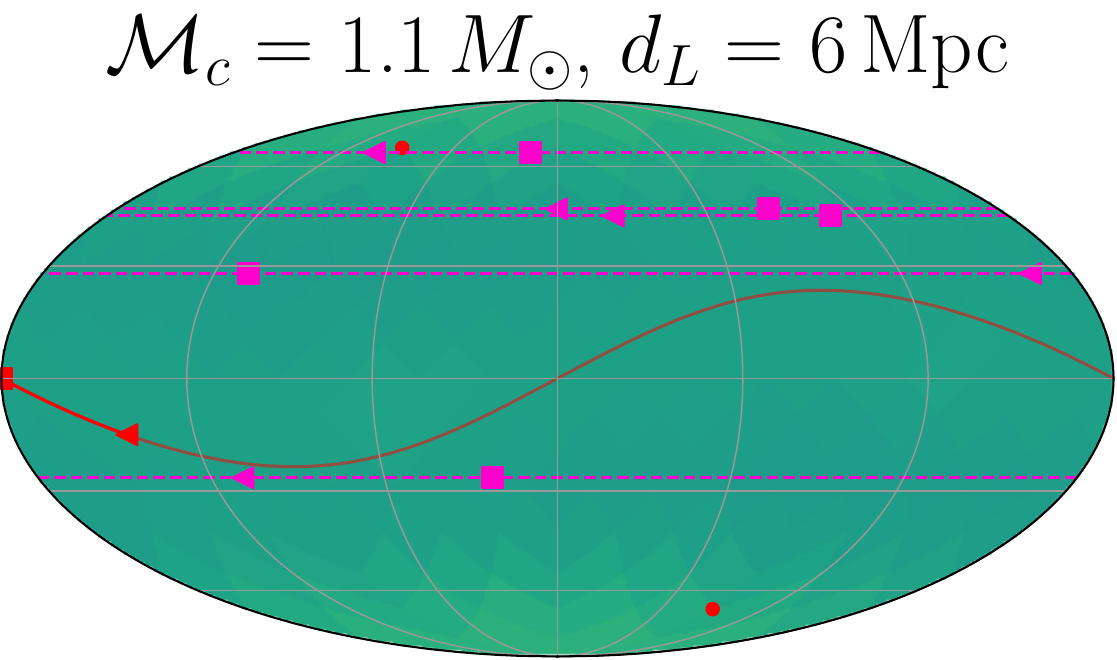}
   \includegraphics[trim=3.6cm 5.8cm 3.9cm 0cm, clip, width=0.32\linewidth]{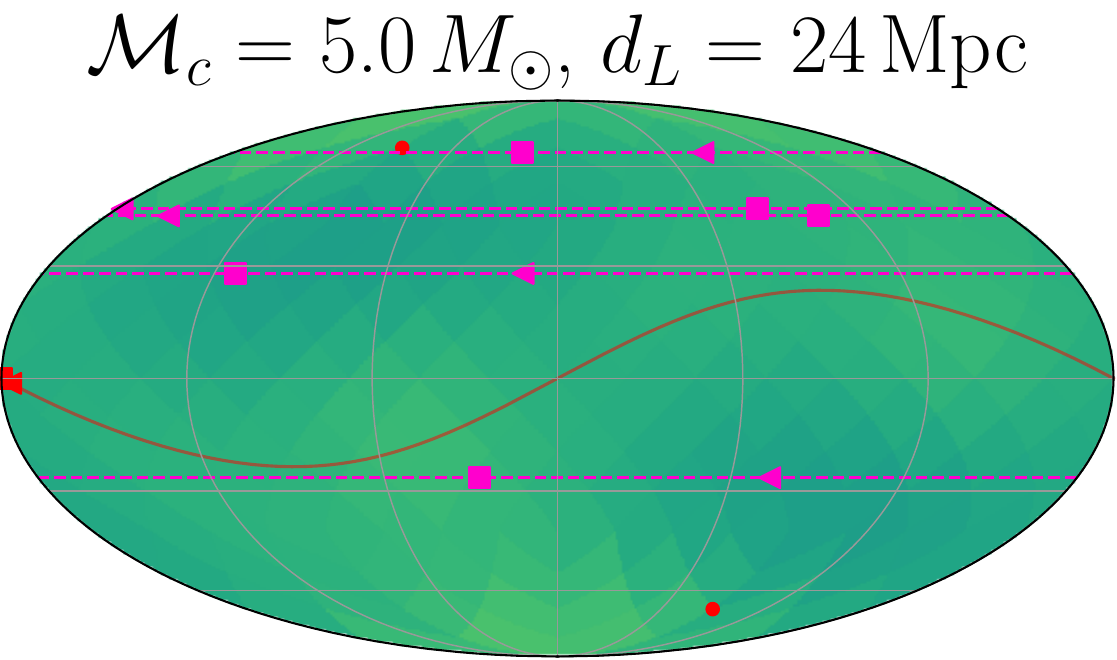}
   \includegraphics[trim=3.6cm 5.8cm 3.9cm 0cm, clip, width=0.32\linewidth]{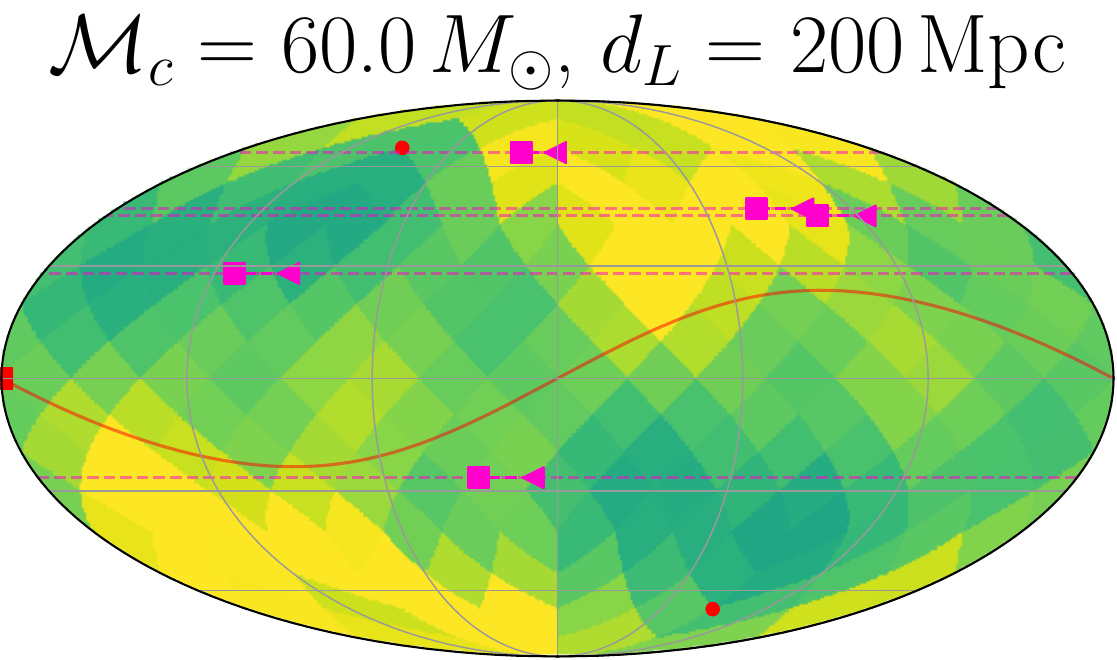}
   \vspace{0.1cm}
   
   \includegraphics[trim=3.6cm 5.8cm 3.9cm 0cm, clip, width=0.32\linewidth]{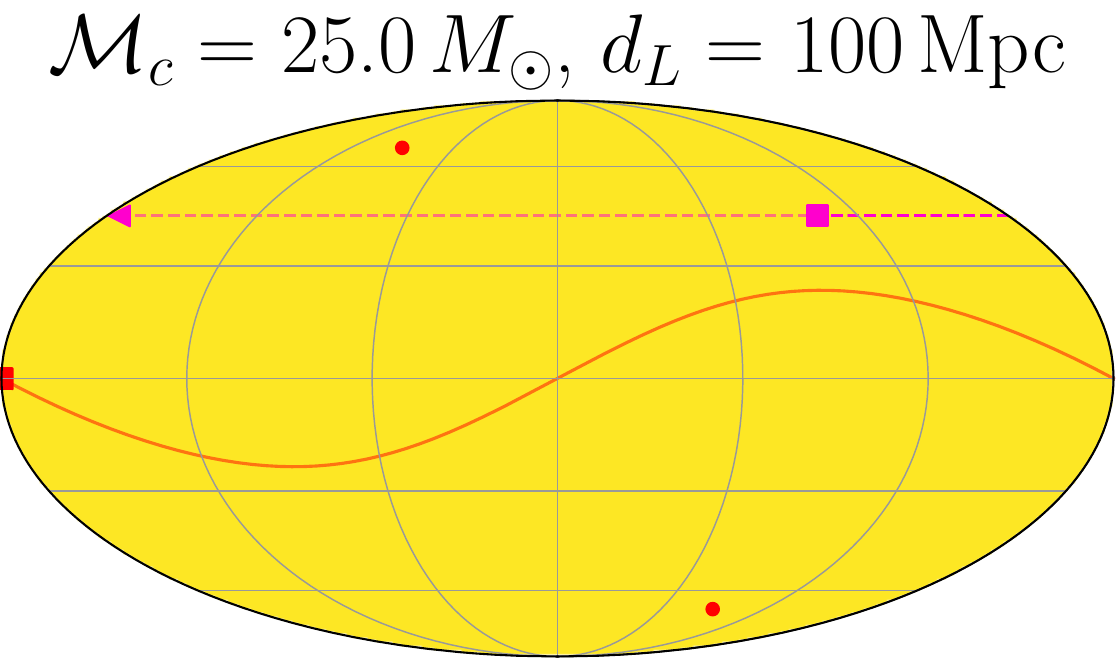}
   \includegraphics[trim=3.6cm 5.8cm 3.9cm 0cm, clip, width=0.32\linewidth]{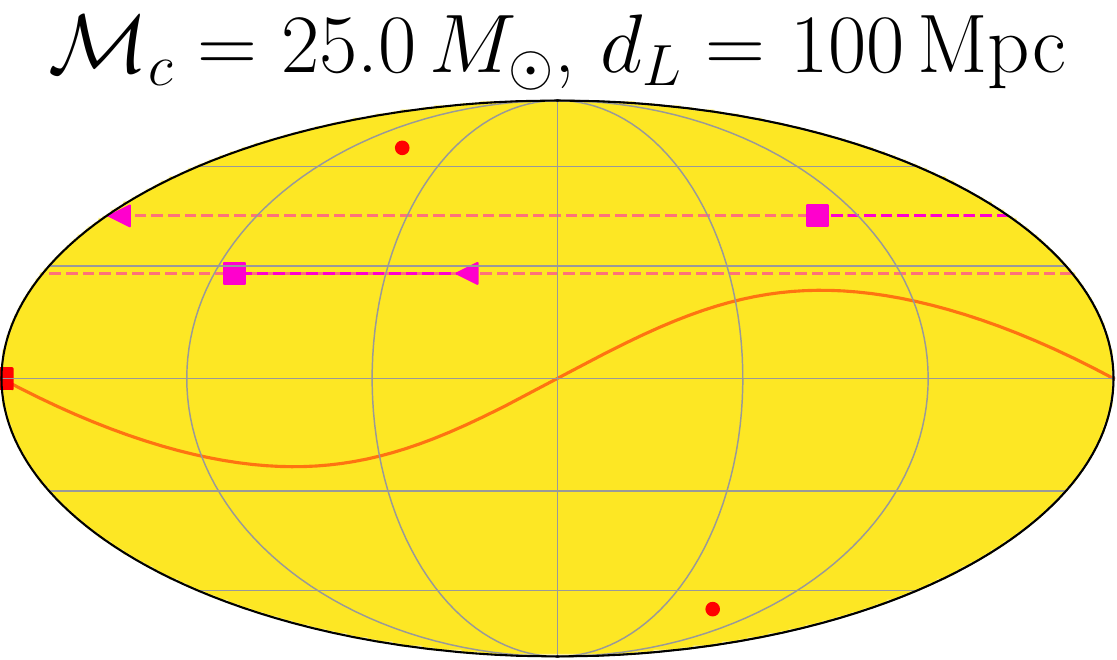}
   \includegraphics[trim=3.6cm 5.8cm 3.9cm 0cm, clip, width=0.32\linewidth]{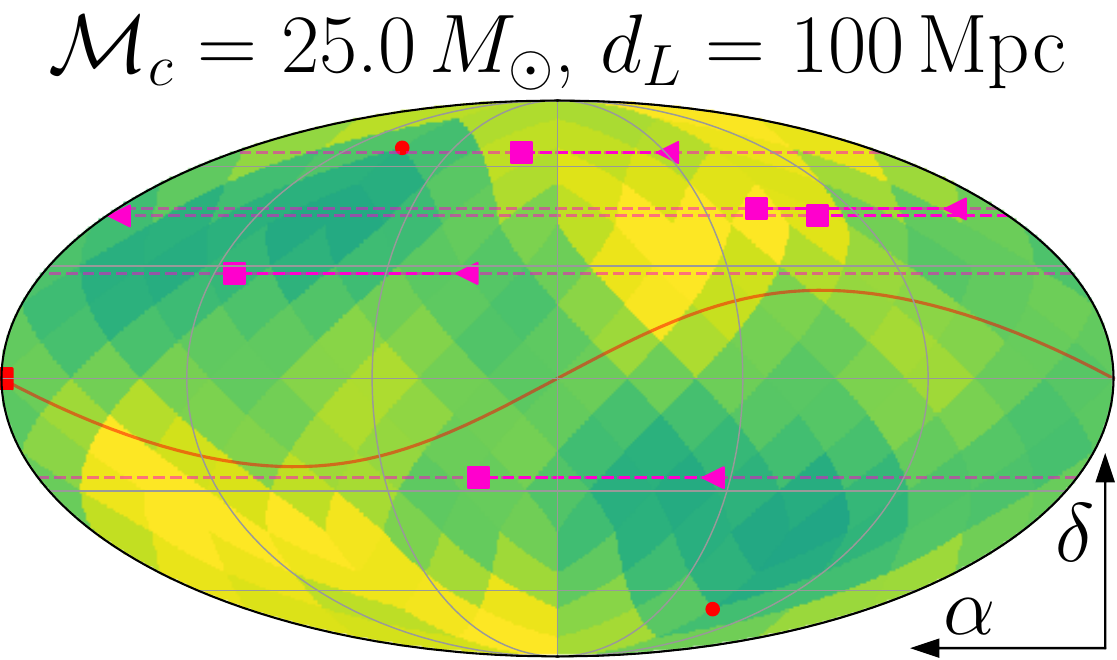}

   \includegraphics[trim=0cm 0cm 0cm 11.4cm, clip, width=0.45\linewidth]{Figs/skymap_Lres_Ground_Mc25_c}
   \vspace{-0.5cm}
   \caption{{\bf \boldmath Uncertainty in the direction of the binary's angular momentum ($\sqrt{\sigma_{\Omega_L}}$)} [see Eq.~\eqref{eq:OmegaLUncertaintyDefinition}] for ``MAGIS-1\,km'', otherwise the same as Fig.~\ref{fig:skymap_dL_Ground}.}
   \label{fig:skymap_Lres_Ground}
\end{figure}

\begin{figure}
   \centering
   \text{MAGIS-space, ecliptic orbit}
   \vspace{0.2cm}
   
   \includegraphics[trim=0.9cm 4.6cm 1.1cm 0cm, clip, width=0.4\linewidth]{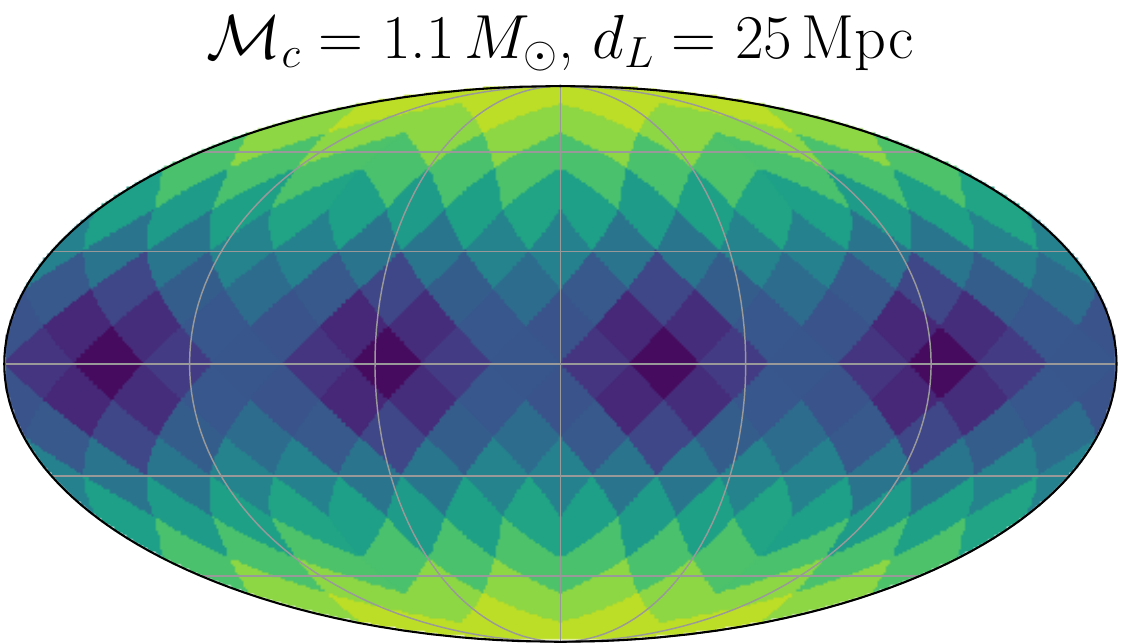}
   \hspace{1cm}
   \includegraphics[trim=0.9cm 4.6cm 1.1cm 0cm, clip, width=0.4\linewidth]{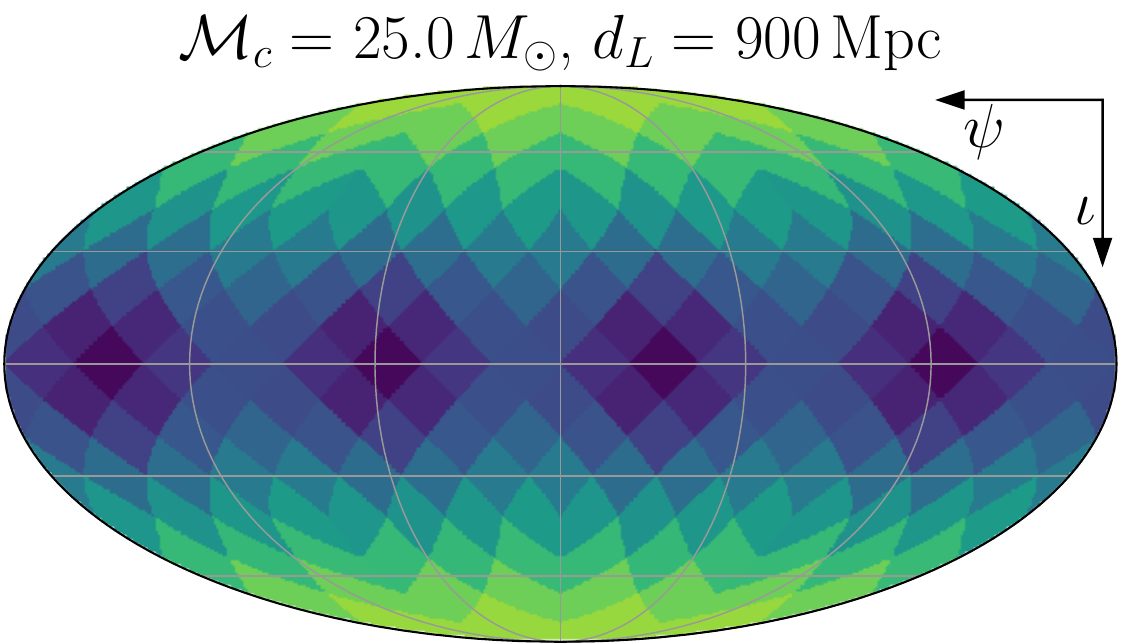}
   \vspace{0.1cm}
   
   \includegraphics[trim=0.9cm 4.6cm 1.1cm 0cm, clip, width=0.4\linewidth]{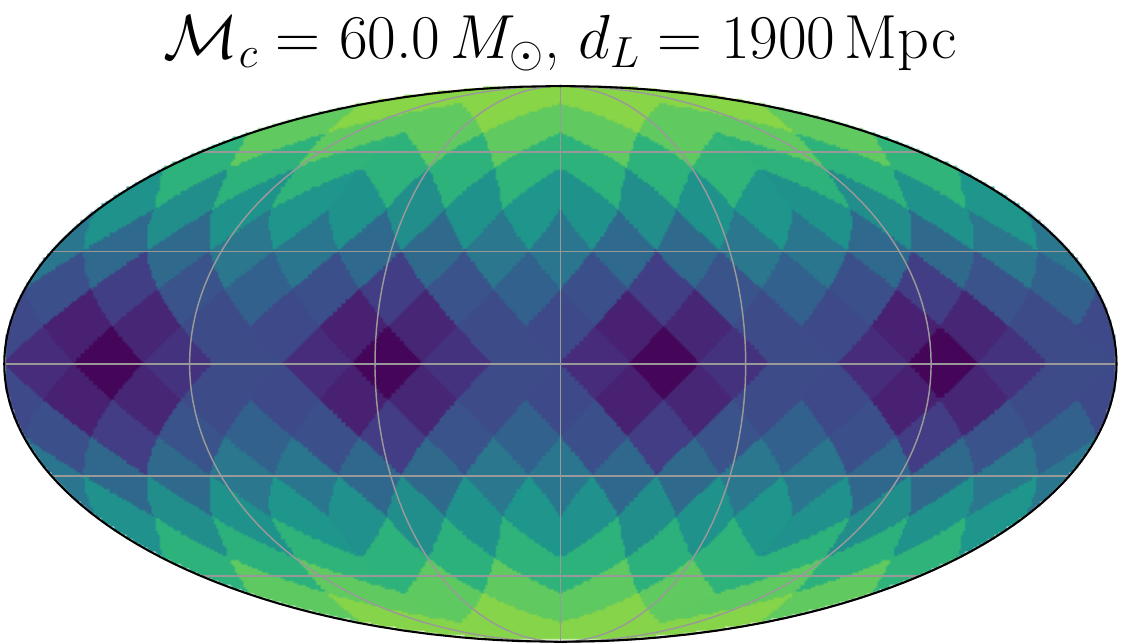}
   \hspace{1cm}
   \includegraphics[trim=0.9cm 4.6cm 1.1cm 0cm, clip, width=0.4\linewidth]{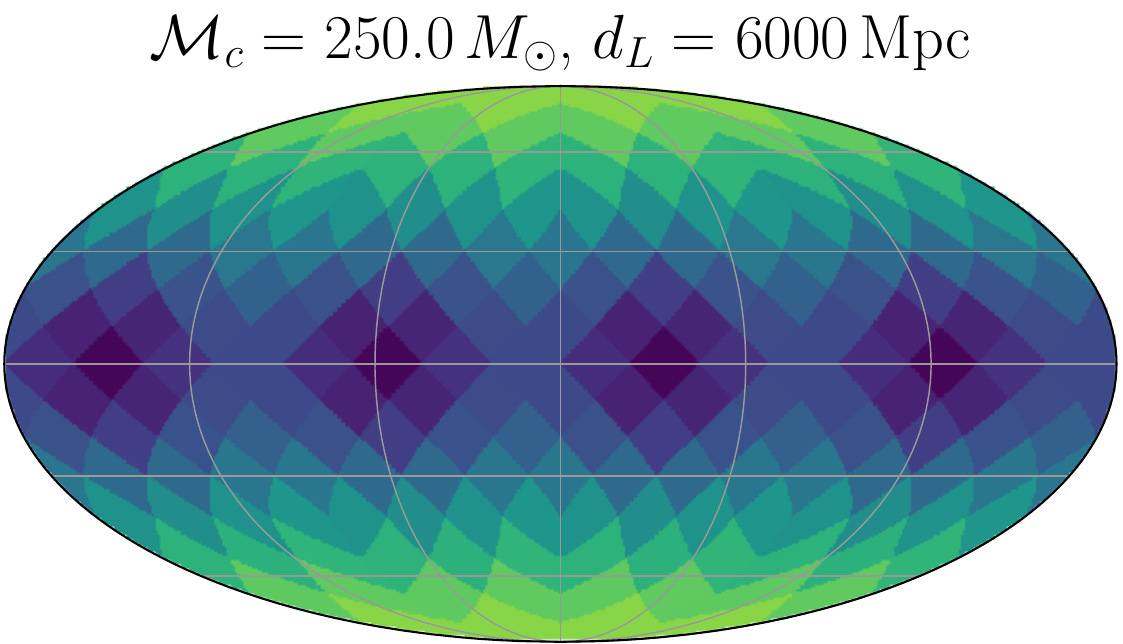}
   
   \includegraphics[trim=0cm 0cm 0cm 11.5cm, clip, width=0.43\linewidth]{Figs/Lmap_SNR_Space_Mc60_0deg}
   \vspace{-0.5cm}
   \caption{{\bf Signal-to-noise ratio (SNR)} as a function of the orientation of the binary's orbital momentum for a ``MAGIS-space'' detector in geocentric orbit parallel to the ecliptic plane. All plots are Mollweide projections. The polarization angle ($\psi$) increases from right to left, the inclination angle ($\iota$) increases from top to bottom, and the origin ($\psi = \iota = 0$) is in the middle of the top of each map. The different panels are for difference choices of the chirp mass and the luminosity distance as denoted at the top of each panel; note that these choices are the same as in Figs.~\ref{fig:skymap_SNR_Space},~\ref{fig:skymap_angRes_Space},~\ref{fig:skymap_dL_Space}, and~\ref{fig:skymap_Lres_Space}. All other parameters are set to the benchmark values given in Tab.~\ref{tab:Parameters}.}
   \label{fig:Lmap_SNR_Space}
   \vspace{0.3cm}
   
   \includegraphics[trim=1.3cm 4.7cm 0.9cm 0cm, clip, width=0.4\linewidth]{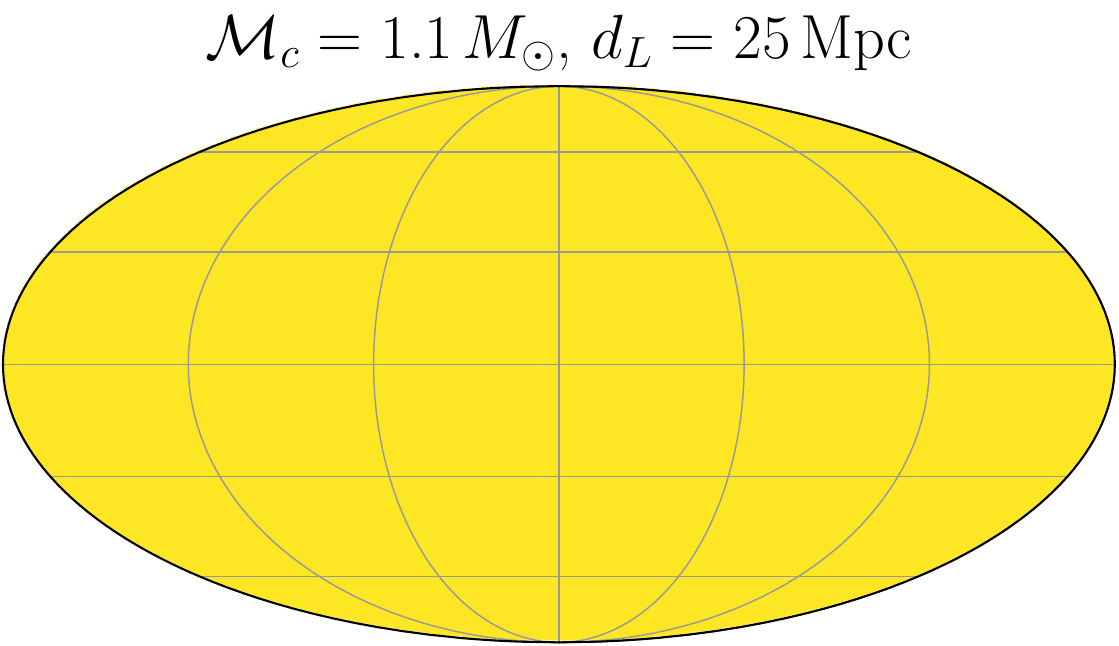}
   \hspace{1cm}
   \includegraphics[trim=1.3cm 4.7cm 0.9cm 0cm, clip, width=0.4\linewidth]{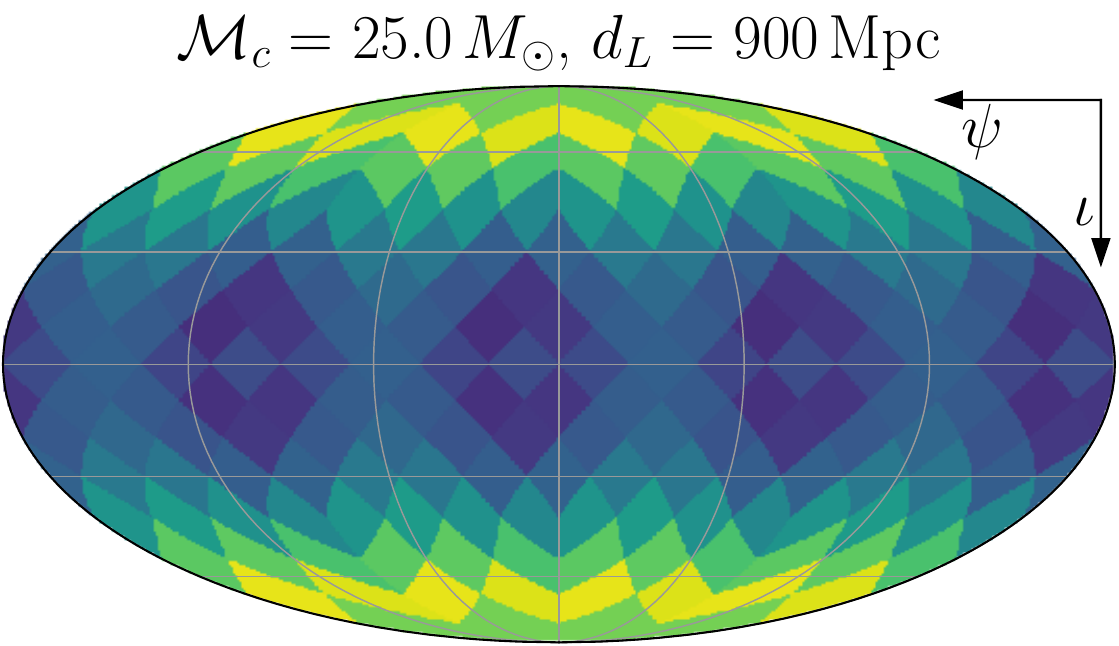}
   \vspace{0.1cm}
   
   \includegraphics[trim=1.3cm 4.7cm 0.9cm 0cm, clip, width=0.4\linewidth]{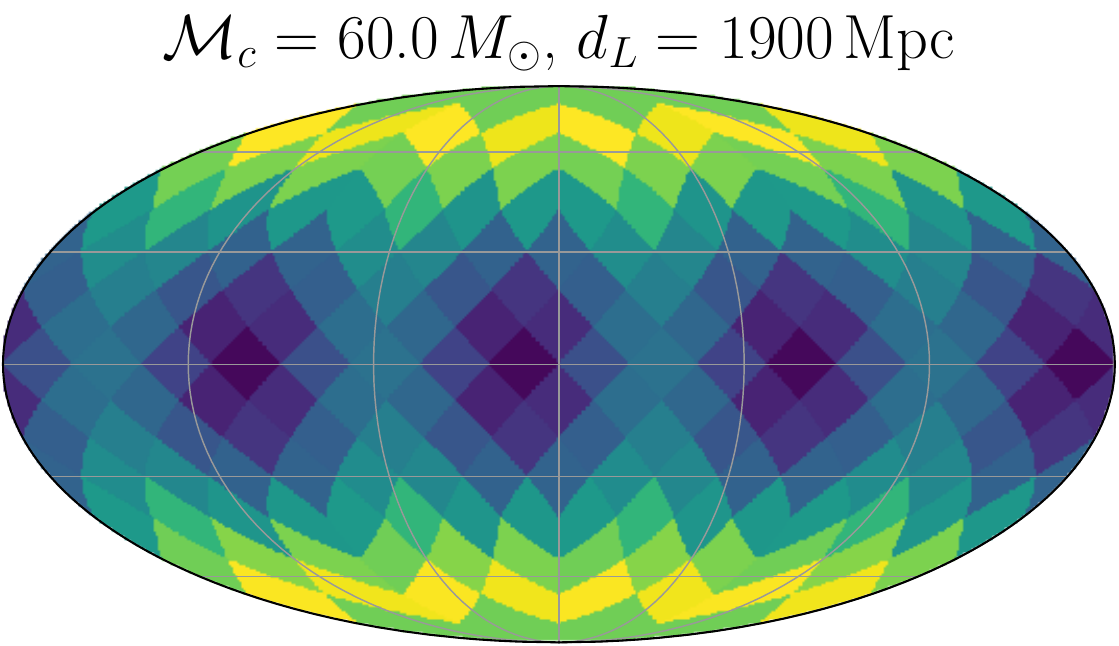}
   \hspace{1cm}
   \includegraphics[trim=1.3cm 4.7cm 0.9cm 0cm, clip, width=0.4\linewidth]{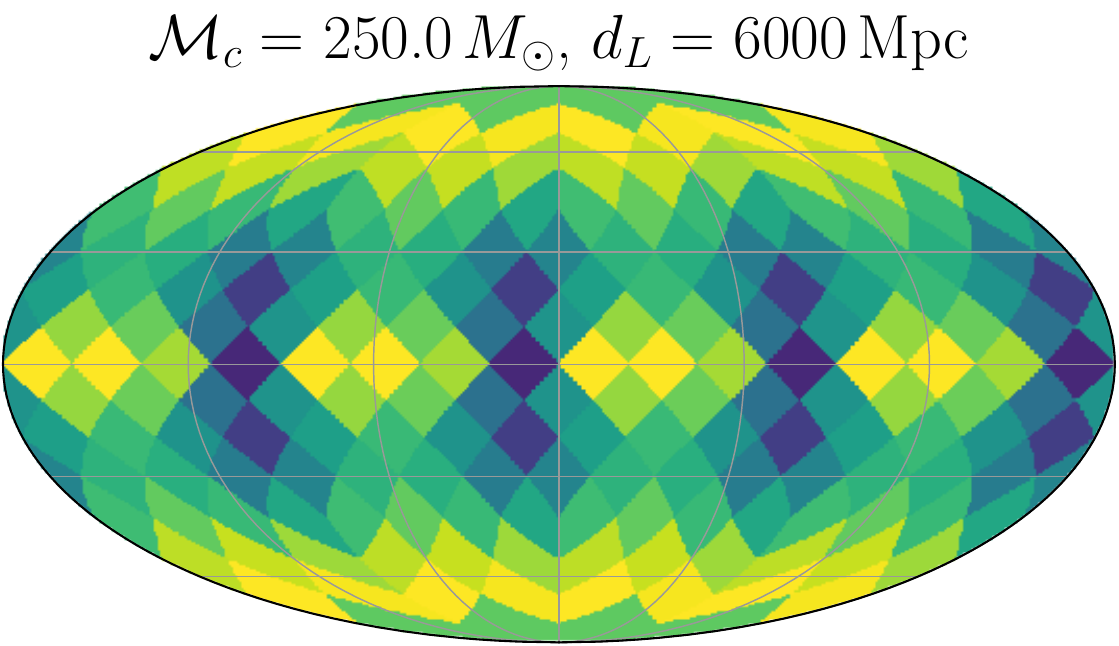}

   \includegraphics[trim=0cm 0cm 0cm 11.5cm, clip, width=0.43\linewidth]{Figs/Lmap_dL_Space_Mc250_0deg}
   \vspace{-0.5cm}
   \caption{{\bf \boldmath Relative luminosity distance error [$\sigma(d_L)/d_L$]} as a function of the orientation of the binary's orbital momentum for a ``MAGIS-space'' detector, otherwise the same as Fig.~\ref{fig:Lmap_SNR_Space}.}
   \label{fig:Lmap_dL_Space}
\end{figure}

\begin{figure}
   \centering
   \text{$5 \times \left(\text{MAGIS-1\,km}\right)$}
   \vspace{0.2cm}
   
   \includegraphics[trim=0.9cm 4.6cm 1.1cm 0cm, clip, width=0.4\linewidth]{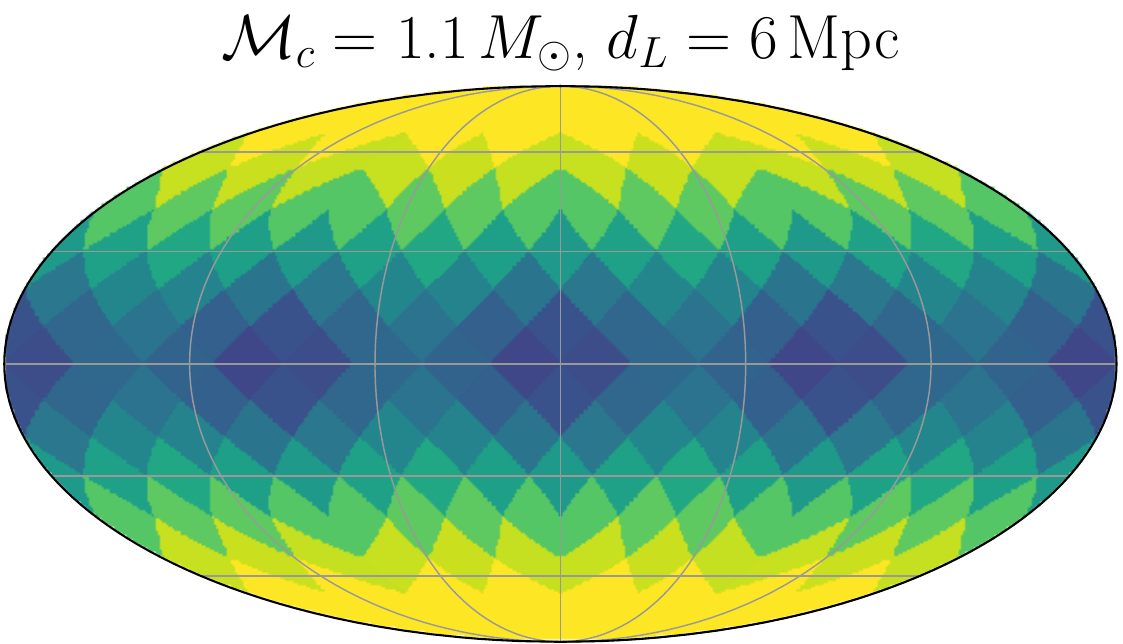}
   \hspace{1cm}
   \includegraphics[trim=0.9cm 4.6cm 1.1cm 0cm, clip, width=0.4\linewidth]{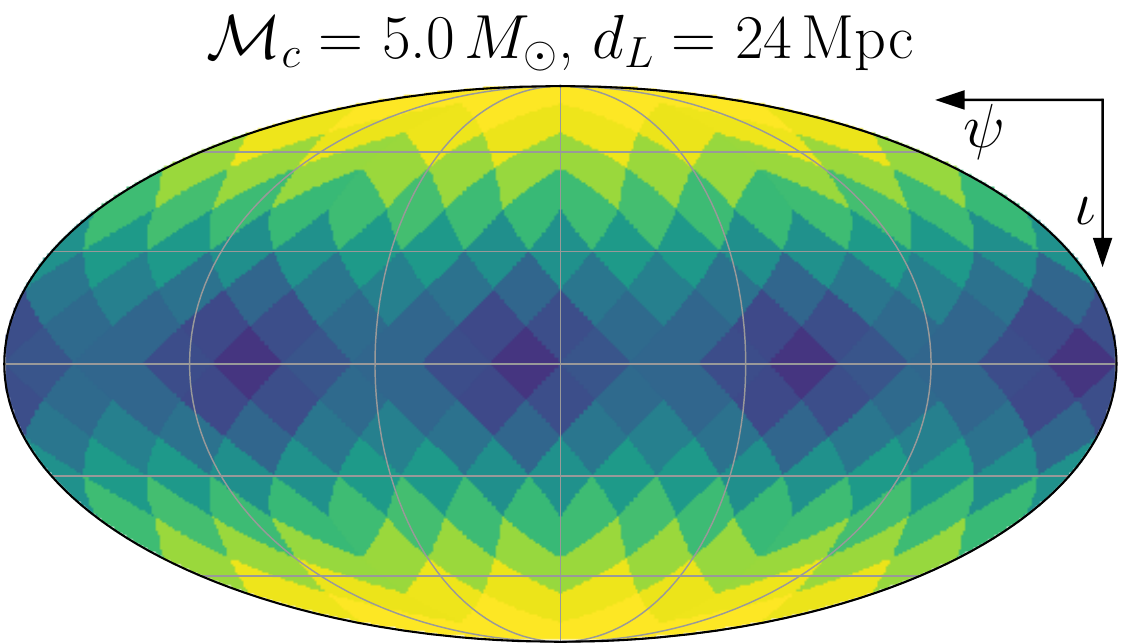}
   \vspace{0.1cm}
   
   \includegraphics[trim=0.9cm 4.6cm 1.1cm 0cm, clip, width=0.4\linewidth]{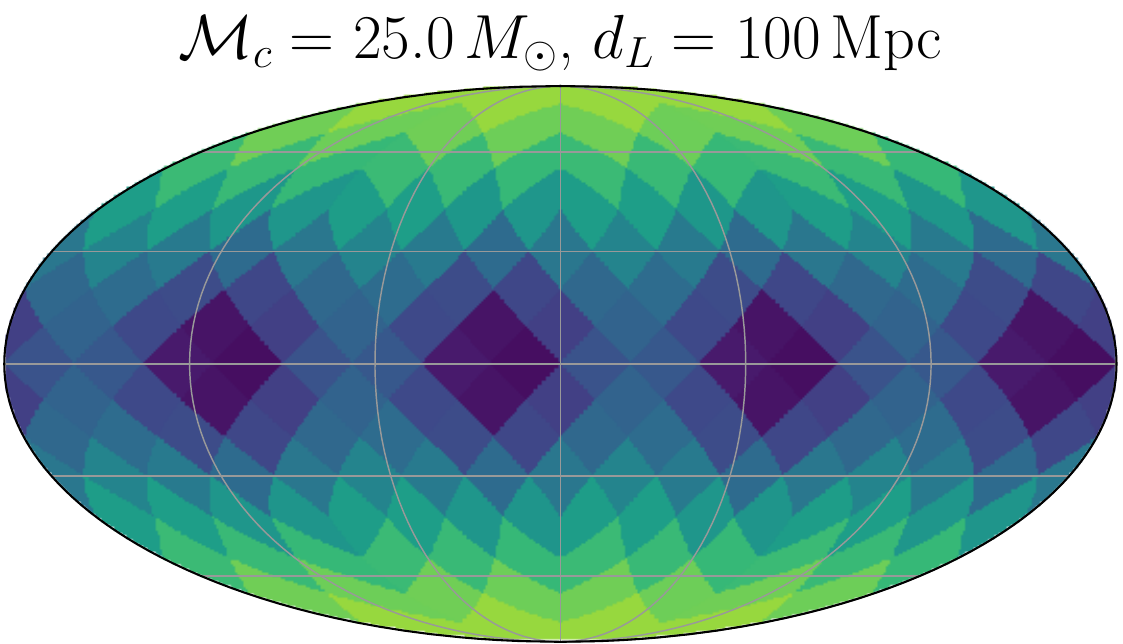}
   \hspace{1cm}
   \includegraphics[trim=0.9cm 4.6cm 1.1cm 0cm, clip, width=0.4\linewidth]{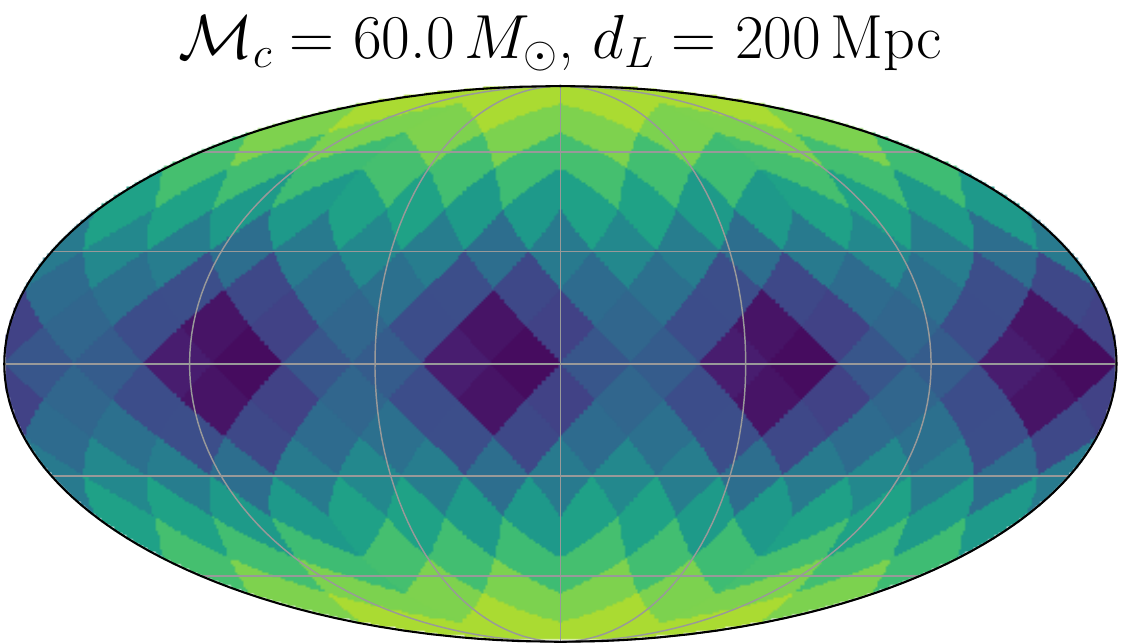}
   
   \includegraphics[trim=0cm 0cm 0cm 11.5cm, clip, width=0.43\linewidth]{Figs/Lmap_SNR_Ground_Mc60_f}
   \vspace{-0.5cm}
   \caption{{\bf Signal-to-noise ratio (SNR)} as a function of the orientation of the binary's orbital momentum for a terrestrial \TMN network. All plots are Mollweide projections. The polarization angle ($\psi$) increases from right to left, the inclination angle ($\iota$) increases from top to bottom, and the origin ($\psi = \iota = 0$) is in the middle of the top of each map. The different panels are for difference choices of the chirp mass and the luminosity distance as denoted at the top of each panel; note that these choices are the same as in Figs.~\ref{fig:skymap_SNR_Ground},~\ref{fig:skymap_angRes_Ground},~\ref{fig:skymap_dL_Ground}, and~\ref{fig:skymap_Lres_Ground}. All other parameters are set to the benchmark values given in Tab.~\ref{tab:Parameters}.}
   \label{fig:Lmap_SNR_Ground}
   \vspace{0.2cm}
   
   \includegraphics[trim=1.3cm 4.7cm 0.9cm 0cm, clip, width=0.4\linewidth]{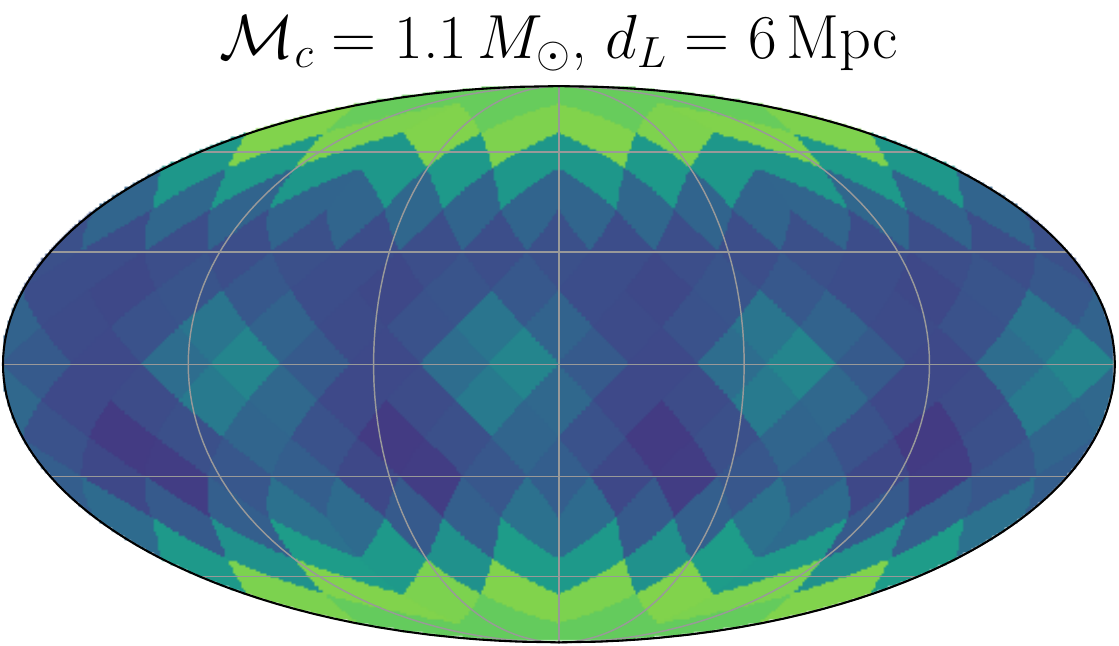}
   \hspace{1cm}
   \includegraphics[trim=1.3cm 4.7cm 0.9cm 0cm, clip, width=0.4\linewidth]{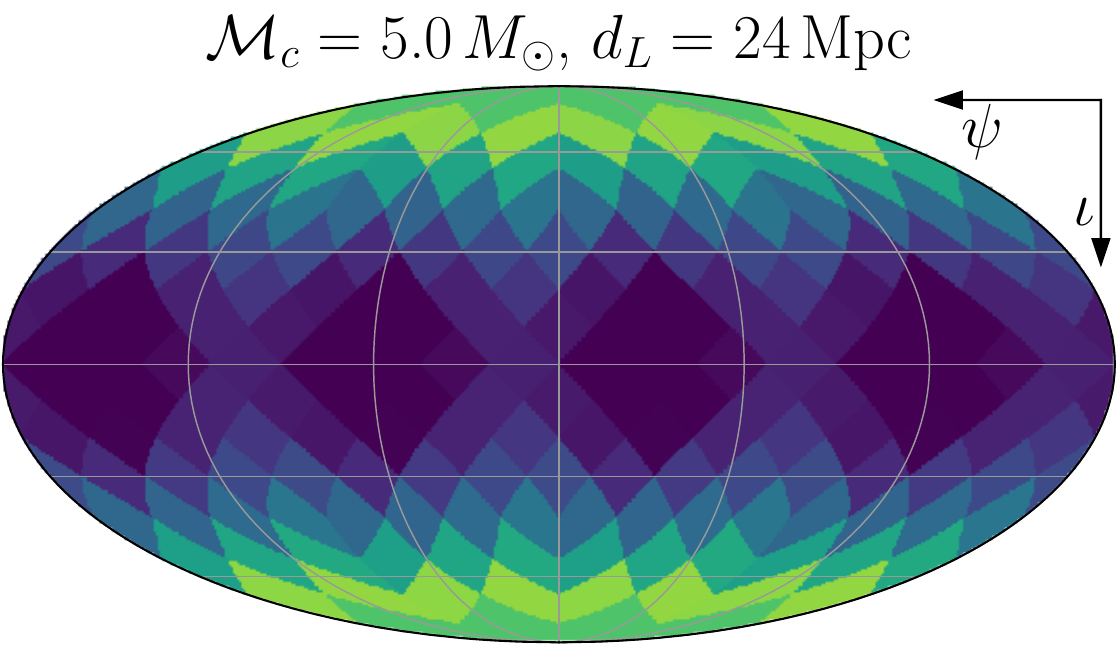}
   \vspace{0.1cm}
   
   \includegraphics[trim=1.3cm 4.7cm 0.9cm 0cm, clip, width=0.4\linewidth]{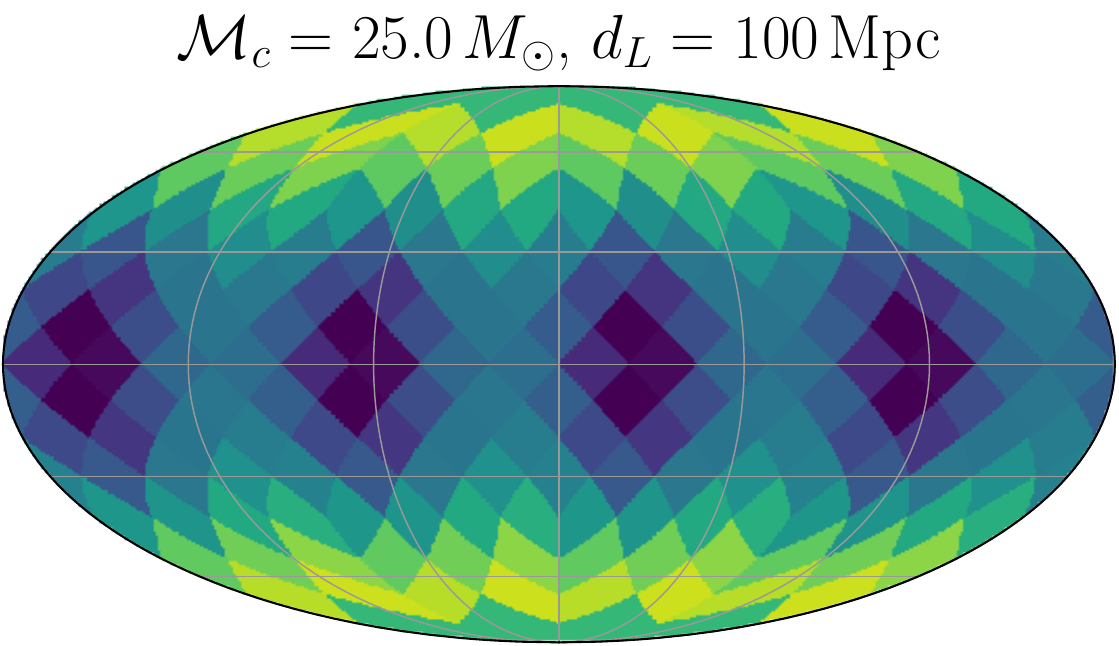}
   \hspace{1cm}
   \includegraphics[trim=1.3cm 4.7cm 0.9cm 0cm, clip, width=0.4\linewidth]{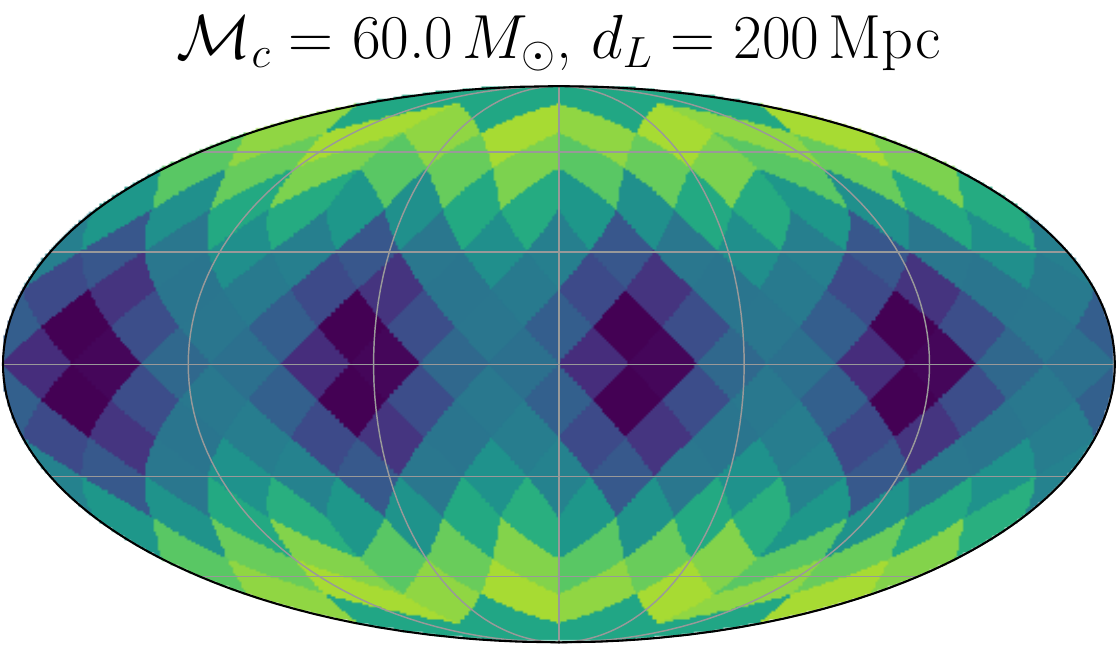}
   
   \includegraphics[trim=0cm 0cm 0cm 11.5cm, clip, width=0.43\linewidth]{Figs/Lmap_dL_Ground_Mc60_f}
   \vspace{-0.5cm}
   \caption{{\bf \boldmath Relative luminosity distance error [$\sigma(d_L)/d_L$]} as a function of the orientation of the binary's orbital momentum for a \TMN network, otherwise the same as Fig.~\ref{fig:Lmap_SNR_Ground}.} 
   \label{fig:Lmap_dL_Ground}
\end{figure}

In Fig.~\ref{fig:ChirpMass_tc_Scans}, we show the uncertainty with which the merger time can be measured in the $\mathcal{M}_c$--$d_L$ plane. Note the similarity of the behavior apparent in Fig.~\ref{fig:ChirpMass_tc_Scans} with the scaling of sky localization uncertainty discussed in Sec.~\ref{sec:ResultsDiscussion}; see Fig.~\ref{fig:ChirpMass_Omegan_Scans}. As we discussed in Sec.~\ref{sec:Code}, in our computation, $t_c$ is defined in an inertial frame (co-moving with the solar system) located at the position of the center of Earth at the mid-point in time between the signal entering and leaving the sensitivity band of the detector to compute the GW signal. The detector measures a quantity related to $t_c$ but defined in its accelerated frame; the relation of that quantity to $t_c$ involves a frame transformation that requires knowing the sky-position of the source. This leads to degeneracy between the time of merger and the source's location in the sky, translating the behavior of Fig.~\ref{fig:ChirpMass_Omegan_Scans} into that of Fig.~\ref{fig:ChirpMass_tc_Scans}. Nonetheless, for any source that can be seen, the merger time can be predicted with $\sigma_{t_c} \lesssim 10\,$sec.

We next consider the dependence of the luminosity distance uncertainty on the location of a source in the sky, which is plotted in Fig.~\ref{fig:skymap_dL_Space} for ``MAGIS-space'' and Fig.~\ref{fig:skymap_dL_Ground} for terrestrial ``MAGIS-1\,km'' detectors. The primary behavior observable in these plots is simply the result of variation in SNR; see Figs.~\ref{fig:skymap_SNR_Space} and~\ref{fig:skymap_SNR_Ground}. Beyond this SNR dependence, the most notable feature is the significantly increased luminosity distance uncertainty for the $\mathcal{M}_c = 1.1\,M_\odot$ source. As we discuss in App.~\ref{app:Signal}, this is due to a degeneracy between $\iota$ and $d_L$. Recall that ``MAGIS-space'' is sensitive to smaller GW frequencies than ``MAGIS-1\,km'' and, hence, observes signals generated deeper in the inspiral phase of a binary with given $\mathcal{M}_c$. While higher-order PN effects do break the $\iota$--$d_L$ degeneracy, such PN corrections to the signal are suppressed the deeper in the inspiral phase the GW signal is generated. Hence, the ability of a detector to measure the luminosity to a source from signals generated during the inspiral phase depreciates for signals with small $\mathcal{M}_c$, leading to the behavior we can observe in the top-left panel of Fig.~\ref{fig:skymap_dL_Space}.

Finally, we turn to a property of the GW source to which we have paid little attention in the rest of this work: the orientation of the binary itself, parameterized by the direction of its angular momentum ${\bm L}$ (via the angle $\iota$ relative to the line of sight to the binary, and the angle $\psi$ around it; see App.~\ref{app:Signal}). We first present the solid angle uncertainty on this orientation as a function of the source's location in the sky, with the results for a ``MAGIS-space'' detector in Fig.~\ref{fig:skymap_Lres_Space} and for terrestrial ``MAGIS-1\,km'' detectors in Fig.~\ref{fig:skymap_Lres_Ground}. Essentially the only observable features in both sets of plots correspond to variation in the SNR. Note that for the different terrestrial detector networks compared in the bottom row of Fig.~\ref{fig:skymap_Lres_Ground}, there are additional effects arising from growing degeneracies between various parameters for a signal that cannot be measured along a sufficient number of independent baselines.

We can likewise consider the dependence of other parameter uncertainties on the binary orientation. We first show the dependence of SNR on this orientation, with the results for a ``MAGIS-space'' in Fig.~\ref{fig:Lmap_SNR_Space} and for a terrestrial \TMN network in Fig.~\ref{fig:Lmap_SNR_Ground}. Both cases exhibit essentially identical behavior, which is easily understood: SNR is maximal when $\iota$ is far from $\pi/2$, since the GW amplitude at the detector is minimal at $\iota=\pi/2$; see Eq.~\eqref{eq:Amplitude}. At $\iota=0$ and $\iota=\pi$, the angle $\psi$ becomes unphysical, and thus it has little effect for extreme values of $\iota$. Conversely, for $\iota\sim\pi/2$, some dependence on $\psi$ appears due to the varying orientation of the observable GW polarization (namely $+$, in our parameterization) relative to the baseline orientation, with a clearly visible $\psi \to \psi + \pi/2$ periodicity that can be understood from noting that the only effect of rotating ${\bm L}$ by $90^\circ$ in the plane orthogonal to the direction of the source on the sky (${\bm n}$) is to change the phase of the GW signal by $\pi/2$.

Almost all of the binary parameters' uncertainties depend on $\bm L$ only through its effect on the SNR, discussed above. The sole exception to this is the luminosity distance; those results are shown in Figs.~\ref{fig:Lmap_dL_Space} and~\ref{fig:Lmap_dL_Ground} for ``MAGIS-space'' and terrestrial ``MAGIS-1\,km'' detectors, respectively. Again, the behavior here is easily interpretable: while the SNR is largest when $\iota$ is far from $\pi/2$, the $\iota$--$d_L$ degeneracy discussed multiple times throughout this work is most severe as $\iota \to 0$ and $\iota \to \pi$; see the leading-order expression of the polarization waveform amplitudes in Eq.~\eqref{eq:Amplitude}. 

Note that our benchmark choices for $\iota$ and $\psi$ we use in the rest of the paper lead to fairly typical values of SNR and the parameter uncertainties, picking neither a particularly ideal nor disadvantageous orientation of the binary.

\bibliographystyle{JHEP.bst}
\bibliography{main}

\end{document}